\documentclass{MYpinchcrn}    

\usepackage{amsmath}
\usepackage{amssymb} 
\usepackage{pstricks}
\def\bc{\begin{center}}
\def\ec{\end{center}}
\newcommand{\be}{\begin{equation}}
\newcommand{\ee}{\end{equation}}
\newcommand{\ba}{\begin{array}{c}}
\def\bat{\begin{array}{cc}}
\newcommand{\ea}{\end{array}}
\newcommand{\beqn}{\begin{eqnarray}}
\newcommand{\eeqn}{\end{eqnarray}}

\newcommand{\bi}{\begin{itemize}}
\newcommand{\ei}{\end{itemize}}
\def\eqn#1{(\ref{#1})}
\def\bel#1{\be\label{#1}}
\def\bi{\begin{itemize}}
\def\ei{\end{itemize}}
\newcommand{\lrder}{\stackrel{\leftrightarrow}{\partial}}
\def\cA{{\mathcal{A}}}
\def\cD{{\mathcal{D}}}
\def\cF{{\mathcal{F}}}
\def\cG{{\mathcal{G}}}
\def\cK{{\mathcal{K}}}
\def\cL{{\mathcal{L}}}
\def\cM{{\mathcal{M}}}
\def\cO{{\mathcal{O}}}
\def\cP{{\mathcal{P}}}
\def\cQ{{\mathcal{Q}}}
\def\cT{{\mathcal{T}}}
\def\cY{{\mathcal{Y}}}
\newcommand{\dg}{\dagger}
\newcommand{\no}{\nonumber}
\newcommand{\lsim}{\stackrel{<}{_\sim}}

\def\e{{\mathrm{e}}}
\def\toG{\,\stackrel{G}{\longrightarrow\,}}
\newcommand{\ket}{\,\rangle}
\newcommand{\bra}{\langle \,}
\title{Effective Field Theory with Nambu--Goldstone Modes}

\author{A. Pich}
\affiliation{Departament de F\'{\i}sica Te\`orica, IFIC, Universitat de Val\`encia –- CSIC
Apt.  Correus 22085, E-46071 Val\`encia, Spain}

\begin{document}

\begin{titlepage}
\mbox{}
\vspace{2cm}

\begin{center}
{\huge\bf Effective Field Theory with\\[10pt] Nambu--Goldstone Modes}
\vskip 1cm
{\large\bf Antonio Pich}
\vskip .5cm
Departament de F\'{\i}sica Te\`orica, IFIC, Universitat de Val\`encia –- CSIC\\
Apt.  Correus 22085, E-46071 Val\`encia, Spain
\vskip 2cm
{\bf Abstract}
\end{center}

These lectures provide an introduction to the low-energy dynamics of Nambu--Goldstone fields, associated with some spontaneous (or dynamical) symmetry breaking, using the powerful methods of effective field theory. The generic symmetry properties of these massless modes are described in detail and two very relevant phenomenological applications are worked out: chiral perturbation theory, the low-energy effective theory of QCD, and the (non-linear) electroweak
effective theory. The similarities and differences between these two effective theories are emphasized, and their current status is reviewed. Special attention is given to the short-distance dynamical information encoded in the low-energy couplings of the effective Lagrangians. The successful methods developed in QCD could help us to uncover fingerprints of new physics scales from future measurements of the electroweak effective theory couplings.

\end{titlepage}

\tableofcontents

\maintext

\chapter{Effective Field Theory with Nambu--Goldstone Modes}

Field theories with spontaneous 
symmetry breaking (SSB) provide an ideal environment to apply the techniques of effective field theory (EFT). They contain massless Nambu--Goldstone modes,  separated from the rest of the spectrum by an energy gap.
The low-energy dynamics of the massless fields can then be analysed through an expansion in powers of momenta over some characteristic mass scale. Owing to the Nambu--Goldstone nature of the light fields, the resulting effective theory is highly constrained by the pattern of symmetry breaking.

Quantum Chromodynamics (QCD) and the electroweak Standard Model (SM) are two paradigmatic field theories where symmetry breaking plays a critical role. If quark masses are neglected, the QCD Lagrangian has a global chiral symmetry that gets dynamically broken through a non-zero vacuum expectation value of the $\bar q q$ operator. With $n_f=2$ light quark flavours, there are three associated Nambu--Goldstone modes that can be identified with the pion multiplet. The symmetry breaking mechanism is quite different in the SM case, where the electroweak gauge theory is spontaneously broken through a scalar potential with non-trivial minima. Once a given choice of the (non-zero) scalar vacuum expectation value is adopted, the excitations along the flat directions of the potential give rise to three massless Nambu--Goldstone modes, which in the unitary gauge become the longitudinal polarizations of the $W^\pm$ and $Z$ gauge bosons. In spite of the totally different underlying dynamics (non-perturbative dynamical breaking versus perturbative spontaneous symmetry breaking), the low-energy interactions of the Nambu--Goldstone modes are formally identical in the two theories because they share the same pattern of symmetry breaking.

These lectures provide an introduction to the effective field theory description of the Nambu--Goldstone fields, and some important phenomenological applications. A toy model incorporating the relevant symmetry properties is first studied in Section~\ref{sec:sigma}, and the different symmetry realizations, Wigner--Weyl and Nambu--Goldstone,  are discussed in Section~\ref{sec:symmetry}. Section~\ref{sec:ChiralSymmetry} analyses the chiral symmetry of massless QCD. The corresponding Nambu--Goldstone EFT is developed next in Section~\ref{sec:NG_Lagrangian}, using symmetry properties only, while Section~\ref{sec:ChPT} discusses the explicit breakings of chiral symmetry and presents a detailed description of chiral perturbation theory ($\chi$PT), the low-energy effective realization of QCD. A few phenomenological applications are presented in Section~\ref{sec:phenomenology}. The quantum chiral anomalies are briefly touched in Section~\ref{sec:anomalies}, and Sections~\ref{sec:MassiveFields} and \ref{sec:LargeNC} are devoted to the dynamical understanding of the $\chi$PT couplings.

The electroweak symmetry breaking is analysed in Section~\ref{sec:EWSB}, which discusses the custodial symmetry of the SM scalar potential and some relevant phenomenological consequences. Section~\ref{sec:EWET} presents the electroweak effective theory formalism, while the short-distance information encoded in its couplings is studied in Section~\ref{sec:fingerprints}. A few summarizing comments are finally given in Section~\ref{sec:Summary}.

To prepare these lectures I have made extensive use of my previous reviews on effective field theory \cite{Pich:1998xt}, $\chi$PT \cite{Pich:1995bw,Pich:2002xy,Pich:2004vn} and electroweak symmetry breaking \cite{Pich:2015lkh}. Complementary information can be found in many excellent reports \cite{Bernard:2006gx,Bijnens:2006zp,Brivio:2017vri,Contino:2010rs,Ecker:1994gg,Feruglio:1992wf,Georgi:1994qn,Manohar:1996cq,Manohar:1998xv,Scherer:2002tk}
and books \cite{Donoghue:1992dd,Scherer:2012xha}, covering related subjects.

\section{A toy Lagrangian: the linear sigma model}
\label{sec:sigma}

\begin{figure}[t]
\begin{center}
\includegraphics[width=5cm]{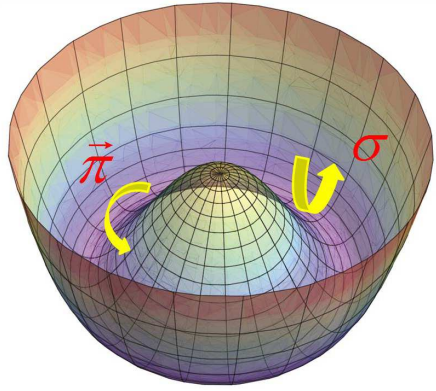}
\end{center}
\caption{Sigma-model potential.}
\label{fig:SigmaModel}
\end{figure}

Let us consider a multiplet of four real scalar fields $\Phi(x)^T \equiv (\vec \pi , \sigma)$, described by the Lagrangian
\be\label{eq:sigma}
\cL_\sigma\, =\, \frac{1}{2}\,\partial_\mu\Phi^T\partial^\mu\Phi - \frac{\lambda}{4}\,
\left(\Phi^T\Phi - v^2\right)^2\, .
\ee
$\cL_\sigma$ remains invariant under ($x^\mu$-independent) $SO(4)$ rotations of the four scalar components. If $v^2$ were negative, this global symmetry would be realised in the usual Wigner--Weyl way, with four degenerate states of mass $m_\Phi^2=-\lambda v^2$.
However, for $v^2>0$, the potential has a continuous set of minima, occurring for all field configurations with $\Phi^T\Phi = v^2$. 
This is illustrated in Fig.~\ref{fig:SigmaModel} that shows the analogous three-dimensional potential.
These minima correspond to degenerate ground states, which transform into each other under $SO(4)$ rotations. Adopting the vacuum choice
\be\label{eq:choice}
\langle 0|\sigma|0\rangle = v\, , \qquad\qquad\qquad
\langle 0|\vec{\pi}|0\rangle = 0\, ,
\ee
and making the field redefinition $\hat{\sigma} = \sigma - v$, the Lagrangian takes the form
\be\label{eq:sigma2}
\cL_\sigma  =  {1\over 2}
\left[ \partial_\mu\hat{\sigma}\partial^\mu\hat{\sigma}
- 2 \lambda v^2 \hat{\sigma}^2
+ \partial_\mu\vec{\pi}\partial^\mu\vec{\pi} \right]
 - \lambda v\, \hat{\sigma} \left( \hat{\sigma}^2 + \vec{\pi}^2\right)
- {\lambda\over 4} \left( \hat{\sigma}^2 + \vec{\pi}^2\right)^2 ,
\ee
which  shows that the three $\vec{\pi}$ fields are massless Nambu--Goldstone modes, corresponding to excitations along the three flat directions of the potential, while $\hat{\sigma}$ acquires a mass $M^2 = 2 \lambda v^2$.

The vacuum choice \eqn{eq:choice} is only preserved by $SO(3)$ rotations among the 
first three field components, leaving the fourth one untouched. Therefore, the potential triggers an $SO(4)\to SO(3)$ spontaneous symmetry breaking, and there are three ($\frac{4\times 3}{2}-\frac{3\times 2}{2}$) broken generators that do not respect the adopted vacuum. 
The three Nambu--Goldstone fields are associated with these broken generators.

To better understand the role of symmetry on the Goldstone dynamics, it is useful to rewrite the sigma-model Lagrangian with different field variables. Using the $2\times 2$ matrix notation
\be\label{eq:notation}
\Sigma(x) \,\equiv\, \sigma(x)\, I_2 + i\, \vec{\tau}\,\vec{\pi}(x)\, ,
\ee
with $\vec\tau$ the three Pauli matrices and $I_2$ the identity matrix,
the Lagrangian (\ref{eq:sigma}) can be compactly written as
\be\label{eq:sigma3}
\cL_\sigma\, =\, {1\over 4}\, \langle \partial_\mu\Sigma^\dagger\partial^\mu\Sigma\rangle
- {\lambda \over 16}
\left( \langle \Sigma^\dagger\Sigma\rangle - 2 v^2 \right)^2 ,
\ee
where $\langle A \rangle$ denotes the trace of the matrix $A$.
In this notation, $\cL_\sigma$ is explicitly invariant under
global $G\equiv SU(2)_L\otimes SU(2)_R$ transformations,
\be
\Sigma \, \stackrel{G}{\longrightarrow} \,
g_R^{\phantom{\dagger}} \,\Sigma\, g_L^\dagger\, , \qquad\qquad
g_{L,R}^{\phantom{\dagger}}  \in SU(2)_{L,R}\, ,
\ee
while the vacuum choice $\langle 0|\Sigma |0\rangle = v\, I_2$ only remains invariant under those transformations satisfying $g_L^{\phantom{\dagger}} = g_R^{\phantom{\dagger}}$, {\it i.e.}, under the diagonal subgroup $H\equiv SU(2)_{L+R}$. Therefore, the pattern of symmetry breaking is
\bel{eq:SymBreak}
SU(2)_L\otimes SU(2)_R\,\longrightarrow\, SU(2)_{L+R}\, .
\ee
The change of field variables just makes manifest the equivalences of the groups $SO(4)$ and
$SO(3)$ with $SU(2)_L\otimes SU(2)_R$ and $SU(2)_{L+R}$, respectively. The physics content is of course the same.

We can now make the polar decomposition
\be\label{eq:polar}
\Sigma(x) \, = \, \left[ v + S(x) \right] \, U(\vec{\phi\,})\, , 
\qquad\qquad
 U(\vec{\phi\,}) \, = \, \exp{\left\{ i \,\frac{\vec{\tau}}{v}\, \vec{\phi}(x) \right\} } ,\quad
\ee
in terms of an Hermitian scalar field $S(x)$ and three pseudoscalar variables $\vec{\phi}(x)$, normalized with the scale $v$ in order to preserve the canonical dimensions.
$S(x)$ remains invariant under the symmetry group, while the matrix $U(\vec{\phi\,})$ inherits the chiral transformation of $\Sigma(x)$:
\be\label{eq:transf}
S \, \stackrel{G}{\longrightarrow} \, S \, ,
\qquad\qquad\qquad
U(\vec{\phi\,}) \, \stackrel{G}{\longrightarrow} \, g_R^{\phantom{\dagger}}\, U(\vec{\phi\,})\, g_L^\dagger\, .
\ee
Obviously, the fields $\vec{\phi}(x)$ within the exponential
transform non-linearly.
The sigma-model Lagrangian takes then a very enlightening form:
\be\label{eq:sigma4}
\cL_\sigma\,  =\,  {v^2\over 4} \left ( 1 + {S\over v} \right)^2
\langle \partial_\mu U^\dagger \partial^\mu U \rangle
 + {1\over 2}
\left( \partial_\mu S \,\partial^\mu S - M^2 S^2\right)
- \frac{M^2}{2 v} \, S^3 - \frac{M^2}{8 v^2}\, S^4\, .
\ee
This expression shows explicitly the following important properties:
\bi
\item The massless Nambu--Goldstone bosons $\vec\phi$, parametrized through the matrix 
$U(\vec{\phi\,})$, have purely derivative couplings. Therefore, their scattering amplitudes vanish at zero momenta. This was not so obvious in eqn~(\ref{eq:sigma2}), and implies that this former expression of $\cL_\sigma$ gives rise to exact (and not very transparent) cancellations among different momentum-independent contributions. The two functional forms of the Lagrangian should of course give the same physical predictions.

\item The potential only depends on the radial variable $S(x)$, which describes a massive field with $M^2=2\lambda v^2$. In the limit $\lambda \gg 1$, the scalar field $S$ becomes very heavy and can be integrated out from the Lagrangian. The linear sigma model then reduces to
\be\label{eq:universal}
\cL_2\, =\, \frac{v^2}{4}\,
\langle \partial_\mu U^\dagger \partial^\mu U \rangle \, ,
\ee
which contains an infinite number of interactions among the $\vec\phi$ fields, owing to the non-linear functional form of $U(\vec{\phi\,})$. As we will see later, $\cL_2$ is a direct consequence of the pattern of SSB in \eqn{eq:SymBreak}. It represents a universal (model-independent) interaction of the Nambu--Goldstone modes at very low energies.

\item In order to be sensitive to the particular dynamical structure
of the potential, and not just to its symmetry properties, one needs to test the model-dependent part involving the scalar field $S$. At low momenta ($p << M$), the dominant tree-level corrections originate from $S$ exchange, which generates the four-derivative term
\be\label{eq:sigma5}
\cL_4\, =\, {v^2\over 8 M^2}\,
\langle \partial_\mu U^\dagger \partial^\mu U \rangle^2 .
\ee
The corresponding contributions to the low-energy scattering amplitudes are suppressed by a factor $p^2/M^2$ with respect to the leading contributions from \eqn{eq:universal}.
\ei

One can easily identify $\cL_\sigma$ with the (non-gauged) scalar Lagrangian of the electroweak SM. However, the non-linear sigma model was originally suggested   to describe the low-energy dynamics of the QCD pions \cite{GellMann:1960np,Schwinger:1957em}. Both theories have the pattern of symmetry breaking displayed in eqn~\eqn{eq:SymBreak}.

\section{Symmetry realizations}
\label{sec:symmetry}

Noether's theorem guarantees the existence of conserved quantities associated with any continuous symmetry of the action. If a group G of field transformations leaves the Lagrangian invariant, for each generator of the group $T^a$, there is a conserved current $j^\mu_a(x)$ such that $\partial_\mu j^\mu_a=0$ when the fields satisfy the Euler--Lagrangian equations of motion. The space integrals of the time-components $j^0_a(x)$ are then conserved charges, independent of the time coordinate:
\bel{eq:Noether}
\cQ_a\, =\, \int d^3x\; j^0_a(x)\, ,
\qquad\qquad\qquad
\frac{d}{dt}\,\cQ_a\, =\, 0\, .
\ee
In the quantum theory, the conserved charges become symmetry generators that implement the group of transformations through the unitary operators $U = \exp{\{i\theta^a \cQ_a\}}$, being $\theta^a$ the continuous parameters characterizing the transformation. These unitary operators commute with the Hamiltonian, {\it i.e.}, $U H U^\dagger = H$.

In the usual Wigner--Weyl realization of the symmetry, the charges annihilate the vacuum, $\cQ_a |0\rangle = 0$, so that it remains invariant under the group of transformations: $U |0\rangle = |0\rangle$.
This implies the existence of degenerate multiplets in the spectrum. Given a state
$|A\rangle = \phi_A^\dagger |0\rangle$, the symmetry transformation $U\phi_A^{\phantom{\dagger}} U^\dagger = \phi_B^{\phantom{\dagger}}$ generates another state $|B\rangle = \phi_B^\dagger |0\rangle = U |A\rangle $ with the same energy:
\be
E_B\, =\, \langle B | H | B \rangle \, =\,\langle A | U^\dagger H U | A \rangle \, =\,
\langle A | H | A \rangle \, =\, E_A\, .
\ee 

The previous derivation is no-longer valid when the vacuum is not invariant under some group transformations. Actually, if a charge does not annihilate the vacuum, $\cQ_a |0\rangle$ is not even well defined because
\be
\langle 0 |\cQ_a\cQ_b | 0\rangle\, =\,\int d^3 x\; \langle 0 |j^0_a(x)\cQ_b | 0\rangle\, =\,\langle 0 |j^0_a(0)\cQ_b | 0\rangle\;\int d^3 x \, =\,\infty\, ,
\ee
where we have made use of the invariance under translations of the space-time coordinates, which implies
\bel{eq:TransInv}
j^\mu_a(x)\, =\, \e^{i P_\mu x^\mu}\, j^\mu_a(0)\,\e^{-i P_\mu x^\mu}\, ,
\ee
with $P_\mu$ the four-momentum operator that satisfies $[ P_\mu, \cQ_b ] = 0$ and $P_\mu |0\rangle = 0$. Thus, one needs to be careful and state the vacuum properties of $Q_a$ in terms of commutation relations that are mathematically well defined. One can easily proof the following important result \cite{Goldstone:1961eq,Goldstone:1962es,Nambu:1960xd,Nambu:1960tm,Nambu:1961tp,Nambu:1961fr}.

\medskip
\noindent {\bf Nambu--Goldstone theorem:} 
Given a conserved current $j^\mu_a(x)$ and its corresponding conserved charge $\cQ_a$, if there exists some operator $\cO$ such that $v_a\equiv \langle 0 | [ \cQ_a , \cO ] | 0\rangle \not= 0$, then the spectrum of the theory contains a massless state $|\phi_a\rangle$  that couples both to $\cO$ and $j_a^0$, {\it i.e.},
$\langle 0 | \cO | \phi_a\rangle\,\langle \phi_a | j_a^0(0) | 0\rangle\not= 0$.

\begin{proof}
Using \eqn{eq:Noether}, \eqn{eq:TransInv} and the completeness relation $\sum_n |n\rangle \langle n | = 1$, where the sum is over the full spectrum of the theory, the non-zero vacuum expectation value can be written as
\beqn
v_a\, & =\, &  \sum_n\int d^3x\;\left\{
\langle 0|j_a^0(x)|n\rangle\langle n|\cO|0\rangle\, - \,
\langle 0|\cO|n\rangle\langle n|j_a^0(x)|0\rangle\right\}
\no\\ 
& =\, &
\sum_n\int d^3x\;\left\{ \e^{-ip_n\cdot x}\,
\langle 0|j_a^0(0)|n\rangle\langle n|\cO|0\rangle \, - \,
\e^{ip_n\cdot x}\,\langle 0|\cO|n\rangle\langle n|j_a^0(0)|0\rangle\right\}
\no\\ 
& =\, &
(2\pi)^3\sum_n \delta^{(3)}(\vec{p}_n)\,\left\{\e^{-iE_nt}\,
\langle 0|j_a^0(0)|n\rangle\langle n|\cO|0\rangle \, - \,
\e^{iE_nt}\,\langle 0|\cO|n\rangle\langle n|j_a^0(0)|0\rangle\right\}
\,\not=\, 0\, .
\no\eeqn
Since $\cQ_a$ is conserved, $v_a$ should be time independent. Therefore, taking a derivative with respect to $t$,
\begin{displaymath}
0\, =\, -i (2\pi)^3\sum_n \delta^{(3)}(\vec{p}_n)\, E_n\,\left\{ \e^{-iE_nt}\,
\langle 0|j_a^0(0)|n\rangle\langle n|\cO|0\rangle
+ \e^{iE_nt}\,\langle 0|\cO|n\rangle\langle n|j_a^0(0)|0\rangle\right\} .
\end{displaymath}
The two equations can only be simultaneously true if there exist a state 
$|n\rangle \equiv |\phi_a\rangle$ such that $\delta^{(3)}(\vec{p}_n)\, E_n = 0$ ({\it i.e.}, a massless state) and 
$\langle 0 | \cO | n\rangle\,\langle n | j_a^0(0) | 0\rangle\not= 0$.
\end{proof}

The vacuum expectation value $v_a$ is called an order parameter of the symmetry breaking. Obviously, when $\cQ_a |0\rangle = 0$ the parameter $v_a$ is trivially zero for all operators of the theory. Notice that we have proved the existence of massless Nambu--Goldstone modes without making use of any perturbative expansion. Thus, the theorem applies to any (Poincare-invariant) physical system where a continuous symmetry of the Lagrangian is broken by the vacuum, either spontaneously or dynamically.

\section{Chiral symmetry in massless QCD}
\label{sec:ChiralSymmetry}
 
Let us consider $n_f$ flavours of massless quarks, collected in a vector field in flavour space: $q^T = (u,d,\ldots)$. Colour indices are omitted, for simplicity. The corresponding QCD Lagrangian can be compactly written in the form:
\bel{eq:LQCD}
{\cL}_{\mathrm{QCD}}^0 = -{1\over 4}\, G^a_{\mu\nu} G^{\mu\nu}_a
 + i \bar q_L^{\phantom{\dagger}} \gamma^\mu D_\mu q_L^{\phantom{\dagger}}  + i \bar q_R^{\phantom{\dagger}} \gamma^\mu D_\mu q_R^{\phantom{\dagger}}\, ,
\ee
with the gluon interactions encoded in the flavour-independent covariant derivative $D_\mu$. In the absence of a quark mass term, the left and right quark chiralities separate into two different sectors that can only communicate through gluon interactions. The QCD Lagrangian is then invariant under
independent global $G\equiv SU(n_f)_L\otimes SU(n_f)_R$
transformations of the left- and right-handed quarks in flavour space:\footnote{ 
Actually, the Lagrangian \eqn{eq:LQCD}
has a larger $U(n_f)_L\otimes U(n_f)_R$ global symmetry. However, the
$U(1)_A$ part is broken by quantum effects (the $U(1)_A$ anomaly),
while the quark-number symmetry $U(1)_V$ is trivially realised 
in the meson sector.}
%
\bel{eq:qrot}
q_L^{\phantom{\dagger}}  \toG  g_L^{\phantom{\dagger}} \, q_L^{\phantom{\dagger}}\,  , \qquad
q_R^{\phantom{\dagger}}  \toG  g_R^{\phantom{\dagger}} \, q_R^{\phantom{\dagger}}\,  , \qquad
g_{L,R}^{\phantom{\dagger}} \in SU(n_f)_{L,R} \, .
\ee
The Noether currents associated with the chiral group $G$ are:
\bel{eq:noether_currents}
J^{a\mu}_X\, =\, \bar q_X^{\phantom{\dagger}} \gamma^\mu 
T^a q_X^{\phantom{\dagger}} \, ,
\qquad\qquad (X = L,R;\quad a = 1,\,\ldots,\, n_f^2-1) ,
\ee
where $T^a$ denote the $SU(n_f)$ generators that fulfil the Lie algebra
$[ T^a , T^b ] = i f_{abc} T^c$. 
The corresponding Noether charges $\cQ^a_X$
satisfy the commutation relations
\bel{eq:commutation_relations}
[\cQ_X^a,\cQ_Y^b] \, =\, i\, \delta_{XY}\, f_{abc}\, \cQ^c_X ,
\ee
involving the $SU(n_f)$ structure constants $f_{abc}$.
These algebraic relations were the starting point of the successful Current-Algebra
methods of the sixties \cite{Adler:1968xxx,deAlfaro:1973zz}, before the development of QCD.

The chiral symmetry \eqn{eq:qrot}, which should be approximately good in the light quark sector ($u$,$d$,$s$), is however not seen in the hadronic spectrum. Since parity exchanges left and right, a normal Wigner--Weyl realization of the symmetry would imply degenerate mirror multiplets with opposite chiralities. However, although hadrons can be nicely classified in $SU(3)_V$ representations, degenerate multiplets with the opposite parity do not exist. Moreover, the octet of pseudoscalar mesons is much  lighter than all the other hadronic states. These empirical facts clearly indicate that the vacuum is not symmetric under the full chiral group. Only those transformations with $g_R^{\phantom{\dagger}} = g_L^{\phantom{\dagger}}$ remain a symmetry of the physical QCD vacuum.
Thus, the $SU(3)_L \otimes SU(3)_R$ symmetry dynamically breaks down to
$SU(3)_{L+R}$.

Since there are eight broken axial generators, $\cQ^a_A = \cQ^a_R - \cQ^a_L$, there should be eight pseudoscalar Nambu--Goldstone states
$|\phi^a\rangle$, which we can identify with the eight lightest hadronic states
($\pi^+$, $\pi^-$, $\pi^0$, $\eta$, $K^+$, $K^-$, $K^0$ 
and $\bar{K}^0$). Their small masses are generated by the quark-mass matrix,
which explicitly breaks the global chiral symmetry of the QCD Lagrangian.
The quantum numbers of the Nambu--Goldstone bosons are dictated by those of the broken axial currents $J^{a\mu}_A$ and the operators $\cO^b$ that trigger the needed non-zero vacuum expectation values, because
$\langle 0|J^{a0}_A|\phi^a\rangle \, \langle \phi^a|\cO^b|0\rangle\not= 0$. Therefore $\cO^b$ must be pseudoscalar operators. The simplest possibility is $\cO^b = \bar q \gamma_5 \lambda^b q$, with $\lambda^b$ the set of eight $3\times 3$ Gell-Mann matrices, which satisfy
\bel{eq:vev_relation}
\langle 0|[\cQ^a_A, \bar q \gamma_5 \lambda^b q] |0\rangle \, =\,
-{1\over 2} \,\langle 0|\bar q \{\lambda^a,\lambda^b\} q |0\rangle\, =\,
-{2\over 3} \,\delta_{ab} \,\langle 0|\bar q q |0\rangle \, .
\ee
The quark condensate
\bel{eq:quark_condensate}
\langle 0|\bar u u |0\rangle\, =\,
\langle 0|\bar d d |0\rangle\, =\,
\langle 0|\bar s s |0\rangle \not\, =\, 0
\ee
is then the natural order parameter of the dynamical chiral symmetry breaking ($\chi$SB). The $SU(3)_V$ symmetry of the vacuum guarantees that this order parameter is flavour independent.

With $n_f = 2$, $q^T = (u, d)$, one recovers the pattern of $\chi$SB in eqn~\eqn{eq:SymBreak}. The corresponding three Nambu--Goldstone bosons are obviously identified with the pseudoscalar pion multiplet.

\section{Nambu--Goldstone effective Lagrangian}
\label{sec:NG_Lagrangian}

Since there is a mass gap separating the Nambu--Goldstone bosons from the
rest of the spectrum, we can build an EFT containing only the massless modes.
Their Nambu--Goldstone nature implies strong constraints on their interactions, which can be most easily analysed on the basis of an effective Lagrangian, expanded in powers of momenta over some characteristic scale, with the only assumption of the pattern of symmetry breaking $G\to H$. In order to proceed we need first to choose a good parametrization of the fields.

\subsection{Coset-space coordinates}
\label{subsec:CosetSpace}

Let us consider the $O(N)$ sigma model, described by the Lagrangian \eqn{eq:sigma}, where now $\Phi(x)^T\equiv (\varphi_1,\varphi_2,\cdots,\varphi_N)$ is an $N$-dimensional vector of real scalar fields. The Lagrangian has a global $O(N)$ symmetry, under which $\Phi(x)$ transforms as an $O(N)$ vector, and a degenerate ground-state manifold composed by all field configurations satisfying $|\Phi|^2 = \sum_i \varphi_i^2 = v^2$. This vacuum manifold is the $N-1$ dimensional sphere $S^{N-1}$, in the $N$-dimensional space of scalar fields. Using the $O(N)$ symmetry, we can always rotate the vector $\langle 0|\Phi|0\rangle$ to any given direction, which can be taken to be
\bel{eq:ONvev}
\Phi_0^T\,\equiv\,\langle 0|\Phi|0\rangle^T\, =\, (0,0,\cdots,0,v)\, .
\ee
This vacuum choice only remains invariant under the $O(N-1)$ subgroup, acting on the first $N-1$ field coordinates. Thus, there is an $O(N)\to O(N-1)$ SSB. Since $O(N)$ has $N(N-1)/2$ generators, while $O(N-1)$ only has $(N-1)(N-2)/2$, there are $N-1$ broken generators $\widehat{T}^a$. The $N-1$ Nambu--Goldstone bosons parametrize the corresponding rotations of $\Phi_0$ over the vacuum manifold $S^{N-1}$. 

Taking polar coordinates, we can express the $N$-component field $\Phi(x)$ in the form
\be
\Phi(x)\, =\, \left(1+\frac{S(x)}{v}\right)\, U(x)\,\Phi_0\, ,
\ee
with $S(x)$ the radial excitation of mass $M^2= 2 \lambda v^2$, and the $N-1$ Nambu--Goldstone fields $\phi_a(x)$ encoded in the matrix
\bel{eq:NGparam} 
U(x)\,=\,\exp{\left\{ i\, \widehat{T}^a\, \frac{\phi_a(x)}{v}\right\}}\, .
\ee
%

\begin{figure}[t]
\begin{center}
\begin{minipage}[t]{5cm}
\begin{center}
\includegraphics[width=4.5cm,clip]{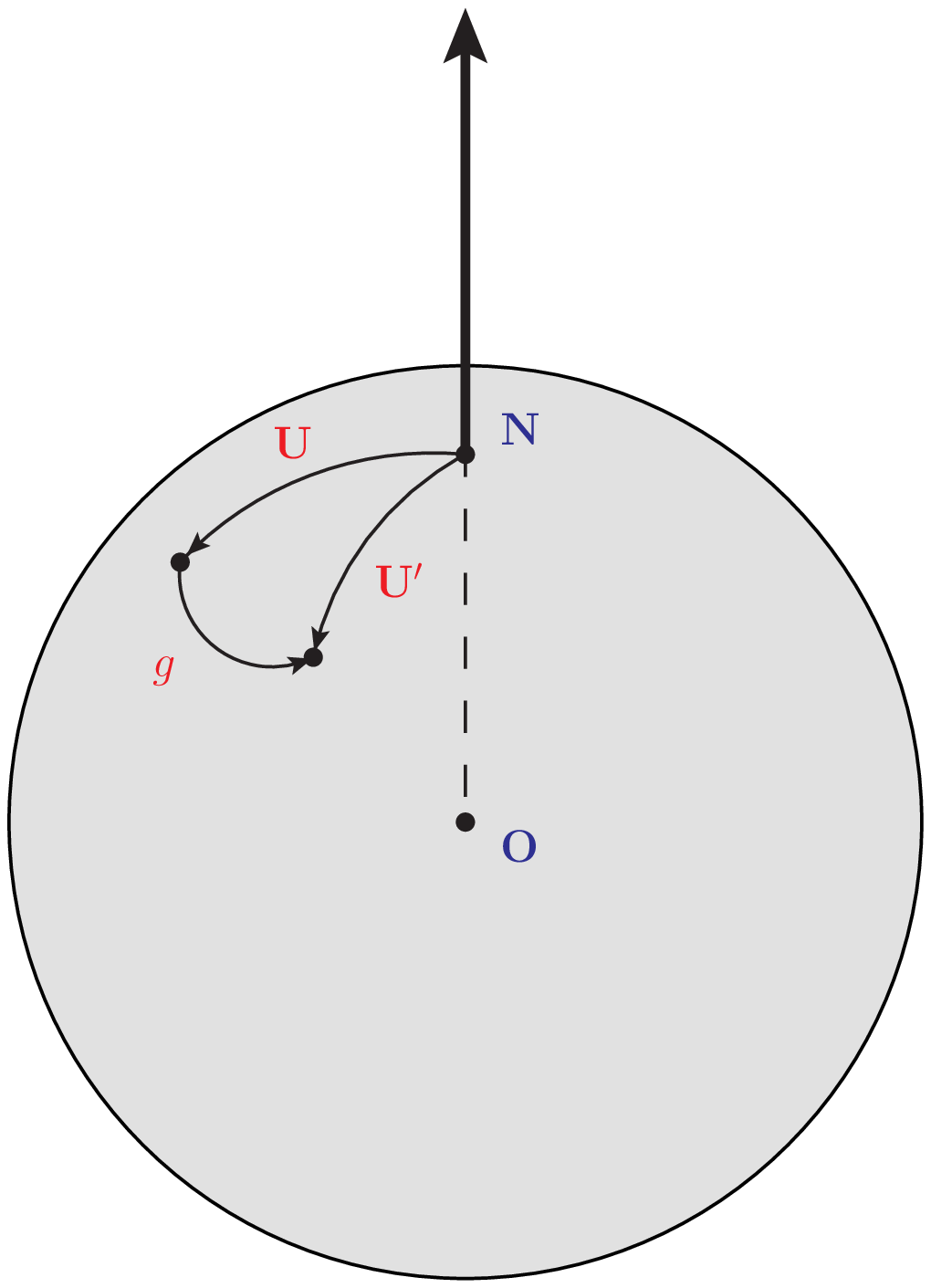}
\end{center}
\caption{Geometrical representation of the vacuum manifold $S^{N-1}$. The arrow indicates the chosen vacuum direction $\Phi_0$.}
\label{fig:NorthPole}
\end{minipage}
\hskip .75cm
\begin{minipage}[t]{7cm}
\begin{center}
\includegraphics[width=6.5cm,clip]{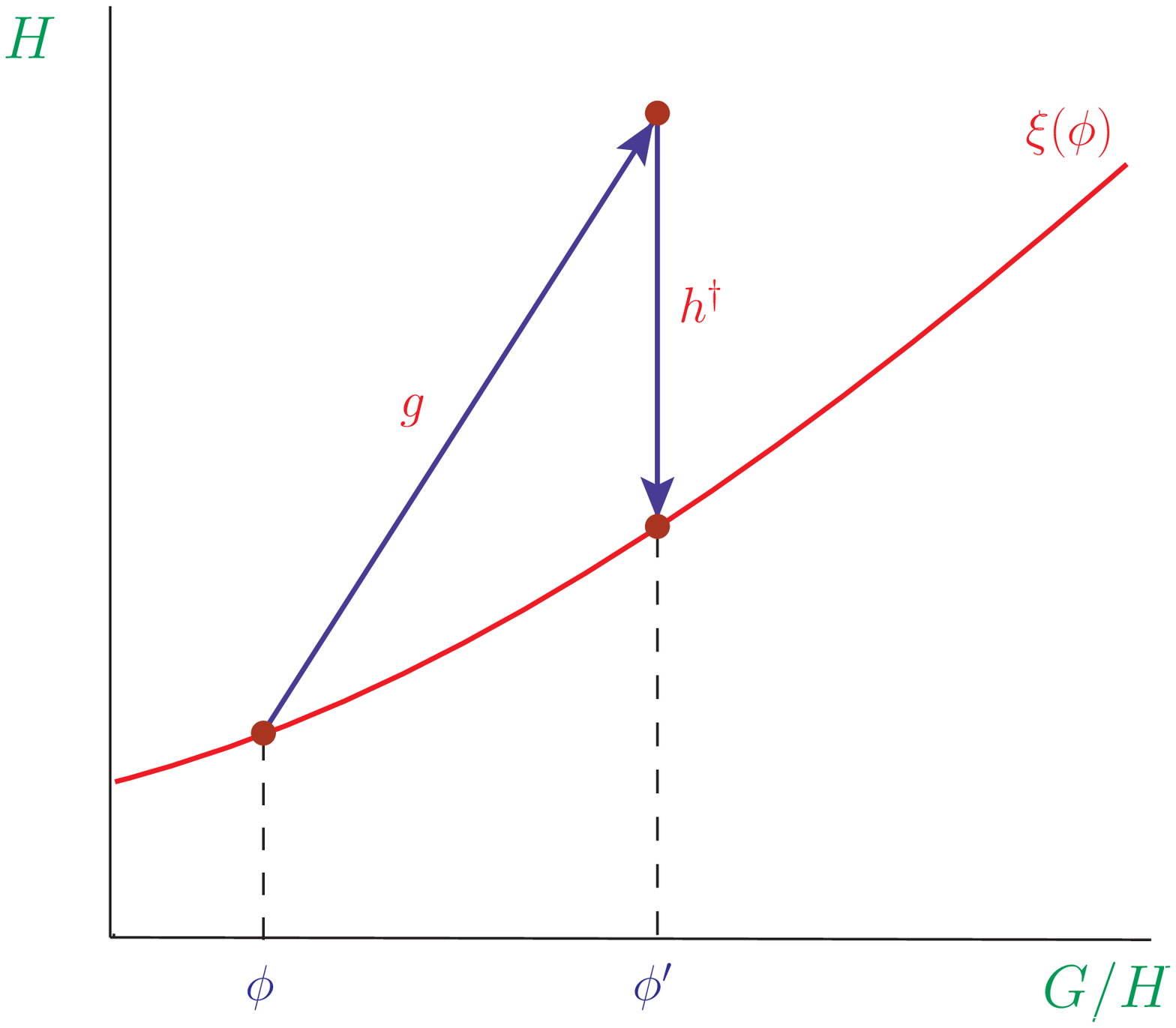}
\end{center}
\caption{
Under the action of $g\in G$, the coset representative $\xi(\vec{\phi\,})$ transforms into some element of the $\vec{\phi'}$ coset. A compensating $h(\vec\phi,g)\in H$ transformation is needed to go back to $\xi(\vec{\phi'})$.}
\label{fig:CosetSpace}
\end{minipage}
\end{center}
\end{figure}

Figure~\ref{fig:NorthPole} displays a geometrical representation of the vacuum manifold $S^{N-1}$, with the north pole of the sphere indicating the vacuum choice $\Phi_0$. This direction remains invariant under any transformation $h\in O(N-1)$, {\it i.e.}, $h\,\Phi_0 = \Phi_0$, while the broken generators $\widehat{T}^a$ induce rotations over the surface of the sphere. Thus, the matrix $U(x)$ provides a general parametrization of the $S^{N-1}$ vacuum manifold. Under a global symmetry transformation $g\in O(N)$, $U(x)$ is transformed into a new matrix $g\, U(x)$ that corresponds to a different point on $S^{N-1}$. However, in general, the matrix $g\, U(x)$ is not in the standard form \eqn{eq:NGparam}, the difference being a  transformation $h\in O(N-1)$:
\bel{eq:Utransf}
g\; U(x)\, =\, U'(x)\; h(g,U)\, .
\ee
This is easily understood in three dimensions: applying two consecutive rotations $g\, U$ over an object in the north pole $N$ is not equivalent to directly making the rotation $U'$; an additional rotation around the ON axis is needed to reach the same final result.
The two matrices $g\, U$ and $U'$ describe the same Nambu--Goldstone configuration over the sphere $S^{N-1}$, but they correspond to different choices of the Goldstone coordinates in the coset space $O(N)/O(N-1)\equiv S^{N-1}$. The compensating transformation $h(g,U)$ is non-trivial because the vacuum manifold is curved; it depends on both the transformation $g$ and the original configuration $U(x)$.

In a more general situation, we have a symmetry group $G$ and a vacuum manifold that is only invariant under the subgroup $H\subset G$, generating a $G\to H$ SSB with
$N =\mathrm{dim}(G) - \mathrm{dim}(H)$ Nambu--Goldstone fields $\vec\phi \equiv (\phi_1,\cdots,\phi_N)$, corresponding to the number of broken generators. The action of the symmetry group on these massless fields is given by some mapping
\bel{eq:Gtransform} 
\vec\phi (x)\,\toG\,\vec{\phi'}(x)\,  = \vec\cF(g,\vec{\phi\,})\, ,
\ee
which depends both on the group transformation $g\in G$ and the vector field $\vec\phi (x)$. This mapping should satisfy $\vec\cF(e,\vec{\phi\,}) = \vec\phi$, where $e$ is the identity element of the group, and the group composition law
$\vec\cF(g_1,\vec\cF(g_2,\vec{\phi\,})) = \vec\cF(g_1 g_2,\vec{\phi\,})$.

Once a vacuum choice $\vec \phi_0^{\phantom{1}}$ has been adopted, the Nambu--Goldstone fields correspond to quantum excitations along the full vacuum manifold. Since the vacuum is invariant under the unbroken subgroup, $\vec\cF(h,\vec\phi_0^{\phantom{1}}) = \vec\phi_0^{\phantom{1}}$ for all $h\in H$. Therefore,
\bel{eq:NGfields}
\vec\phi(x)\, =\, \vec\cF(g,\vec\phi_0^{\phantom{1}})\, =\,\vec\cF(gh,\vec\phi_0^{\phantom{1}})
\qquad\qquad
\forall h\in H\, .
\ee
Thus, the function $\vec \cF$ represents a mapping between the members of the (left) coset equivalence class $gH = \{ g h\; |\; h\in H\}$ and the corresponding field configuration $\vec\phi(x)$. Since the mapping is isomorphic and invertible,\footnote{
If two elements of the coset space $g_1^{\phantom{1}} H$ and $g_2^{\phantom{1}} H$ are mapped into the same field configuration, {\it i.e.}, $\vec\cF(g_1^{\phantom{1}},\vec\phi_0^{\phantom{1}})\, =\, \vec\cF(g_2^{\phantom{1}},\vec\phi_0^{\phantom{1}})$, then  $\vec\cF(g_1^{-1} g_2^{\phantom{1}},\vec\phi_0^{\phantom{1}})\, =\,\vec\phi_0^{\phantom{1}}$, implying that $g_1^{-1} g_2^{\phantom{1}}\in H$ and therefore $g_2^{\phantom{1}}\in g_1^{\phantom{1}} H$.} 
the Nambu--Goldstone fields can be then identified with the elements of the coset space $G/H$.

For each coset, and therefore for each field configuration $\vec\phi(x)$, one can choose an arbitrary group element to be the coset representative $\xi(\vec{\phi\,})$, as shown in Fig.~\ref{fig:CosetSpace} that visualizes the partition of the group elements (points in the plane) into cosets (vertical lines). Under a transformation $g\in G$, $\vec\phi(x)$ changes as indicated in \eqn{eq:Gtransform}, but the group element representing the original field does not get necessarily transformed into the coset representative of the new field configuration $\vec{\phi'}(x)$. In general, one needs a compensating transformation $h(\vec\phi,g)\in H$ to get back to the chosen coset representative \cite{Callan:1969sn,Coleman:1969sm}:
\bel{eq:CCWZ}
\xi(\vec{\phi\,})\,\toG\, \xi(\vec{\phi'})\, =\, g\;\xi(\vec{\phi\,})\; h^\dagger(\vec\phi,g)\, .
\ee
In the $O(N)$ model, the selected coset representative was the matrix $U(x)$ in \eqn{eq:NGparam}, which only involves the broken generators.

\subsection{Chiral symmetry formalism}

Let us now particularize the previous discussion to the $\chi$SB
\bel{eq:scsb}
G \equiv SU(n_f)_L\otimes SU(n_f)_R \;
\longrightarrow\; H \equiv SU(n_f)_V ,
\ee
with $n_f^2-1$ Nambu--Goldstone fields $\phi_a(x)$, and a choice of coset representative $\xi(\vec{\phi\,})\equiv (\xi_L^{\phantom{\dagger}}(\vec{\phi\,}),\xi_R^{\phantom{\dagger}}(\vec{\phi\,}))\in G$. Under a chiral transformation $g\equiv (g_L^{\phantom{\dagger}},g_R^{\phantom{\dagger}})\in G$,
the change of the field coordinates in the coset space $G/H$ is given by
\bel{eq:h_def}
\xi_L^{\phantom{\dagger}}(\vec{\phi\,}) \toG g_L^{\phantom{\dagger}}\,\xi_L^{\phantom{\dagger}}(\vec{\phi\,})\, h^\dagger(\vec\phi,g)\,  ,
\qquad\quad
\xi_R^{\phantom{\dagger}}(\vec{\phi\,}) \toG g_R^{\phantom{\dagger}}\,\xi_R^{\phantom{\dagger}}(\vec{\phi\,})\, h^\dagger(\vec\phi,g)\,  .
\ee
The compensating transformation $h(\vec\phi,g)$ is the same in the two chiral sectors because they are related by a parity transformation that leaves $H$ invariant.

We can get rid of $h(\vec\phi,g)$ by combining the two chiral relations in \eqn{eq:h_def} into the simpler form
\bel{eq:u_def}
U(\vec{\phi\,})\,\equiv\,\xi_R^{\phantom{\dagger}}(\vec{\phi\,})\,\xi_L^\dagger(\vec{\phi\,}) \;
\toG\; g_R^{\phantom{\dagger}}\, U(\vec{\phi\,})\, g_L^\dagger \, .
\ee
We will also adopt the canonical choice of coset representative 
$\xi_R^{\phantom{\dagger}}(\vec{\phi\,}) = \xi_L^\dagger(\vec{\phi\,}) \equiv u(\vec{\phi\,})$, involving only the broken axial generators.
The $n_f\times n_f$ unitary matrix
\be
U(\vec{\phi\,})\, = \, u(\vec{\phi\,})^2\, =\,
\exp{\left\{i\sqrt{2}\,\frac{\Phi}{F}\right\}}\, ,
\qquad\qquad
\Phi(x)\, \equiv\, \sqrt{2}\, \widehat T^a \phi_a(x)\, ,
\label{eq:u_parametrization}
\ee
gives a very convenient parametrization of the Nambu--Goldstone modes, with $F$ some characteristic scale that is needed to compensate the dimension of the scalar fields. With $n_f = 3$,
\be
\Phi (x) \equiv {\vec{\lambda}\over\sqrt 2} \, \vec{\phi}\, = \, 
\begin{pmatrix}
{1\over\sqrt 2}\pi^0 \, + 
\, {1\over\sqrt 6}\eta_8^{\phantom{0}}
 & \pi^+ & K^+ \cr
\pi^- & - {1\over\sqrt 2}\pi^0 \, + \, {1\over\sqrt 6}\eta_8^{\phantom{0}}   
 & K^0 \cr K^- & \bar{K}^0 & - {2 \over\sqrt 6}\eta_8^{\phantom{0}} 
 \end{pmatrix} .
\label{eq:phi_matrix}
\ee
The corresponding $n_f = 2$ representation, $\Phi (x) \equiv \vec{\tau}\,\vec{\phi}/\sqrt{2}$, reduces to the upper-left $2\times 2$ submatrix, with the pion fields only. The given field labels are of course arbitrary in the fully symmetric theory, but they will correspond (in QCD) to the physical pseudoscalar mass eigenstates, once the symmetry-breaking quark masses will be taken into account. Notice that
$U(\vec{\phi\,})$ transforms linearly under the chiral group, but the induced transformation on the Nambu--Goldstone fields $\vec{\phi}$ is non-linear.

In QCD, we can intuitively visualize the matrix $U(\vec{\phi\,})_{ij}$ as parametrizing the zero-energy excitations over the quark vacuum condensate $\langle 0 |\bar q^j_L q^i_R |0\rangle \propto \delta_{ij}$, where $i,j$ denote flavour indices.

\subsection {Effective Lagrangian}

In order to obtain a model-independent description of the Nambu--Goldstone dynamics at low energies, we should write the most general Lagrangian involving the matrix $U(\vec{\phi\,})$, which is consistent with the chiral symmetry \eqn{eq:scsb}, {\it i.e.}, invariant under the transformation \eqn{eq:u_def}. We can organise the Lagrangian as an expansion in powers of momenta or, equivalently, in terms of an increasing number of derivatives. Owing to parity conservation, the number of derivatives should be even:
\be
\cL_{\mathrm{eff}}(U) \, = \, \sum_n \cL_{2n} \, .
\label{eq:l_series}
\ee
The terms with a minimum number of derivatives will dominate at low energies.

The unitarity of the $U$ matrix, $U^\dagger U = I$, implies that all possible invariant operators without derivatives are trivial constants because $\langle (U^\dagger U)^m\rangle = \langle\, I\,\rangle = n_f$, where $\langle A\rangle$ denotes the flavour trace of the matrix $A$. Therefore, one needs at least two derivatives to generate a non-trivial interaction. To lowest order (LO), there is only one independent chiral-symmetric structure:
\be
\cL_2\, =\, {F^2\over 4}\;
\langle \partial_\mu U^\dagger \partial^\mu U \rangle \, .
\label{eq:l2}
\ee
This is precisely the operator \eqn{eq:universal} that we found with the toy sigma model discussed in Section~\ref{sec:sigma}, but now we have derived it without specifying any particular underlying Lagrangian (we have only used chiral symmetry). Therefore, \eqn{eq:l2} is a universal low-energy interaction associated with the $\chi$SB \eqn{eq:scsb}.

Expanding $U(\vec{\phi\,})$ in powers of $\Phi$, the Lagrangian $\cL_2$ gives the kinetic terms plus a tower of interactions involving an increasing number of pseudoscalars. The requirement that the kinetic terms are properly normalized
fixes the global coefficient $F^2/4$ in eqn~\eqn{eq:l2}. All interactions are then predicted in terms of the single coupling $F$ that characterizes the dynamics of the Nambu--Goldstone fields:
\be
\cL_2 \, = \, {1\over 2} \,\langle\partial_\mu\Phi\,
\partial^\mu\Phi\rangle
\, + \, {1\over 12 F^2} \,\langle (\Phi\!\lrder_{\!\mu}\!\Phi) \,
(\Phi\!\buildrel \leftrightarrow \over {\partial^\mu}\!\Phi)
\rangle \, + \, \cO (\Phi^6/F^4) \, ,
\label{eq:l2_expanded}
\ee
where $(\Phi\!\lrder_{\!\mu}\!\Phi)\equiv
\Phi\, (\partial_\mu\Phi)-(\partial_\mu\Phi)\, \Phi$.

The calculation of scattering amplitudes becomes now a trivial perturbative exercise.
For instance, for the $\pi^+\pi^0$ elastic scattering, one gets the tree-level amplitude \cite{Weinberg:1966kf}
\be
T(\pi^+\pi^0\to\pi^+\pi^0)\, =\, {t\over F^2} \, ,
\label{eq:WE1}
\ee
in terms of the Mandelstam variable $t\equiv (p_+' - p_+^{\phantom{'}})^2$ that obviously vanishes at zero momenta.
Similar results can be easily obtained for $\pi\pi\to 4\pi, 6\pi, 8\pi \ldots $
The non-linearity of the effective Lagrangian relates processes with different numbers of pseudoscalars, allowing for absolute predictions in terms of the scale $F$.
 
The derivative nature of the Nambu--Goldstone interactions is a generic feature associated with the SSB mechanism, which is related to the existence of a symmetry under the shift  transformation $\phi_a'(x) = \phi_a(x) + c_a$. This constant shift amounts to a global rotation of the whole vacuum manifold that leaves the physics unchanged.
 
The next-to-leading order (NLO) Lagrangian contains four derivatives:%
\be\label{eq:l4NG}
\cL_4^{SU(3)}  =\,
L_1 \,\langle \partial_\mu U^\dagger \partial^\mu U\rangle^2  +
L_2 \,\langle \partial_\mu U^\dagger \partial_\nu U\rangle
   \langle \partial^\mu U^\dagger \partial^\nu U\rangle
+ L_3 \,\langle \partial_\mu U^\dagger \partial^\mu U \partial_\nu U^\dagger
\partial^\nu U\rangle\, .
\ee
We have particularized the Lagrangian to $n_f=3$, where there are three independent chiral-invariant structures.\footnote{
Terms such as 
$\langle \Box U^\dagger \Box U\rangle$ or $\langle \partial_\mu\partial_\nu U^\dagger \partial^\mu\partial^\nu U\rangle$ 
can be eliminated through partial integration and the use of the $\cO(p^2)$ equation of motion: 
$U^\dagger\Box U - (\Box U^\dagger) U = 0$. 
Since the loop expansion is a systematic expansion around the classical solution, the LO equation of motion can be consistently applied to simplify higher-order terms in the Lagrangian.}
 In the general $SU(n_f)_L\otimes SU(n_f)_R$ case, with $n_f>3$, one must also include the term 
$\langle \partial_\mu U^\dagger \partial_\nu U \partial^\mu U^\dagger
\partial^\nu U\rangle$. 
However, for $n_f=3$, this operator can be expressed as a combination of the three chiral structures in \eqn{eq:l4NG}, applying the Cayley-Hamilton relation
\bel{eq:CayleyHamilton}
\langle ABAB\rangle\, =\, -2\,\langle A^2 B^2\rangle + \frac{1}{2}\,\langle A^2\rangle\langle B^2\rangle + \langle A B\rangle^2\, ,
\ee
which is valid for any pair of traceless, Hermitian $3\times 3$ matrices, to the matrices $A= i (\partial_\mu U^\dagger) U$ and $B= i U^\dagger \partial_\mu U$.
For $n_f=2$, the $L_3$ term can also be eliminated with the following algebraic relation among arbitrary $SU(2)$ matrices $a, b, c$ and $d$:
\be 
2\,\langle abcd\rangle \, =\, \langle ab\rangle\langle cd\rangle -\langle ac\rangle\langle bd\rangle + \langle ad\rangle\langle bc\rangle\, .
\ee
Therefore,
\be\label{eq:l4NG-SU2}
\cL_4^{SU(2)}  =\,
\frac{\ell_1}{4} \,\langle \partial_\mu U^\dagger \partial^\mu U\rangle^2  +
\frac{\ell_2}{4} \,\langle \partial_\mu U^\dagger \partial_\nu U\rangle
   \langle \partial^\mu U^\dagger \partial^\nu U\rangle\, .
\ee

While the LO Nambu--Goldstone dynamics is fully determined by symmetry constraints and a unique coupling $F$, three (two) more free parameters $L_i$ ($\ell_i$) appear at NLO for $n_f=3$ ($n_f=2$). These couplings encode all the dynamical information about the underlying `fundamental' theory. The physical predictions of different short-distance Lagrangians, sharing the same pattern of SSB, only differ at long distances in the particular values of these low-energy couplings (LECs). The $SO(4)$ sigma model discussed in Section~\ref{sec:sigma}, for instance, is characterized at tree level by the couplings $F=v$, $\ell_1 = v^2/(2M^2)=1/(4\lambda)$ and $\ell_2=0$, where $v$ and $\lambda$ are the parameters of the potential.

\subsection{Quantum loops}
\label{subsec:loops}

The effective Lagrangian defines a consistent quantum field theory, involving the corresponding path integral over all Nambu--Goldstone field configurations. The quantum loops contain massless boson propagators and give rise to logarithmic dependences with momenta, with their corresponding cuts, as required by unitarity. Since the loop integrals are homogeneous functions of the external momenta, we can easily determine the power suppression of a given diagram with a straightforward dimensional counting.

\medskip
\noindent {\bf Weinberg power-counting theorem:} 
Let us consider a connected Feynman diagram $\Gamma$ with $L$ loops, $I$ internal boson propagators, $E$ external boson lines and $N_d$ vertices of $\cO(p^d)$. $\Gamma$ scales with momenta as $p^{d_\Gamma}$, where \cite{Weinberg:1978kz}
\bel{eq:WeinbergPC}
d_\Gamma\, =\, 2L + 2 + \sum_d (d-2) N_d\, .
\ee
\begin{proof}
Each loop integral contributes four powers of momenta, while propagators scale as $1/p^2$. Therefore, $d_\Gamma= 4 L - 2 I + \sum_d d\, N_d$. The number of internal lines is related to the total number of vertices in the diagram, $V=\sum_d N_d$, and the number of loops, through the topological identity $L = I + 1 -V$. Therefore,
\eqn{eq:WeinbergPC} follows.
\end{proof}

Thus, each loop increases the power suppression by two units. This establishes a crucial power counting that allows us to organise the loop expansion as a low-energy expansion in powers of momenta. The leading $\cO(p^2)$ contributions are obtained with $L=0$ and $N_{d>2}=0$. Therefore, at LO we must only consider tree-level diagrams with $\cL_2$ insertions. At $\cO(p^4)$, we must include tree-level contributions with a single insertion of $\cL_4$ ($L=0$, $N_4=1$, $N_{d>4}=0$) plus any number of $\cL_2$ vertices, and one-loop graphs with the LO Lagrangian only ($L=1$, $N_{d>2}=0$). The $\cO(p^6)$ corrections would involve tree-level diagrams with a single insertion of $\cL_6$ ($L=0$, $N_4=0$, $N_6=1$, $N_{d>6}=0$), one-loop graphs with one insertion of $\cL_4$ ($L=1$, $N_4=1$, $N_{d>4}=0$) and two-loop contributions from $\cL_2$ ($L=2$, $N_{d>2}=0$).

The ultraviolet loop divergences need to be renormalized. This can be done order by order in the momentum expansion, thanks to Weinberg's power-counting. Adopting a regularization that preserves the symmetries of the Lagrangian, such as dimensional regularization, the divergences generated by the loops have a symmetric local structure and the needed counter-terms necessarily correspond to operators that are already included in the effective Lagrangian, because $\cL_{\mathrm{eff}}(U)$ contains by construction all terms permitted by the symmetry. Therefore, the loop divergences can be reabsorbed through a renormalization of the corresponding LECs, appearing at the same order in momentum. 

In the usually adopted $\chi$PT renormalization scheme \cite{Gasser:1983yg}, one has at $\cO(p^4)$ 
\bel{eq:LECrenormalization}
L_i\, =\, L_i^r(\mu) + \Gamma_i\,\Delta\, ,
\qquad\qquad\qquad
\ell_i\, =\, \ell_i^r(\mu) + \gamma_i\,\Delta\, ,
\ee
where
\be 
\Delta\, =\, \frac{\mu^{D-4}}{32\pi^2}\,\left\{ \frac{2}{D-4} -\log{(4\pi)} + \gamma_E-1\right\}\, ,
\ee
with $D$ the space-time dimension. Notice that the subtraction constant differs from the $\overline{\mathrm{MS}}$ one by a factor $-1$. The explicit calculation of the one-loop generating functional gives in the $n_f=3$ theory \cite{Gasser:1984gg}
\bel{eq:gammas-su3}
\Gamma_1  = {3\over 32}\, , \qquad \Gamma_2 = \frac{3}{16}\, , 
\qquad  \Gamma_3 = 0 \, ,
\ee
while for $n_f=2$ one finds \cite{Gasser:1983yg}
\bel{eq:gammas-su2}
\gamma_1  = {1\over 3}\, , \qquad \gamma_2 = \frac{2}{3}\, .
\ee

The renormalized couplings $L_i^r(\mu)$ and $\ell_i^r(\mu)$ depend on the arbitrary
scale $\mu$ of dimensional regularization. Their logarithmic running is dictated by \eqn{eq:LECrenormalization}:
\bel{eq:running}
L_i^r(\mu_2)\, =\, L_i^r(\mu_1) + \frac{\Gamma_i}{(4\pi)^2}\,\log{\left(\frac{\mu_1}{\mu_2}\right)}\, ,
\qquad\qquad
\ell_i^r(\mu_2)\, =\, \ell_i^r(\mu_1) + \frac{\gamma_i}{(4\pi)^2}\,\log{\left(\frac{\mu_1}{\mu_2}\right)}\, .
\ee
This renormalization-scale dependence cancels exactly with that of the loop amplitude, in all measurable quantities.

\begin{figure}[t]
\begin{center}
\includegraphics[width=10.5cm]{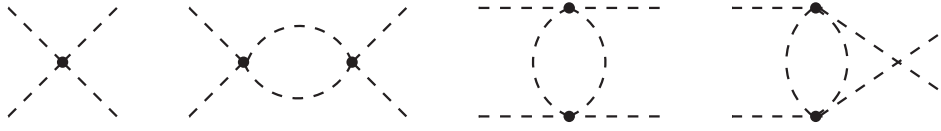}
\end{center}
\caption{Feynman diagrams contributing to $\pi^a \pi^b\to \pi^c \pi^d$ at the NLO.}
\label{fig:2piScattering}
\end{figure}

A generic $\cO (p^4)$ amplitude consists of a non-local (non-polynomial)
loop contribution, plus a polynomial in momenta that depends on the unknown constants $L_i^r(\mu)$ or $\ell_i^r(\mu)$. Let us consider, for instance, the elastic scattering of two Nambu--Goldstone particles in the $n_f=2$ theory:
\bel{eq:PionScattering-1L}
 \cA(\pi^a \pi^b\to \pi^c \pi^d)\, =\,
A(s,t,u)\;\delta_{ab}\,\delta_{cd} + A(t,s,u)\;\delta_{ac}\,\delta_{bd} +
A(u,t,s)\;\delta_{ad}\,\delta_{bc}\, .
\ee
Owing to crossing symmetry, the same analytic function governs the $s$, $t$ and $u$ channels, with the obvious permutation of the three Mandelstam variables. At $\cO(p^4)$, we must consider the one-loop Feynman topologies shown in Fig.~\ref{fig:2piScattering}, with $\cL_2$ vertices, plus the tree-level contribution from $\cL_4^{SU(2)}$. One obtains the result \cite{Gasser:1983yg}:
\beqn\label{eq:PionScattering-1L-b}
A(s,t,u) & = & \frac{s}{F^2} \, +\,
\frac{1}{F^4}\,\left[ 2\, \ell_1^r(\mu)\, s^2 + \ell_2^r(\mu)\, (t^2+u^2) \right]
\no\\[3pt]
& + & \frac{1}{96\pi^2 F^4}\,\left\{ \frac{4}{3}\, s^2 + \frac{7}{3}\, (t^2+u^2)
+ \frac{1}{2}\, (s^2-3 t^2-u^2)\,\log{\left(\frac{-t}{\mu^2}\right)}
\right.\no\\[3pt] && \hskip 1.25cm\left.\mbox{}
+ \frac{1}{2}\, (s^2- t^2-3 u^2)\,\log{\left(\frac{-u}{\mu^2}\right)}
-3\, s^2\,\log{\left(\frac{-s}{\mu^2}\right)}
\right\}\, ,\quad
\eeqn
which also includes the leading $\cO(p^2)$ contribution.
Using \eqn{eq:running}, it is straightforward to check that the scattering amplitude is independent of the renormalization scale $\mu$, as it should.

The non-local piece contains the so-called chiral logarithms that are fully predicted as a function of the LO coupling $F$. This chiral structure can be easily understood in terms of dispersion relations. The non-trivial analytic behaviour associated with physical intermediate states (the absorptive contributions) can be calculated with the LO Lagrangian $\cL_2$. Analyticity then allows us to reconstruct the full amplitude, through
a dispersive integral, up to a subtraction polynomial. The effective theory satisfies unitarity and analyticity, therefore, it generates perturbatively the correct dispersion integrals and organises the subtraction polynomials in a derivative expansion. In addition, the symmetry embodied in the effective Lagrangian implies very strong constraints that relate the scattering amplitudes of different processes, {\it i.e.}, all subtraction polynomials are determined in terms of the LECs.

\section{Chiral perturbation theory} 
\label{sec:ChPT}
 
So far, we have been discussing an ideal theory of massless Nambu--Goldstone bosons where the symmetry is exact. However, the physical pions have non-zero masses because chiral symmetry is broken explicitly by the quark masses. Moreover, the pion dynamics is sensitive to the electroweak interactions that also break chiral symmetry. In order to incorporate all these sources of explicit symmetry breaking, it is useful to introduce external classical fields coupled to the quark currents.

Let us consider an extended QCD Lagrangian, with the quark currents coupled to external Hermitian matrix-valued fields
$v_\mu$, $a_\mu$, $s$, $p$\/ :
\be
\cL_{\mathrm{QCD}}\, =\, \cL^0_{\mathrm{QCD}} +
\bar q \gamma^\mu (v_\mu + \gamma_5 a_\mu ) q -
\bar q (s - i \gamma_5 p) q \, ,
\label{eq:extendedqcd}
\ee
where $\cL^0_{\mathrm{QCD}}$ is the massless QCD Lagrangian \eqn{eq:LQCD}.
The external fields can be used to parametrize the different breakings of chiral symmetry through the identifications
\beqn\label{eq:breaking}
r_\mu &\equiv & v_\mu + a_\mu \, = \, - e \cQ A_\mu\, , 
\no\\
\ell_\mu & \equiv & v_\mu - a_\mu \, =
\,  - e \cQ A_\mu - {e\over\sqrt{2}\sin{\theta_W}}
(W_\mu^\dagger T_+ + {\rm h.c.})\, ,  
\no\\
s & = & \cM\, ,  
\no\\ 
p &=& 0\, ,
\eeqn
with $\cQ$ and $\cM$ the quark charge and mass matrices ($n_f=3$), respectively,
\be
\cQ = {1\over 3}\, \hbox{\rm diag}(2,-1,-1)\, , \qquad\qquad
\cM = \hbox{\rm diag}(m_u,m_d,m_s) \, .
\label{eq:q_m_matrices}
\ee
The $v_\mu$ field contains the electromagnetic interactions, while the scalar source $s$ accounts for the quark masses. The charged-current couplings of the $W^\pm$ bosons, which govern semileptonic weak transitions, are incorporated into $\ell_\mu$, with the $3\times 3$ matrix
\be
T_+ \, = \, 
\begin{pmatrix}
0 & V_{ud} & V_{us} \cr 0 & 0 & 0 \cr 0 & 0 & 0
\end{pmatrix}
\label{eq:t_matrix}
\ee
carrying the relevant quark mixing factors. One could also add the $Z$ couplings into $v_\mu$ and $a_\mu$, and the Higgs Yukawa interaction into $s$. More exotic quark couplings to other vector, axial, scalar or pseudoscalar fields, outside the Standard Model framework, could also be easily included in a similar way.

The Lagrangian \eqn{eq:extendedqcd} is invariant under local
$SU(3)_L\otimes SU(3)_R$ transformations, provided the external fields
are enforced to transform in the following way:
\beqn
q_L^{\phantom{\dagger}}  &\longrightarrow &  g_L^{\phantom{\dagger}} \, q_L^{\phantom{\dagger}} \, , \qquad\quad
q_R^{\phantom{\dagger}}  \,\longrightarrow\,  g_R^{\phantom{\dagger}} \, q_R^{\phantom{\dagger}} \, , \qquad\quad
s + i p  \,\longrightarrow\,  g_R^{\phantom{\dagger}} \, (s + i p) \, g_L^\dagger \, ,
\no\\
\ell_\mu &\longrightarrow &  g_L^{\phantom{\dagger}} \, \ell_\mu \, g_L^\dagger \, + \,
i g_L^{\phantom{\dagger}} \partial_\mu g_L^\dagger \, ,
\qquad\qquad\quad
r_\mu  \,\longrightarrow\,  g_R^{\phantom{\dagger}} \, r_\mu \, g_R^\dagger \, + \,
i g_R^{\phantom{\dagger}} \partial_\mu g_R^\dagger \, .\quad
\label{eq:symmetry}
\eeqn
This formal symmetry can be used to build a generalised effective Lagrangian, in the presence of external sources. In order to respect the local invariance, the gauge fields $v_\mu$ and $a_\mu$ can only appear through the covariant derivatives
\bel{eq:CovDer}
D_\mu U = \partial_\mu U - i r_\mu U + i U \ell_\mu \, ,
\qquad\quad
D_\mu U^\dagger = \partial_\mu U^\dagger  + i U^\dagger r_\mu
- i \ell_\mu U^\dagger ,
\ee
and through the field strength tensors
\bel{eq:StrengthTensors}
F^{\mu\nu}_L =
\partial^\mu \ell^\nu - \partial^\nu \ell^\mu 
- i\, [ \ell^\mu , \ell^\nu ] \, ,
\qquad\quad
F^{\mu\nu}_R =
\partial^\mu r^\nu - \partial^\nu r^\mu - i\, [ r^\mu , r^\nu ] .
\ee

At LO in derivatives and number of external fields, the most general effective Lagrangian consistent with Lorentz invariance and the local chiral symmetry \eqn{eq:symmetry} takes the form \cite{Gasser:1984gg}:
\be
\cL_2\, =\, {F^2\over 4}\,
\langle D_\mu U^\dagger D^\mu U \, + \, U^\dagger\chi  \, 
+  \,\chi^\dagger U
\rangle \, ,
\label{eq:lowestorder}
\ee
with
\be
\chi \, = \, 2 B_0 \, (s + i p) \, .
\label{eq:chi}
\ee
The first term is just the universal LO Nambu--Goldstone interaction, but now with covariant derivatives that include the external vector and axial-vector sources.
The scalar and pseudoscalar fields, incorporated into $\chi$, give rise to a second invariant structure with a coupling $B_0$, which, like $F$, cannot be fixed with
symmetry requirements alone.
 
Once the external fields are frozen to the particular values in \eqn{eq:breaking},
the symmetry is of course explicitly broken. However, the choice of a special direction in the flavour space breaks chiral symmetry in the effective Lagrangian \eqn{eq:lowestorder}, in exactly the same way as it does in the fundamental short-distance Lagrangian \eqn{eq:extendedqcd}. Therefore, \eqn{eq:lowestorder} provides the correct low-energy realization of QCD, including its symmetry breakings.
 
The external fields provide, in addition, a powerful tool to compute the effective realization of the chiral Noether currents. The Green functions of quark currents are obtained as functional derivatives of the generating functional
$Z[v,a,s,p]$, defined via the path-integral formula
\be\label{eq:generatingfunctional}
\quad\exp{\{i Z\}} \, = \,  \int  \cD q \,\cD \bar q
\,\cD G_\mu \,
\exp{\left\{i \int d^4x\, \cL_{\mathrm{QCD}}\right\}}  
\, =\,
\int  \cD U \,
\exp{\left\{i \int d^4x\, \cL_{\mathrm{eff}}\right\}} .\quad
\ee
This formal identity provides a link between the fundamental and effective  theories. At lowest order in momenta, the generating functional reduces to the
classical action $S_2 = \int d^4x \,\cL_2$.
Therefore, the low-energy realization of the QCD currents can be easily computed by taking the appropriate derivatives with respect to the external fields:
\beqn\label{eq:l_r_currents}
J^\mu_L \, =\, \bar q_L^{\phantom{\dagger}}\gamma^\mu q_L^{\phantom{\dagger}} \,
\doteq\, {\delta S_2\over \delta \ell_\mu} \, = \, & \displaystyle
 \frac{i}{2}\, F^2\, D_\mu U^\dagger U\, =\,
\hphantom{-}{F\over\sqrt{2}}\, D_\mu \Phi -
{i\over 2}  \left(\Phi
\stackrel{\leftrightarrow}{D^\mu}\Phi\right) +
\cO (\Phi^3/F)\, , 
\no\\ 
J^\mu_R \, =\, \bar q_R^{\phantom{\dagger}}\gamma^\mu q_R^{\phantom{\dagger}} \,
\doteq\, {\delta S_2\over \delta r_\mu} \, = \, &\displaystyle
 {i\over 2}\, F^2\, D_\mu U U^\dagger\, =\,
-{F\over\sqrt{2}}\, D_\mu \Phi -
{i\over 2} \left(\Phi
\stackrel{\leftrightarrow}{D^\mu}\Phi\right) +
\cO (\Phi^3/F)\, . 
\no\\ &&\eeqn
Thus, at $\cO (p^2)$, the fundamental chiral coupling $F$ equals the pion decay constant, $F = F_\pi = 92.2$ MeV, defined as
\be
\langle 0 | (J^\mu_A)^{12} | \pi^+\rangle
 \,\equiv\, i \sqrt{2} F_\pi\, p^\mu\, .
\label{f_pi}
\ee
Taking derivatives with respect to the external scalar and pseudoscalar sources,
\beqn\label{eq:s_p_currents}
\bar q^j_L q^i_R\,
\doteq\, -{\delta S_2\over \delta (s-ip)^{ji}} \, = &\displaystyle\, 
-{F^2\over 2}\, B_0 \; U(\vec{\phi\,})_{ij} ,
\no\\
\bar q^j_R q^i_L\,
\doteq\, -{\delta S_2\over \delta (s+ip)^{ji}} \, = &\displaystyle\, 
-{F^2\over 2}\, B_0 \; U^\dagger(\vec{\phi\,})_{ij} ,
\eeqn
we also find that the coupling $B_0$ is related to the quark vacuum condensate:
\be
\langle 0 | \bar q^j q^i|0\rangle\, =\, -F^2 B_0 \,\delta^{ij} .
\label{eq:b0}
\ee
The Nambu--Goldstone bosons, parametrized by the matrix $U(\vec{\phi\,})$, represent indeed the zero-energy excitations over this vacuum condensate that triggers the dynamical breaking of chiral symmetry.

\subsection{Pseudoscalar meson masses at lowest order}

With $s = \cM$ and $p=0$, the non-derivative piece of the Lagrangian~\eqn{eq:lowestorder} generates a quadratic mass term for the pseudoscalar bosons, plus $\Phi^{2n}$ interactions proportional to the quark masses.
Dropping an irrelevant constant, one gets:
\be
{F^2\over 4}\, 2 B_0 \;\langle \cM (U + U^\dagger) \rangle\,
= \, B_0 \left\{ - \langle \cM\Phi^2\rangle
+ {1\over 6 F^2}\, \langle \cM \Phi^4\rangle
+ \cO\left({\Phi^6\over F^4} \right) \right\} .
\label{eq:massterm}
\ee
The explicit evaluation of the trace in the quadratic term provides
the relation between the masses of the physical mesons and the quark masses:
\beqn
M_{\pi^\pm}^2 \, & = &\,  2 \hat{m}\, B_0 \, , 
\qquad\qquad\qquad\qquad\qquad
M_{\pi^0}^2\,  =\,   2 \hat{m}\, B_0 - \varepsilon +
\cO (\varepsilon^2) \, , 
\no\\
M_{K^\pm}^2  & = &\,   (m_u + m_s)\, B_0 \, , 
\qquad\qquad\qquad\;\;
M_{K^0}^2\,  =\,   (m_d + m_s)\, B_0 \, , 
\label{eq:masses} 
\no\\
M_{\eta_{\raisebox{-1pt}{$\scriptscriptstyle 8$}}}^2 \; & = &\,   
{2\over 3}\, (\hat{m} + 2 m_s)\,  B_0 + \varepsilon +
\cO (\varepsilon^2) \, ,   
\eeqn
with\footnote{   
The $\cO (\varepsilon)$ corrections to $M_{\pi^{\raisebox{-1pt}{$\scriptscriptstyle 0$}}}^2$
and $M_{\eta_{\raisebox{-1pt}{$\scriptscriptstyle 8$}}}^2$
originate from a small mixing term between the
$\phi_3^{\phantom{.}}$ and $\phi_8^{\phantom{.}}$ fields:
$ \quad
- B_0 \langle \cM\Phi^2\rangle \longrightarrow
- (B_0/\sqrt{3})\, (m_u - m_d)\, \phi_3^{\phantom{.}}\phi_8^{\phantom{.}} \, .
\quad $
The diagonalization of the quadratic $\phi_3^{\phantom{.}}$, $\phi_8^{\phantom{.}}$
mass matrix, gives the mass eigenstates,
$\pi^0 = \cos{\delta} \,\phi_3^{\phantom{.}} + \sin{\delta} \,\phi_8^{\phantom{.}}$
and
$\eta_{\raisebox{-1pt}{$\scriptscriptstyle 8$}} = -\sin{\delta} \,\phi_3^{\phantom{.}} + \cos{\delta} \,\phi_8^{\phantom{.}}$,
where
$
\tan{(2\delta)} = \sqrt{3} (m_d-m_u)/\left( 2 (m_s-\hat{m})\right) .
$
\label{foot:PiEtaMixing}}
%
\be
\hat m\, =\, {1\over 2}\, (m_u + m_d) \, , \qquad\qquad
\varepsilon\, =\, {B_0\over 4}\, {(m_u - m_d)^2\over  (m_s - \hat m)} \, .
\label{eq:mhat}
\ee
Owing to chiral symmetry, the meson masses squared are proportional to a single power of the quark masses, the proportionality constant being related to the vacuum quark condensate \cite{GellMann:1968rz}:
\be
F^2_\pi M_{\pi^\pm}^2\, =\, -\hat m \,\langle 0|\bar u u + \bar d d|0\rangle\, .
\label{eq:gmor}
\ee

Taking out the common proportionality factor $B_0$, the relations~\eqn{eq:masses} imply the old Current-Algebra mass ratios \cite{GellMann:1968rz,Weinberg:1977xxx},
\be
{M^2_{\pi^\pm}\over 2 \hat m}\, =\, \frac{M^2_{K^+}}{m_u+m_s} \, =\,
\frac{M_{K^0}^2}{m_d+m_s}
\,\approx\, \frac{3 M^2_{\eta_{\raisebox{-1pt}{$\scriptscriptstyle 8$}}}}{2 \hat m + 4 m_s} \, ,
\label{eq:mratios}
\ee
and, up to $\cO (m_u-m_d)$ corrections, the Gell-Mann--Okubo \cite{GellMann:1962xb,Okubo:1961jc} mass relation,
\be
3 M^2_{\eta_{\raisebox{-1pt}{$\scriptscriptstyle 8$}}}\, =\, 4 M_K^2 - M_\pi^2 \, .
\label{eq:gmo}
\ee

Chiral symmetry alone cannot fix the absolute values of the quark masses, because they are short-distance parameters that depend on QCD renormalization conventions. The renormalization scale and scheme dependence cancels out in the products $m_q\bar q q\sim m_q B_0$, which are the relevant combinations governing the pseudoscalar masses. Nevertheless, $\chi$PT provides information about quark mass ratios, where the dependence on $B_0$ drops out (QCD is flavour blind). Neglecting the tiny $\cO (\varepsilon)$ corrections, one gets the relations
\be
{m_d - m_u \over m_d + m_u} \, = \,
{(M_{K^0}^2 - M_{K^+}^2) - (M_{\pi^0}^2 - M_{\pi^+}^2)
\over M_{\pi^0}^2}
\, = \, 0.29 \, , 
\label{eq:ratio1}
\ee\be
{m_s -\hat m\over 2 \hat m} \, = \,
{M_{K^0}^2 - M_{\pi^0}^2\over M_{\pi^0}^2}
\, = \, 12.6 \, . 
\label{eq:ratio2}
\ee
In the first equation, we have subtracted the electromagnetic pion mass-squared difference to account for the virtual photon contribution to the meson self-energies.
In the chiral limit ($m_u=m_d=m_s=0$), this correction is proportional
to the square of the meson charge and it is the same for 
$K^+$ and $\pi^+$.\footnote{This result, known as Dashen's theorem \cite{Dashen:1969eg}, can be easily proved using the external sources $\ell_\mu$ and $r_\mu$ in \eqn{eq:breaking}, with formal electromagnetic charge matrices $\cQ_L$ and $\cQ_R$, respectively, transforming as $\cQ_X\to g_X^{\phantom{\dagger}} \cQ_X g_X^\dagger$. A quark (meson) electromagnetic self-energy involves a virtual photon propagator between two interaction vertices.
Since there are no external photons left, the LO chiral-invariant operator with this structure is $e^2\,\langle \cQ_R U \cQ_L U^\dagger\rangle = - 2 e^2\,
(\pi^+\pi^- + K^+ K^-)/F^2 + \cO(\phi^4/F^4)$.}
The mass formulae \eqn{eq:ratio1} and \eqn{eq:ratio2}
imply the quark mass ratios advocated by Weinberg \cite{Weinberg:1977xxx}:
\be
m_u : m_d : m_s\, =\, 0.55 : 1 : 20.3 \, .
\label{eq:Weinbergratios}
\ee
Quark mass corrections are therefore dominated by the strange quark mass $m_s$, which is much larger than $m_u$ and $m_d$. The light-quark mass difference $m_d-m_u$ is not small compared with the individual up and down quark masses. In spite of that, isospin turns out to be a very good symmetry, because isospin-breaking effects are governed by the small ratio $(m_d-m_u)/m_s$.

The $\Phi^4$ interactions in eqn~\eqn{eq:massterm} introduce mass corrections to the $\pi\pi$ scattering amplitude \eqn{eq:WE1},
\be
T(\pi^+\pi^0\to\pi^+\pi^0)\, =\, {t - M_\pi^2\over F^2}\, ,
\label{eq:WE2}
\ee
showing that it vanishes at $t = M_\pi^2$ \cite{Weinberg:1966kf}. This result is now an absolute prediction of chiral symmetry, because the scale $F=F_\pi$ has been already fixed from pion decay. 
 
The LO chiral Lagrangian \eqn{eq:lowestorder} encodes in a very compact way all the Current-Algebra results, obtained in the sixties \cite{Adler:1968xxx,deAlfaro:1973zz}. These successful phenomenological predictions corroborate the pattern of $\chi$SB in \eqn{eq:scsb} and the explicit breaking incorporated by the QCD quark masses. Besides its elegant simplicity, the EFT formalism provides a powerful technique to estimate higher-order corrections in a systematic way.

\subsection{Higher-order corrections}
\label{subsec:ChPT-NLO}

In order to organise the chiral expansion, we must first establish a well-defined power counting for the external sources. Since $p_\phi^2 = M_\phi^2$, the physical pseudoscalar masses scale in the same way as the external on-shell momenta. This implies that the field combination $\chi$ must be counted as $\cO (p^2)$, because $B_0 m_q\propto M^2_\phi$. The left and right sources, $\ell_\mu$ and $r_\mu$, are part of the covariant derivatives \eqn{eq:CovDer} and, therefore, are of $\cO(p)$. Finally, the field strength tensors \eqn{eq:StrengthTensors} are obviously $\cO(p^2)$ structures. Thus:
\be\label{eq:PowerCounting}
U(\vec{\phi\,})\, \sim\,\cO(p^0)\, ; \qquad\quad
D_\mu\, ,\, \ell_\mu\, , \, r_\mu\, \sim\,\cO(p^1)\, ;\qquad\quad
F_L^{\mu\nu}\, ,\, F_R^{\mu\nu}\, ,\, \chi\, \sim\,\cO(p^2)\, .
\ee
The full LO Lagrangian \eqn{eq:lowestorder} is then of $\cO(p^2)$ and, moreover, Weinberg's power counting \eqn{eq:WeinbergPC} remains valid in the presence of all these symmetry-breaking effects.

At $\cO (p^4)$, the most general Lagrangian, invariant under Lorentz symmetry,
parity, charge conjugation and the local chiral transformations \eqn{eq:symmetry},
is given by \cite{Gasser:1984gg}
\beqn\label{eq:l4}
\cL_4 & = & 
L_1 \,\langle D_\mu U^\dagger D^\mu U\rangle^2  + 
L_2 \,\langle D_\mu U^\dagger D_\nu U\rangle\,
   \langle D^\mu U^\dagger D^\nu U\rangle
+ L_3 \,\langle D_\mu U^\dagger D^\mu U D_\nu U^\dagger
D^\nu U\rangle
\no\\ & + &\mbox{}
 L_4 \,\langle D_\mu U^\dagger D^\mu U\rangle\,
   \langle U^\dagger\chi +  \chi^\dagger U \rangle
+ L_5 \,\langle D_\mu U^\dagger D^\mu U \left( U^\dagger\chi +
\chi^\dagger U \right)\rangle
\no\\ & + &\mbox{}
 L_6 \,\langle U^\dagger\chi +  \chi^\dagger U \rangle^2
+ L_7 \,\langle U^\dagger\chi -  \chi^\dagger U \rangle^2
+ L_8 \,\langle\chi^\dagger U \chi^\dagger U
+ U^\dagger\chi U^\dagger\chi\rangle
\no\\ & - &\mbox{}
 i L_9 \,\langle F_R^{\mu\nu} D_\mu U D_\nu U^\dagger +
     F_L^{\mu\nu} D_\mu U^\dagger D_\nu U\rangle
+  L_{10} \,\langle U^\dagger F_R^{\mu\nu} U F_{L\mu\nu} \rangle
\no\\ & + &\mbox{}
H_1 \,\langle F_{R\mu\nu} F_R^{\mu\nu} + F_{L\mu\nu} F_L^{\mu\nu}\rangle 
+  H_2 \,\langle \chi^\dagger\chi\rangle \, .
\eeqn
The first three terms correspond to the $n_f=3$ Lagrangian \eqn{eq:l4NG}, changing the normal derivatives by covariant ones. The second line contains operators with two covariant derivatives and one insertion of $\chi$, while the operators in the third line involve two powers of $\chi$ and no derivatives. The fourth line includes operators with field strength tensors. The last structures proportional to $H_1$ and $H_2$ are just needed for renormalization purposes; they only contain external sources and, therefore, do not have any impact on the pseudoscalar meson dynamics.

Thus, at $\cO (p^4)$, the low-energy behaviour of the QCD Green functions is determined by ten chiral couplings $L_i$. They renormalize the one-loop divergences, as indicated in \eqn{eq:LECrenormalization}, and their logarithmic dependence with the renormalization scale is given by eqn~\eqn{eq:running} with \cite{Gasser:1984gg} 
\beqn
\Gamma_1 &=& {3\over 32}\, , \quad\;\, \Gamma_2 = \frac{3}{16}\, , 
\quad\;\,  \Gamma_3 = 0 \, ,
\quad\;\, \Gamma_4 = {1\over 8} \, , \quad\;\,
\Gamma_5 = {3\over 8} \, , \quad\;\, \Gamma_6 = \frac{11}{144} \, ,
\no\\
\Gamma_7 &=& 0 \, ,
\quad\; \Gamma_8 = {5\over 48}\, , \quad\;
\Gamma_9 = {1\over 4} \, ,\quad\; \Gamma_{10} = -\frac{1}{4} \, , 
\quad\; \widetilde\Gamma_1 = -{1\over 8} \, , \quad\;
\widetilde\Gamma_2 = {5\over 24}\, ,\quad\quad
\label{eq:d_factors}
\eeqn
where $\widetilde\Gamma_1$ and $\widetilde\Gamma_2$ are the corresponding quantities for the two unphysical couplings $H_1$ and $H_2$. 

The structure of the $\cO(p^6)$ $\chi$PT Lagrangian has been also thoroughly analysed. It contains $90+4$ independent chiral structures of even intrinsic parity (without Levi-Civita pseudotensors) \cite{Bijnens:1999sh}, the last four containing external sources only, and $23$ operators of odd intrinsic parity \cite{Bijnens:2001bb,Ebertshauser:2001nj}:\footnote{The $n_f=2$ theory contains $7+3$ independent operators at $\cO(p^4)$ \cite{Gasser:1983yg}, while at $\cO(p^6)$ it has $52+4$ structures of even parity  \cite{Bijnens:1999sh,Haefeli:2007ty} plus 5 odd-parity terms (13 if a singlet vector source is included) \cite{Bijnens:2001bb,Ebertshauser:2001nj}.}
\bel{eq:ChPT-p6}
\cL_6\, =\, \sum_{i=1}^{94} C_i\; O_i^{p^6}\, +\, \sum_{i=1}^{23} \widetilde C_i\; \widetilde O_i^{p^6}\, .
\ee
The complete renormalization of the $\chi$PT generating functional has been already accomplished at the two-loop level \cite{Bijnens:1999hw}, which determines the renormalization group equations for the renormalized $\cO(p^6)$ LECs.

$\chi$PT is an expansion in powers of momenta over some typical hadronic
scale $\Lambda_\chi$, associated with the $\chi$SB, which can be expected to be of the order of the (light-quark) resonance masses. The variation of the loop amplitudes under a rescaling of $\mu$, by say $e$, provides a natural order-of-magnitude estimate of the $\chi$SB scale: $\Lambda_\chi\sim 4\pi F_\pi\sim 1.2\,{\rm GeV}$ \cite{Manohar:1983md,Weinberg:1978kz}.

At $\cO(p^2)$, the $\chi$PT Lagrangian is able to describe all QCD Green functions with only two parameters, $F$ and $B_0$, a quite remarkable achievement. However, with $p \lsim M_K \, (M_\pi)$, we expect $\cO (p^4)$ contributions to the LO amplitudes at the level of $p^2/\Lambda_\chi^2 \lsim 20\% \, (2\% )$. In order to increase the accuracy of the $\chi$PT predictions beyond this level, the inclusion of NLO corrections is mandatory, which introduces ten additional unknown LECs. Many more free parameters ($90+23$) are needed to account for $\cO(p^6)$ contributions. Thus, increasing the precision reduces the predictive power of the effective theory. 

The LECs  parametrize our ignorance about the details of the underlying QCD dynamics. They are, in principle, calculable functions of $\Lambda_{\mathrm{QCD}}$ and the heavy-quark masses, which can be analysed with lattice simulations. However, at present, our main source of information about these couplings is still low-energy phenomenology.
At $\cO(p^4)$, the elastic $\pi\pi$ and $\pi K$ scattering amplitudes are sensitive to $L_{1,2,3}$. The two-derivative couplings $L_{4,5}$ generate mass corrections to the meson decay constants (and mass-dependent wave-function renormalizations), while the pseudoscalar meson masses get modified by the non-derivative terms $L_{6,7,8}$. $L_9$ is mainly responsible for the charged-meson electromagnetic radius and $L_{10}$ only contributes to amplitudes with at least two external vector or axial-vector fields, like the radiative semileptonic decay $\pi\to e\nu\gamma$.

Table~\ref{tab:Lcouplings} summarises our current knowledge on the $\cO(p^4)$ constants $L_i$. The quoted numbers correspond to the renormalized couplings, at a scale $\mu = M_\rho$. The second column shows the LECs extracted from $\cO(p^4)$ phenomenological analyses \cite{Bijnens:2014lea}, without any estimate of the uncertainties induced by the missing higher-order contributions. In order to assess the possible impact of these corrections, the third column shows the results obtained from a global $\cO(p^6)$ fit  \cite{Bijnens:2014lea}, which incorporates some theoretical priors (prejudices) on the unknown $\cO(p^6)$ LECs. In view of the large number of uncontrollable parameters, the $\cO(p^6)$ numbers should be taken with caution, but they can give a good idea of the potential uncertainties. The
$\cO(p^6)$ determination of $L_{10}^r(M_\rho)$ has been directly extracted from hadronic $\tau$ decay data \cite{Rodriguez-Sanchez:2016jvw}.
For comparison, the fourth column shows the results of lattice simulations with $2+1+1$ dynamical fermions by the HPQCD colaboration \cite{Dowdall:2013rya}. Similar results with $2+1$ fermions have been obtained by the MILC collaboration \cite{Bazavov:2010hj}, although the quoted errors are larger. An analogous compilation of LECs for the $n_f=2$ theory can be found in Refs.~\cite{Aoki:2016frl,Bijnens:2014lea}.

\begin{table}[tb]
\tableparts
{
\caption{Phenomenological determinations of the renormalized couplings $L_i^r(M_\rho)$ from $\cO(p^4)$ and $\cO(p^6)$ $\chi$PT analyses, and from lattice simulations (fourth column). The last two columns show the R$\chi$T predictions of Section~\ref{sec:MassiveFields}, without (column 5) and with (column 6) short-distance information.
Values labeled with $\dagger$ have been used as inputs.}
\label{tab:Lcouplings}
}
{
\begin{tabular}{ccccccc}
\hline\\[-10pt]
 & \multicolumn{5}{c}{$L_i^r(M_\rho) \times 10^3$} &  
\\[1.1pt] \cline{2-6}\\[-10pt]
$i$ &  $\cO(p^4)$ \cite{Bijnens:2014lea} & $\cO(p^6)$ \cite{Bijnens:2014lea} & Lattice \cite{Dowdall:2013rya} & R$\chi$T \cite{Ecker:1988te} &
R$\chi$T${}_{\mathrm{SD}}$ \cite{Ecker:1989yg,Pich:2002xy}
\\[1.1pt]
\hline
1 & $\hphantom{-}1.0\pm 0.1$ & $\hphantom{-}0.53\pm 0.06$ &&
$\hphantom{-}0.6^{\phantom{\dagger}}$ & $\hphantom{-}0.9$
\\
2 & $\hphantom{-}1.6\pm 0.2$ & $\hphantom{-}0.81\pm 0.04$ &&
$\hphantom{-}1.2^{\phantom{\dagger}}$ & $\hphantom{-}1.8$
\\
3 & $-3.8\pm 0.3$ & $-3.07\pm 0.20$ &&
$-2.8^{\phantom{\dagger}}$ & $-4.8$
\\
4 & $\hphantom{-}0.0\pm 0.3$ &  $0.3$ (fixed) & $0.09\pm 0.34$ &
$\hphantom{-}0.0^{\phantom{\dagger}}$ & $\hphantom{-}0.0$
\\
5 & $\hphantom{-}1.2\pm 0.1$ & $\hphantom{-}1.01\pm 0.06$ & $1.19\pm 0.25$ &
$\hphantom{-}1.2^\dagger$ & $\hphantom{-}1.1$
\\
6 & $\hphantom{-}0.0\pm 0.4$ & $\hphantom{-}0.14\pm 0.05$ & $0.16\pm 0.20$ &
$\hphantom{-}0.0^{\phantom{\dagger}}$  & $\hphantom{-}0.0$
\\
7 & $-0.3\pm 0.2$ & $-0.34\pm 0.09$ & &
$-0.3^{\phantom{\dagger}}$ & $-0.3$
\\
8 & $\hphantom{-}0.5\pm 0.2$ & $\hphantom{-}0.47\pm 0.10$ & $0.55\pm 0.15$ &
$\hphantom{-}0.5^\dagger$ & $\hphantom{-}0.4$
\\
9 & $\hphantom{-}6.9\pm 0.7$ & $\hphantom{-}5.9\pm 0.4$ &&
$\hphantom{-}6.9^\dagger$ & $\hphantom{-}7.1$
\\
10 & $-5.2\pm 0.1$ & $-4.1\pm 0.4$ &&
$-5.8^{\phantom{\dagger}}$ & $-5.3$
\\ \hline
\end{tabular}
}
\end{table}

The values quoted in the table are in good agreement with the expected size of the couplings $L_i$ in terms of the scale of $\chi$SB:
\be
L_i \,\sim\, {F_\pi^2/4 \over \Lambda_\chi^2} \,\sim\, {1\over 4 (4 \pi)^2}
\sim 2\times 10^{-3} .
\label{eq:l_size}
\ee
We have just taken as reference values the normalization of $\cL_2$ and
$\Lambda_\chi \sim 4 \pi F_\pi$. 
Thus, all $\cO(p^4)$ couplings have the right order of magnitude, which implies a good convergence of the momentum expansion below the  resonance region, {\it i.e.}, for $p < M_\rho$. The table displays, however, a clear dynamical hierarchy with some couplings being large while others seem compatible with zero. 
 
$\chi$PT allows us to make a good book-keeping of phenomenological information in terms of some LECs. Once these couplings have been fixed, we can predict other quantities. In addition, the information contained in Table~\ref{tab:Lcouplings}
is very useful to test QCD-inspired models or non-perturbative approaches.
Given any particular theoretical framework aiming to correctly describe QCD at low
energies, we no longer need to make an extensive phenomenological analysis to test its reliability; it suffices to calculate the predicted LECs and compare them with their phenomenological values in Table~\ref{tab:Lcouplings}.
For instance, the linear sigma model discussed in Section~\ref{sec:sigma} has the right chiral symmetry and, therefore, leads to the universal Nambu--Goldstone Lagrangian \eqn{eq:l2} at LO. However, its dynamical content fails to reproduce the data at NLO because, as shown in eqn~\eqn{eq:sigma5}, it only generates a single $\cO(p^4)$  LEC, $L_1$, in complete disagreement with the pattern displayed by the table.

\section{QCD phenomenology at very low energies} 
\label{sec:phenomenology}

Current $\chi$PT analyses have reached an $\cO(p^6)$ precision. This means that the most relevant observables are already known at the two-loop level. Thus, at $\cO(p^6)$, the leading double logarithmic corrections are fully known in terms of $F$ and the meson masses, while single chiral logarithms involve the $L_i^r(\mu)$ couplings through one loop corrections with one insertion of $\cL_4$. The main pending problem is the large number of unknown LECs, $C_i^r(\mu)$ and $\widetilde{C}_i^r(\mu)$, from tree-level contributions with one $\cL_6$ insertion.

To maximise the available information, one makes use of lattice simulations (with $\chi$PT relations implemented into the lattice analyses), unitarity constraints, crossing and analyticity, mainly in the form of dispersion relations. The limit of an infinite number of QCD colours turns also to be a very useful tool to estimate the unknown LECs.

An exhaustive description of the chiral phenomenology is beyond the scope of these lectures. Instead, I will just present a few examples at $\cO(p^4)$ to illustrate both the power and limitations of $\chi$PT.

\subsection{Meson decay constants}

\begin{figure}[t]
\begin{center}
\includegraphics[width=10.5cm]{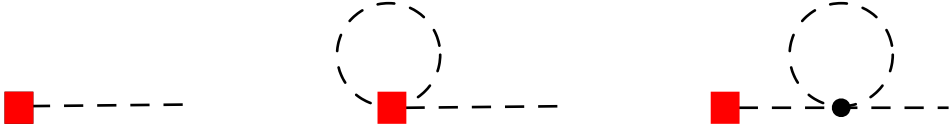}
\end{center}
\caption{Feynman diagrams contributing to the meson decay constants at the NLO.}
\label{fig:Fpi}
\end{figure}

The low-energy expansion of the pseudoscalar-meson decay constants in powers of the light quark masses is known to next-to-next-to-leading order (NNLO) \cite{Amoros:1999dp}. We only show here the NLO results \cite{Gasser:1984gg}, which originate from the Feynman topologies displayed in Fig.~\ref{fig:Fpi}. The red square indicates an insertion of the axial current, while the black dot is an $\cL_2$ vertex. The tree-level diagram involves the NLO expression of the axial current, which is obtained by taking the derivative of $\cL_4$ with respect to the axial source $a_\mu$. Obviously, only the $L_4$ and $L_5$ operators in \eqn{eq:l4} can contribute to the one-particle matrix elements. The middle topology is a one-loop correction with the LO axial current, {\it i.e.}, the $\Phi^3$ term in \eqn{eq:l_r_currents}. The last diagram is a wave-function renormalization correction.

In the isospin limit ($m_u = m_d = \hat m$), the $\chi$PT expressions take the form:
\beqn\label{eq:f_meson}
F_\pi & =  & F\, \left\{ 1 - 2\,\mu_\pi - \mu_K +
    \frac{4 M_\pi^2}{F^2}\, L_5^r(\mu)
    + \frac{8 M_K^2 + 4 M_\pi^2}{F^2}\, L_4^r(\mu)
    \right\}   , 
\no\\
F_K & =  & F\, \left\{ 1 - {3\over 4}\,\mu_\pi - {3\over 2}\,\mu_K
    - {3\over 4}\,\mu_{\eta_{\raisebox{-1pt}{$\scriptscriptstyle 8$}}}
    + \frac{4 M_K^2}{F^2}\, L_5^r(\mu)
    + \frac{8 M_K^2 + 4 M_\pi^2}{F^2}\, L_4^r(\mu)
    \right\}  , 
\no\\
F_{\eta_{\raisebox{-1pt}{$\scriptscriptstyle 8$}}} & =  & F\, \left\{ 1 - 3\,\mu_K +
    \frac{4 M_{\eta_{\raisebox{-1pt}{$\scriptscriptstyle 8$}}}^2}{F^2}\, L_5^r(\mu)
    + \frac{8 M_K^2 + 4 M_\pi^2}{F^2}\, L_4^r(\mu)
    \right\}\! , 
\eeqn
with
\be
\mu_P \,\equiv\, \frac{M_P^2}{32 \pi^2 F^2} \,
\log{\left( \frac{M_P^2}{\mu^2}\right)}\, .
\label{eq:mu_p}
\ee
Making use of \eqn{eq:running} and \eqn{eq:d_factors}, one easily verifies the renormalization-scale independence of these results. 

The $L_4$ contribution generates a universal shift of all pseudoscalar-meson decay constants,
$\delta F = 8 L_4 B_0\, \langle\cM\rangle$,
which can be eliminated taking ratios.
Using the most recent lattice average \cite{Aoki:2016frl} 
\be
F_K/F_\pi = 1.193\pm 0.003 \, ,
\label{eq:f_k_pi_ratio}
\ee
one can then determine $L_5^r(M_\rho)$; this gives the result quoted in
Table~\ref{tab:Lcouplings}. Moreover, one gets the absolute prediction 
\be
F_{\eta_{\raisebox{-1pt}{$\scriptscriptstyle 8$}}}/F_\pi = 1.31 \pm 0.02 \, .
\label{eq:f_eta_pi_ratio}
\ee

The absolute value of the pion decay constant is usually extracted from the measured $\pi^+\to\mu^+\nu_\mu$ decay amplitude, taking $|V_{ud}| = 0.97417\pm 0.00021$ from superallowed nuclear $\beta$ decays \cite{Hardy:2014qxa}. One gets then $F_\pi = (92.21\pm 0.14)$~MeV \cite{Patrignani:2016xqp}. The direct extraction from lattice simulations gives $F_\pi = (92.1\pm 0.6)$~MeV  \cite{Aoki:2016frl}, without any assumption concerning $V_{ud}$.

Lattice simulations can be performed at different (unphysical) values of the quark masses. Approaching the massless limit, one can then be sensitive to the chiral scale $F$. In the $n_f=2$ theory, one finds  \cite{Aoki:2016frl} 
\be 
F_\pi / F\, =\, 1.062\pm 0.007\, .
\ee
The relation between the fundamental scales of $n_f=2$ and $n_f=3$ $\chi$PT is easily obtained from the first equation in \eqn{eq:f_meson}:
\be
F_{\raisebox{-2pt}{$\scriptstyle SU(2)$}}\, =\, F_{\raisebox{-2pt}{$\scriptstyle SU(3)$}}\,\left\{ 1-\bar\mu_K + \frac{8 \bar M_K^2}{F^2}\, L_4^r(\mu)\right\}\, ,
\ee
where barred quantities refer to the limit $m_u=m_d=0$ \cite{Gasser:1984gg}.

\subsection{Electromagnetic form factors}

At LO, the pseudoscalar bosons have the minimal electromagnetic coupling that is generated through the covariant derivative. Higher-order corrections induce a momentum-dependent form factor, which is already known to NNLO \cite{Bijnens:1998fm,Bijnens:2002hp}:
\be
\langle \pi^+\pi^-|J^\mu_{\mathrm{em}}|0\rangle\, =\, (p_+^{\phantom{'}}-p_-^{\phantom{'}})^\mu \; F_V^{\pi}(s)\, ,
\ee
where $J^\mu_{\mathrm{em}} =\frac{2}{3}\,\bar u\gamma^\mu u - \frac{1}{3}\,\bar d\gamma^\mu d -\frac{1}{3}\,\bar s\gamma^\mu s $ is the electromagnetic current carried by the light quarks.
The same expression with the $K^+K^-$ and $K^0\bar K^0$ final states defines the analogous kaon form factors $F^{K^+}_V(s)$ and $F^{K^0}_V(s)$, respectively.
Current conservation guarantees that $F^{\pi}_V(0) = F^{K^+}_V(0)=1$, while
$F^{K^0}_V(0)=0$. Owing to charge-conjugation, there is no corresponding form factor for $\pi^0$ and $\eta$.

\begin{figure}[t]
\begin{center}
\includegraphics[width=12.5cm]{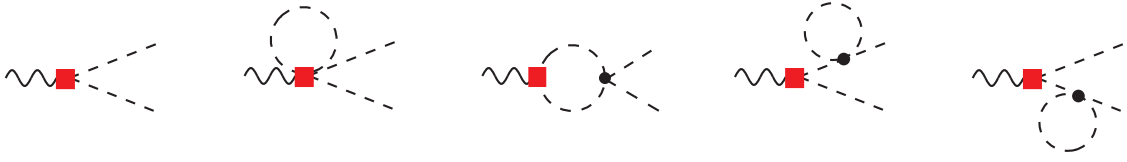}
\end{center}
\caption{Feynman diagrams contributing to the vector form factor at the NLO.}
\label{fig:FV}
\end{figure}

The topologies contributing at NLO to these form factors are displayed in Fig.~\ref{fig:FV}. The red box indicates an electromagnetic-current insertion, at NLO in the tree-level graph and at LO in the one-loop diagrams, while the black dot is an $\cL_2$ vertex. In the isospin limit, one finds the result \cite{Gasser:1984ux}:
\bel{eq:VFF}
F_V^{\pi}(s) \, =\, 1\, +\,\frac{2 L_9^r(\mu)}{F^2}\, s \, -\,
\frac{s}{96\pi^2 F^2}\,
\left[ A\left(\frac{M_\pi^2}{s},\frac{M_\pi^2}{\mu^2}\right) +
{1\over 2}\, A\left(\frac{M_K^2}{s},\frac{M_K^2}{\mu^2}\right)  \right]\, ,
\ee
where
\bel{eq:VFF2}
A\left(\frac{M_P^2}{s},\frac{M_P^2}{\mu^2}\right) \, =\,
\log{\left( \frac{M^2_P}{\mu^2}\right)} + {8 M^2_P\over s} -
\frac{5}{3}  + 
\sigma_P^3 \,\log{\left(\frac{\sigma_P+1}{\sigma_P-1}\right)}\, ,
\ee
with $\sigma_P \equiv \sqrt{1 - 4 M_P^2/s}$. 

The kaon electromagnetic form factors can also be expressed in terms of the same loop functions:
\begin{displaymath}
F_V^{K^0}(s) \, = \,
\frac{s}{192\pi^2 F^2}\,
\left[ A\left(\frac{M_\pi^2}{s},\frac{M_\pi^2}{\mu^2}\right) -
 A\left(\frac{M_K^2}{s},\frac{M_K^2}{\mu^2}\right)  \right]\, ,
\end{displaymath}
\bel{eq:VFF3}
F_V^{K^+}(s) \, =\, F_V^{\pi}(s) + F_V^{K^0}(s) \, .
\ee

At $\cO(p^4)$, there is only one local contribution that originates from the $L_9$ operator. This LEC can then be extracted from the pion electromagnetic radius, defined through the low-energy expansion
\be
F^{\phi^+}_V(s)\, =\, 1 + \frac{1}{6} \,
\langle r^2 \rangle^{\phi^+}_V \, s + \cO(s^2) \, ,
\qquad\qquad
F^{K^0}_V(s)\, =\,  \frac{1}{6} \,
\langle r^2 \rangle^{K^0}_V \, s + \cO(s^2)\, .
\label{eq:ff}
\ee
From \eqn{eq:VFF}, one easily obtains
\bel{eq:PiRadius}
\langle r^2 \rangle^{\pi^\pm}_V \, = \, {12 L^r_9(\mu)\over F^2}
    - {1\over 32 \pi^2 F^2} \left\{
   2 \log{\left({M_\pi^2\over\mu^2}\right)}
    + \log{\left({M_K^2\over\mu^2}\right)} + 3 \right\} \, ,
\ee
while \eqn{eq:VFF3} implies
\be\label{eq:Kradius}
\langle r^2 \rangle^{K^0}_V \, = \, - {1\over 16 \pi^2 F^2}
\,\log{\left({M_K\over M_\pi}\right) } \, ,
\qquad\qquad
\langle r^2 \rangle^{K^\pm}_V \, = \, \langle r^2 \rangle^{\pi^\pm}_V
   + \langle r^2 \rangle^{K^0}_V \, . 
\ee
In addition to the $L_9$ contribution, the meson electromagnetic radius
$\langle r^2 \rangle^{\phi^+}_V$
gets logarithmic loop corrections involving meson masses. The dependence on the renormalization scale $\mu$ cancels exactly between the logarithms and $L_9^r(\mu)$.
The measured electromagnetic pion radius,
$\langle r^2 \rangle^{\pi^\pm}_V = (0.439\pm 0.008) \, {\rm fm}^2$
\cite{Amendolia:1986wj},
is used as input to estimate the coupling $L_9$ in Table~\ref{tab:Lcouplings}.
The numerical value of this observable is dominated by the $L^r_9(\mu)$
contribution, for any reasonable value of $\mu$.

Since neutral bosons do not couple to the photon at tree level,
$\langle r^2 \rangle^{K^0}_V$
only gets a loop contribution, which is moreover finite
(there cannot be any divergence because symmetry forbids the presence of a local operator to renormalize it). The value predicted at $\cO(p^4)$,
$\langle r^2 \rangle^{K^0}_V = -(0.04\pm 0.03) \, \mathrm{fm}^2$, is in
good agreement with the experimental determination
$\langle r^2 \rangle^{K^0}_V = -(0.077\pm 0.010) \, \mathrm{fm}^2$ \cite{Patrignani:2016xqp}.
The measured $K^+$ charge radius,
$\langle r^2 \rangle^{K^\pm}_V = (0.34\pm 0.05) \, {\rm fm}^2$  \cite{Amendolia:1986ui},
has a much larger experimental uncertainty.
Within present errors, it is in agreement with the parameter-free
relation in eqn~\eqn{eq:Kradius}.

The loop function \eqn{eq:VFF2} contains the non-trivial logarithmic dependence on the momentum transfer, dictated by unitarity. It generates an absorptive cut above $s= 4 M_P^2$, corresponding to the kinematical configuration where the two intermediate pseudoscalars in the middle graph of Fig.~\ref{fig:FV} are on-shell.
According to the Watson theorem \cite{Watson:1954uc}, the phase induced by the $\pi\pi$ logarithm coincides with the phase shift of the elastic $\pi\pi$ scattering with $I=J=1$, which at LO is given by
\bel{eq:PhaseShift}
\delta_1^1(s)\, =\, \theta(s-4 M_\pi^2)\;\frac{s}{96\pi F^2}\, \left(1 - 4 M_\pi^2/s\right)^{3/2}\, .
\ee

\subsection[$K_{l3}$ decays]{$\mathbf{K_{\boldsymbol{\ell} 3}}$ decays}

The semileptonic decays $K^+\to\pi^0 \ell^+ \nu_\ell$ and
$K^0\to\pi^- \ell^+ \nu_\ell$ are governed by the corresponding
hadronic matrix elements of the strangeness-changing weak left current. Since the vector and axial components have $J^P=1^-$ and $1^+$, respectively, the axial piece cannot contribute to these $0^-\to 0^-$ transitions.
The relevant vector hadronic matrix element contains two possible Lorentz structures:
\be
\langle \pi| \bar s\gamma^\mu u |K\rangle = C_{K\pi} \,\left\{
\left( p_K + p_\pi\right)^\mu \, f_+^{K\pi}(t) \, + \,
\left( p_K - p_\pi\right)^\mu \, f_-^{K\pi}(t) \right\} ,
\label{eq:vector_matrix}
\ee
where $t\equiv (p_K-p_\pi)^2$,
$C_{K^+\pi^0} = -1/\sqrt{2}$ and $C_{K^0\pi^-} = -1$.
At LO, the two form factors reduce to trivial constants:
$f_+^{K\pi}(t) = 1$ and $f_-^{K\pi}(t) = 0$.
The normalization at $t=0$ is fixed to all orders by the conservation of the vector current, in the limit of equal quark masses. Owing to the Ademollo--Gatto theorem \cite{Ademollo:1964sr,Behrends:1960nf}, the deviations from one are of second order in the symmetry-breaking quark mass difference:
$f_+^{K^0\pi^-}\! (0) = 1 + \cO[(m_s-m_u)^2]$.
There is however a sizeable correction to $f_+^{K^+\pi^0}(t)$,
due to $\pi^0$--$\eta_{\raisebox{-1pt}{$\scriptscriptstyle 8$}}$ mixing, which is proportional to $m_d-m_u$:
\be
f_+^{K^+\pi^0}\! (0) \, = \,
1 + {3\over 4} \, {m_d-m_u\over m_s - \hat m}
\, = \, 1.017 \, .
\label{eq:fp_kp_p0}
\ee
The $\cO (p^4)$ corrections to $f_+^{K\pi}(0)$ can be expressed in
a parameter-free manner in terms of the physical meson masses.
Including those contributions, one obtains the more precise values \cite{Gasser:1984ux}
\be
 f_+^{K^0\pi^-}\! (0) = 0.977 \, , 
\qquad \qquad\qquad
{f_+^{K^+\pi^0}\! (0)\over f_+^{K^0\pi^-}\! (0)} = 1.022 \, .
\label{eq:fp_predictions}
\ee

\begin{figure}[t]
\begin{center}
\includegraphics[width=7.5cm]{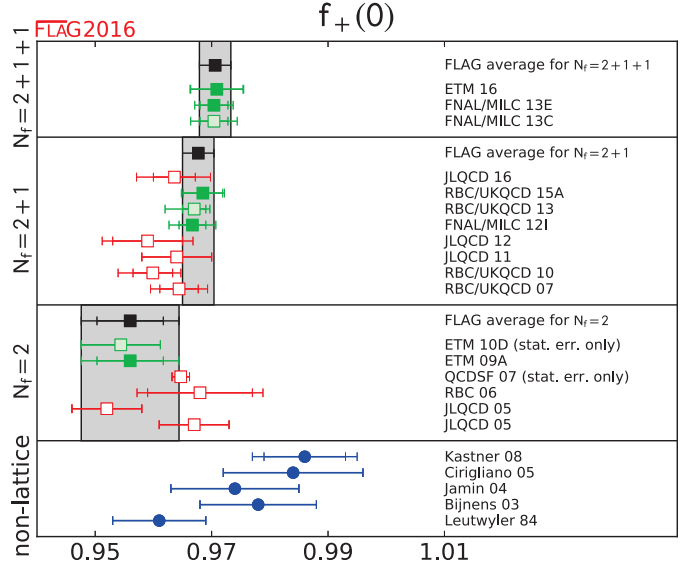}
\end{center}
\caption{Determinations of $f_+^{K^0\pi^-}\! (0)$ from lattice simulations and $\chi$PT analyses \cite{Aoki:2016frl}.}
\label{fig:FpKp}
\end{figure}

From the measured experimental decay rates, one gets a very accurate determination of the product \cite{Antonelli:2010yf,Moulson:2014cra}
\bel{eq:VusF+}
|V_{us}|\; f_+^{K^0\pi^-}\! (0)\, =\, 0.2165\pm 0.0004\, .
\ee
A theoretical prediction of $f_+^{K^0\pi^-}(0)$ with a similar accuracy is needed in order to profit from this information and extract the most precise value of the Cabibbo--Kobayashi--Maskawa matrix element $V_{us}$. The present status is displayed in Fig.~\ref{fig:FpKp}.

Since 1984, the standard value adopted for $f_+^{K^0\pi^-}\! (0)$ has been the $\cO(p^4)$ chiral prediction, corrected with a quark-model estimate of the $\cO(p^6)$ contributions, leading to $f_+^{K^0\pi^-}\! (0) = 0.961\pm 0.008$ \cite{Leutwyler:1984je}. This is however not good enough to match the current experimental precision. The two-loop $\chi$PT corrections, computed in 2003 \cite{Bijnens:2003uy}, turned out to be larger than expected, increasing the predicted value of $f_+^{K^0\pi^-}(0)$ \cite{Bijnens:2003uy,Cirigliano:2005xn,Jamin:2004re,Kastner:2008ch}. The estimated errors did not decrease, unfortunately, owing to the presence of $\cO(p^6)$ LECs that need to be estimated in some way. Lattice simulations were in the past compatible with the 1984 reference value, but the most recent and precise determinations \cite{Bazavov:2013maa,Carrasco:2016kpy}, done with $2+1+1$ active flavours, exhibit a clear shift to higher values, in agreement with the $\cO(p^6)$ $\chi$PT expectations. Taking the present ($2+1+1$) lattice average  \cite{Aoki:2016frl}
\bel{eq:F+(0)}
f_+^{K^0\pi^-}\! (0)\, =\, 0.9706\pm 0.0027\, ,
\ee
one gets:
\be
|V_{us}|\, =\, 0.2231 \pm 0.0007 \, .
\label{eq:v_us}
\ee

\subsection{Meson and quark masses}

The mass relations \eqn{eq:masses} get modified by $\cO (p^4)$ contributions that depend on the LECs $L_4$, $L_5$, $L_6$, $L_7$ and $L_8$.
It is possible, however, to obtain one relation between the
quark and meson masses, which does not contain any $\cO (p^4)$
coupling. The dimensionless ratios
\be
Q_1 \,\equiv\, {M_K^2 \over M_\pi^2} \, , 
\qquad\qquad\qquad
Q_2 \,\equiv\, {(M_{K^0}^2 - M_{K^+}^2)_{\mathrm{QCD}} 
    \over M_K^2 - M_{\pi}^2}\,  ,
\label{eq:q1q2_def}
\ee
get  the same $\cO (p^4)$ correction \cite{Gasser:1984gg}:
\be
Q_1\, =\, {m_s + \hat m \over 2 \hat m} \; \{ 1 + \Delta_M\} ,
\qquad\qquad\quad
Q_2\, =\, {m_d - m_u \over m_s - \hat m} \; \{ 1 + \Delta_M\} ,
\label{eq:q1q2}
\ee
where
\be
\Delta_M\, =\, - \mu_\pi + \mu_{\eta_{\raisebox{-1pt}{$\scriptscriptstyle 8$}}} + {8\over F^2}\,
(M_K^2 - M_\pi^2)\, \left[ 2 L_8^r(\mu) - L_5^r(\mu)\right] .
\label{eq:delta_M}
\ee
Therefore, at this order, the ratio $Q_1/Q_2$ is just given
by the corresponding ratio of quark masses,
\be
Q^2 \,\equiv\, {Q_1\over Q_2}\, =\,
{m_s^2 - \hat m^2 \over m_d^2 - m_u^2} \, .
\label{eq:Q2}
\ee
To a good approximation, \eqn{eq:Q2} can be written as the equation of an ellipse that relates the quark mass ratios:
\be
\left({m_u\over m_d}\right)^2 + {1\over Q^2}\,
\left({m_s\over m_d}\right)^2\, =\, 1 \, .
\label{eq:ellipse}
\ee

The numerical value of $Q$ can be directly extracted from the meson mass ratios \eqn{eq:q1q2_def}, but the resulting uncertainty is dominated by the violations of Dashen's theorem at $\cO (e^2\cM )$, which have been shown to be sizeable. A more precise determination has been recently obtained from a careful analysis of the $\eta\to 3\pi$ decay amplitudes, which leads to $Q= 22.0\pm 0.7$
\cite{Colangelo:2016jmc}.

Obviously, the quark mass ratios \eqn{eq:Weinbergratios},
obtained at $\cO (p^2)$, satisfy the elliptic constraint \eqn{eq:ellipse}.
At $\cO (p^4)$, however, it is not possible to make a separate
estimate of $m_u/m_d$ and $m_s/m_d$ without having additional
information on some LECs.
The determination of the individual quark mass ratios from eqs.~\eqn{eq:q1q2} would require to fix first the constant $L_8$. However, there is no way to find an observable that isolates this coupling. The reason is an accidental symmetry of the effective Lagrangian $\cL_2 + \cL_4$, which remains invariant under the following simultaneous change of the quark mass matrix and some of the chiral couplings  \cite{Kaplan:1986ru}:
\begin{displaymath}
\cM' \, = \, \alpha \,\cM + \beta\, (\cM^\dagger)^{-1} \, \det\cM\, ,
\qquad\qquad\qquad
B_0' \, = \, B_0 / \alpha\, ,\qquad\quad
\end{displaymath}
\be\label{eq:kmsymmetry}
L'_6 \, = \, L_6 - \zeta \, , 
\qquad\qquad
L'_7 \, = \, L_7 - \zeta \, , 
\qquad\qquad
L'_8 \, = \, L_8 + 2 \zeta \, , 
\ee
with $\alpha$ and $\beta$ arbitrary constants, and
$\zeta = \beta f^2 / (32\alpha B_0)$.
The only information on the quark mass matrix $\cM$ that was used
to construct the effective Lagrangian was that it transforms as
$\cM\to g_R^{\phantom{\dagger}} \cM g_L^\dagger$.
The matrix $\cM'$ transforms in the same manner;
therefore, symmetry alone does not allow us to distinguish between
$\cM$ and $\cM'$.
Since only the product $B_0 \cM$ appears in the Lagrangian,
$\alpha$ merely changes the value of the constant $B_0$.
The term proportional to $\beta$ is a correction of $\cO (\cM^2)$;
when inserted in $\cL_2$, it generates a contribution to
$\cL_4$ that gets reabsorbed by the redefinition
of the three $\cO (p^4)$ couplings.
All chiral predictions will be invariant under the transformation
\eqn{eq:kmsymmetry}; therefore it is not possible to
separately determine the values of the quark masses and the
constants $B_0$, $L_6$, $L_7$ and $L_8$.
We can only fix those combinations of chiral couplings and masses
that remain invariant under \eqn{eq:kmsymmetry}.
 
The ambiguity can be resolved with additional information from outside the pseudo\-scalar meson Lagrangian framework. For instance, by analysing 
isospin breaking in the baryon mass spectrum and the $\rho$--$\omega$
mixing, it is possible to fix the ratio  \cite{Gasser:1982ap}
\be
R\,\equiv\, {m_s - \hat m\over m_d - m_u}\, =\, 43.7\pm 2.7 \, .
\label{eq:r}
\ee
This ratio can also be extracted from lattice simulations, the current average being
$R = 35.6\pm 5.16$ \cite{Aoki:2016frl} (with $2+1+1$ dynamical fermions). Inserting this number in \eqn{eq:Q2}, the two separate quark mass ratios can be obtained.
Moreover, one can then determine  $L_8$ from \eqn{eq:q1q2}.

The quark mass ratios can be directly extracted from lattice simulations \cite{Aoki:2016frl}:
\be
\frac{m_s}{\hat m} = 27.30 \pm 0.34 \, ,
\qquad\qquad\qquad
\frac{m_u}{m_d} \, = \, 0.470\pm 0.056 \, .
\label{eq:ms_m_ratio_2}
\ee
The second ratio includes, however, some phenomenological information, in particular on electromagnetic corrections. Using only the first ratio and the numerical value of $Q$ extracted from $\eta\to 3 \pi$, one would predict $m_u/m_d = 0.44\pm 0.03$, in excellent agreement with the lattice determination and with a smaller uncertainty \cite{Colangelo:2016jmc}.

\section{Quantum anomalies}
\label{sec:anomalies}

Until now, we have been assuming that the symmetries of the classical Lagrangian remain valid at the quantum level. However, symmetries with different transformation properties for the left and right fermion chiralities are usually subject to quantum anomalies.  Although the Lagrangian is invariant under local chiral transformations, this is no longer true for the associated generating functional because the path-integral measure transforms non-trivially \cite{Fujikawa:1979ay,Fujikawa:1980eg}. The anomalies of the fermionic determinant break chiral symmetry at the quantum level \cite{Adler:1969gk,Adler:1969er,Bardeen:1969md,Bell:1969ts}.

\subsection{The chiral anomaly}

Let us consider again the $n_f=3$ QCD Lagrangian \eqn{eq:extendedqcd}, with external sources $v_\mu$, $a_\mu$, $s$ and $p$, and its local chiral symmetry  \eqn{eq:symmetry}. 
The fermionic determinant can always be defined with the convention
that $Z[v,a,s,p]$ is invariant under vector transformations.
Under an infinitesimal chiral transformation
\be
g_L^{\phantom{\dagger}}\, =\, 1 + i\, (\alpha - \beta) + \cdots\, ,
\qquad\qquad
g_R^{\phantom{\dagger}} = 1 + i\, (\alpha +\beta) + \cdots\, ,
\label{eq:inf}
\ee
with $\alpha = \alpha_a T^a$ and $\beta = \beta_a T^a$,
the anomalous change of the generating functional
is then given by \cite{Bardeen:1969md}:
\be
\delta Z[v,a,s,p]  \, = \,
-{N_C\over 16\pi^2} \, \int d^4x \,
\langle \beta(x) \,\Omega(x)\rangle \, ,
\label{eq:anomaly}
\ee
where $N_C =3$ is the number of QCD colours,
\be\label{eq:anomaly_b}
\Omega(x) \, = \,\varepsilon^{\mu\nu\sigma\rho} 
 \left[
v_{\mu\nu} v_{\sigma\rho}
+ {4\over 3} \,\nabla_\mu a_\nu \nabla_\sigma a_\rho
+ {2\over 3}\, i \,\{ v_{\mu\nu},a_\sigma a_\rho\}
+ {8\over 3}\, i \, a_\sigma v_{\mu\nu} a_\rho
+ {4\over 3} \, a_\mu a_\nu a_\sigma a_\rho \right]  
\ee
with $\varepsilon_{0123}^{\phantom{\dagger}}=1$, and
\be
v_{\mu\nu} \, = \,
\partial_\mu v_\nu - \partial_\nu v_\mu - i \, [v_\mu,v_\nu] \, ,
\qquad\qquad
\nabla_\mu a_\nu  \, = \,
\partial_\mu a_\nu - i \, [v_\mu,a_\nu] \, .
\label{eq:anomaly_c} 
\ee
Notice that $\Omega(x)$ only depends on the external fields $v_\mu$ and $a_\mu$, which have been assumed to be traceless.\footnote{Since $\langle \sigma^a \{ \sigma^b , \sigma^c\}\rangle = 0$, this non-abelian anomaly vanishes in the $SU(2)_L\otimes SU(2)_R$ theory. However, the singlet currents become anomalous when the electromagnetic interaction is included because the $n_f=2$ quark charge matrix, $\cQ =\mathrm{diag} (\frac{2}{3}, -\frac{1}{3})$ is not traceless \cite{Kaiser:2000ck}.} 
This anomalous variation of $Z$ is an $\cO (p^4)$ effect in the chiral counting.
 
We have imposed chiral symmetry to construct the $\chi$PT Lagrangian. Since this symmetry is explicitly violated by the anomaly at the fundamental QCD level, we need to add to the effective theory a functional $Z_{\cA}$ with the property that its change under a chiral gauge transformation reproduces 
\eqn{eq:anomaly}.
Such a functional was first constructed by Wess and Zumino \cite{Wess:1971yu},
and reformulated in a nice geometrical way by Witten \cite{Witten:1983tw}.
It has the explicit form:
\beqn
S[U,\ell,r]_{\mathrm{WZW}} & = &\mbox{}  
 -\,\dfrac{i N_C}{48 \pi^2} \int d^4 x\;
\varepsilon_{\mu \nu \alpha \beta}\left\{ W (U,\ell,r)^{\mu \nu
\alpha \beta} - W (I_3,\ell,r)^{\mu \nu \alpha \beta} \right\}
\no\\ &&\mbox{}
-\,\dfrac{i N_C}{240 \pi^2}
\int d\sigma^{ijklm}\, \left\langle \Sigma^L_i
\Sigma^L_j \Sigma^L_k \Sigma^L_l \Sigma^L_m \right\rangle \, ,
\label{eq:WZW}
\eeqn
where
\beqn
W (U,\ell,r)_{\mu \nu \alpha \beta} &\, = &\,
\langle U \ell_{\mu} \ell_{\nu} \ell_{\alpha}U^{\dg} r_{\beta} \rangle
+ \frac{1}{4}\,\langle U \ell_{\mu} U^{\dg} r_{\nu} U \ell_\alpha U^{\dg} 
r_{\beta}\rangle
+ i\,\langle U \partial_{\mu} \ell_{\nu} \ell_{\alpha} U^{\dg} r_{\beta}\rangle
\no\\ && \hskip -1.75cm\mbox{}
 +  i\,\langle \partial_{\mu} r_{\nu} U \ell_{\alpha} U^{\dg} r_{\beta}\rangle
- i\,\langle \Sigma^L_{\mu} \ell_{\nu} U^{\dg} r_{\alpha} U \ell_{\beta}\rangle
+ \langle\Sigma^L_{\mu} U^{\dg} \partial_{\nu} r_{\alpha} U \ell_\beta\rangle
- \langle\Sigma^L_{\mu} \Sigma^L_{\nu} U^{\dg} r_{\alpha} U \ell_{\beta}\rangle
\no\\ && \hskip -1.75cm\mbox{}
+ \langle\Sigma^L_{\mu} \ell_{\nu} \partial_{\alpha} \ell_{\beta}\rangle
+ \langle\Sigma^L_{\mu} \partial_{\nu} \ell_{\alpha} \ell_{\beta}\rangle
 - i\, \langle\Sigma^L_{\mu} \ell_{\nu} \ell_{\alpha} \ell_{\beta}\rangle
+ \dfrac{1}{2}\, \langle\Sigma^L_{\mu} \ell_{\nu} \Sigma^L_{\alpha} \ell_{\beta}\rangle
- i\, \langle\Sigma^L_{\mu} \Sigma^L_{\nu} \Sigma^L_{\alpha} \ell_{\beta}\rangle 
\no\\ && \hskip -1.75cm\mbox{} 
 - \left( L \leftrightarrow R \right)\, , 
 \label{eq:WZW2}
\eeqn
\be
\Sigma^L_\mu = U^{\dg} \partial_\mu U \, , 
\qquad\qquad\qquad
\Sigma^R_\mu = U \partial_\mu U^{\dg} \, ,
\label{eq:sima_l_r}
\ee
and
$\left( L \leftrightarrow R \right)$ stands for the interchanges
$U \leftrightarrow U^\dg $, $\ell_\mu \leftrightarrow r_\mu $ and
$\Sigma^L_\mu \leftrightarrow \Sigma^R_\mu $.
The integration in the second term of \eqn{eq:WZW} is over a
five-dimensional manifold whose boundary is four-dimensional 
Minkowski space. The integrand is a surface term; therefore both the first and the second terms of $S_{\mathrm{WZW}}$ are of $\cO (p^4)$, according to the chiral counting rules.
 
The effects induced by the anomaly are completely calculable because they have a short-distance origin. The translation from the fundamental quark-gluon level to the effective chiral level is unaffected by hadronization problems. In spite of its considerable complexity, the anomalous action \eqn{eq:WZW} has no free parameters. The most general solution to the anomalous variation \eqn{eq:anomaly} of the QCD generating functional is given by the Wess--Zumino--Witten (WZW) action \eqn{eq:WZW}, plus the most general chiral-invariant Lagrangian that we have been constructing before, order by order in the chiral expansion. 

The anomaly term does not get renormalized. Quantum loops including insertions of the WZW action generate higher-order divergences that obey the standard Weinberg's power counting and correspond to chiral-invariant structures. They get renormalized by the LECs of the corresponding $\chi$PT operators. 
 
The anomaly functional gives rise to interactions with a Levi-Civita pseudotensor that break the intrinsic parity. This type of vertices are absent in the LO and NLO $\chi$PT Lagrangians because chiral symmetry only allows for odd-parity invariant structures starting at $\cO(p^6)$. Thus, the WZW functional breaks an accidental symmetry of the $\cO(p^2)$ and $\cO(p^4)$ chiral Lagrangians, giving the leading contributions to processes with an odd number of pseudoscalars. In particular, the five-dimensional surface term in the second line of \eqn{eq:WZW} generates interactions among five or more Nambu--Goldstone bosons, such as $K^+K^-\to\pi^+\pi^-\pi^0$.

Taking $v_\mu = -e\cQ A_\mu$ in \eqn{eq:WZW2}, the first line in the WZW action is responsible for the decays $\pi^0\to 2\gamma$ and $\eta\to 2 \gamma$, and the interaction vertices $\gamma 3\pi$ and $\gamma\pi^+\pi^-\eta$. Keeping only the terms with a single pseudoscalar and two photon fields:
\bel{eq:pgg}
\cL_{\mathrm{WZW}}\,\dot=\, - \frac{N_C\alpha}{24\pi F}\;\varepsilon_{\mu\nu\sigma\rho}\; F^{\mu\nu} F^{\sigma\rho}\;\left(\pi^0+\frac{1}{\sqrt{3}}\,\eta_8\right)\, .
\ee
Therefore, the chiral anomaly makes a very strong non-perturbative prediction for the $\pi^0$ decay width,
\bel{eq:pi0-decay}
\Gamma(\pi^0\to\gamma\gamma)\, =\, \left(\frac{N_C}{3}\right)^2\;\frac{\alpha^2 M_\pi^3}{64\pi^3 F^2}\, =\, 7.7\;\mathrm{eV}\, ,
\ee
in excellent agreement with the measured experimental value of $(7.63\pm 0.16)$~eV \cite{Patrignani:2016xqp}.

\subsection{The $\mathbf{U(1)_A}$ anomaly}
\label{subsec:U(1)_A}

With $n_f=3$, the massless QCD Lagrangian has actually a larger $U(3)_L\otimes U(3)_R$ chiral symmetry. One would then expect nine Nambu--Goldstone bosons associated with the $\chi$SB to the diagonal subgroup $U(3)_V$. However, the lightest $SU(3)$-singlet pseudoscalar in the hadronic spectrum corresponds to a quite heavy state: the $\eta'(958)$.

The singlet axial current $J^\mu_5 = \bar q\gamma^\mu\gamma_5 q$
turns out to be anomalous \cite{Adler:1969gk,Adler:1969er,Bell:1969ts},
\bel{eq:AxialAnomaly}
\partial_\mu J^\mu_5\, =\, \frac{g_s^2 n_f}{32\pi^2}\;\varepsilon^{\mu\nu\sigma\rho}\,
G_{\mu\nu}^a G_{a,\sigma\rho}\, ,
\ee
which explains the absence of a ninth Nambu--Goldstone boson, but brings very subtle phenomenological implications. Although the right-hand side of  \eqn{eq:AxialAnomaly} is a total divergence (of a gauge-dependent quantity), the four-dimensional integrals of this term take non-zero values, which have a topological origin and characterize the non-trivial structure of the QCD vacuum \cite{Belavin:1975fg,Callan:1976je,Callan:1977gz,Jackiw:1976dw,Jackiw:1976pf,tHooft:1976snw,tHooft:1976rip}. It also implies the existence of an additional term in the QCD Lagrangian that violates $P$, $T$ and $CP$:
\bel{eq:Ltheta}
\cL_{\theta}\, =\, - \theta_0\,\frac{g^2}{64\pi^2}\; \varepsilon^{\mu\nu\sigma\rho}\,
G_{\mu\nu}^a G_{a,\sigma\rho}\, .
\ee

When diagonalizing the quark mass matrix emerging from the Yukawa couplings of the light quarks, one needs to perform a $U(1)_A$ transformation of the quark fields in order to eliminate the global phase $\mathrm{arg}(\det\cM)$. Owing to the axial anomaly, this transformation generates the Lagrangian $\cL_\theta$ with a $\theta$ angle equal to the rotated phase. The experimental upper bound on the neutron electric dipole moment puts then a very strong constraint on the effective angle
\bel{eq:theta}
|\theta|\,\equiv\, |\theta_0 + \mathrm{arg}(\det\cM)|\, \le\, 10^{-9}\, .
\ee
The reasons why this effective angle is so small remain to be understood (strong CP problem). A detailed discussion of strong CP phenomena within $\chi$PT can be found in Ref.~\cite{Pich:1991fq}.

The $U(1)_A$ anomaly vanishes in the limit of an infinite number of QCD colours
with $\alpha_s N_C$ fixed, {\it i.e.}, choosing the coupling constant $g_s$ to be of $\cO(1/\sqrt{N_C})$ \cite{tHooft:1973alw,tHooft:1974pnl,Witten:1979kh}. This is a very useful limit because the $SU(N_C)$ gauge theory simplifies considerably at $N_C\to\infty$, while keeping many essential properties of QCD. There exist a systematic expansion in powers of $1/N_C$, around this limit, which for $N_C=3$ provides a good quantitative approximation scheme to the hadronic world \cite{Manohar:1998xv} (see Section~\ref{sec:LargeNC}).

In the large-$N_C$ limit, we can then consider a $U(3)_L\otimes U(3)_R\to U(3)_V$ chiral symmetry, with nine Nambu--Goldstone excitations that can be conveniently collected in the $3\times 3$ unitary matrix
\bel{eq:U_tilde}
\widetilde U(\vec{\phi\,}) \,\equiv\,
\exp{\left\{ i{\sqrt{2}\over F}\,\widetilde\Phi \right\}}
\, , 
\qquad\qquad\qquad
\widetilde\Phi\,\equiv\, \frac{\eta_{\raisebox{-1pt}{$\scriptstyle 1$}}}{\sqrt{3}}\, I_3 \, + \,
\frac{\vec{\lambda}}{\sqrt{2}}\,\vec{\phi} \, .
\ee
Under the chiral group, $\widetilde U(\vec{\phi\,})$ transforms as
$\widetilde U\to g_R^{\phantom{\dagger}}\, \widetilde U g_L^\dagger\; $, with $g_{R,L}^{\phantom{\dagger}}\in U(3)_{R,L}$.
The LO interactions of the nine pseudoscalar bosons are described by the Lagrangian \eqn{eq:lowestorder} with $\widetilde U(\vec{\phi\,})$ instead of $U(\vec{\phi\,})$.
Notice that the $\eta_{\raisebox{-1pt}{$\scriptstyle 1$}}$ kinetic term in
$\langle D_\mu\widetilde U^\dagger D^\mu\widetilde U\rangle$ decouples from
the $\vec\phi$ fields and the $\eta_{\raisebox{-1pt}{$\scriptstyle 1$}}$ particle becomes stable in the chiral limit.

To lowest non-trivial order in $1/N_C$, the chiral symmetry breaking
effect induced by the $U(1)_A$ anomaly can be taken into
account in the effective low-energy theory, through the term
\cite{DiVecchia:1980yfw,Rosenzweig:1979ay,Witten:1980sp}
\bel{eq:anom_term}
\cL_{U(1)_A} \, = \, - \frac{F^2}{4} \frac{\tilde a}{N_C} \, 
\left\{ {i \over 2 } \left[\log{(\det{\widetilde U})} - \log{(\det{
\widetilde U^\dagger})}\right] \right\}^2  ,
\ee
which breaks $U(3)_L \otimes U(3)_R$ but preserves
$SU(3)_L \otimes SU(3)_R \otimes U(1)_V$. The
parameter $\,\tilde a \,$ has dimensions of mass squared and, with the factor
$1/N_C$ pulled out, is booked to be of $\cO (1)$ in the large-$N_C$
counting rules. Its value is not fixed by symmetry requirements alone;
it depends crucially on the dynamics of instantons. In the presence
of the term \eqn{eq:anom_term}, the $\eta_{\raisebox{-1pt}{$\scriptstyle 1$}}$ field becomes massive even in the chiral limit:
\bel{eq:M_eta1}
M_{\eta_{\raisebox{-1pt}{$\scriptscriptstyle 1$}}}^2\, =\, 3\, \frac{\tilde a}{N_C} + \cO (\cM) \, .
\ee

Owing to the large mass of the $\eta'$, the effect of the $U(1)_A$ anomaly
cannot be treated as a small perturbation. Rather, one should keep
the term \eqn{eq:anom_term} together with the LO Lagrangian
\eqn{eq:lowestorder}. It is possible to build a consistent combined expansion
in powers of momenta, quark masses and $1/N_C$, by counting the
relative magnitude of these parameters as \cite{Leutwyler:1996sa}:
\bel{eq:U(3)_counting}
\cM \sim 1/N_C \sim p^2 \sim \cO (\delta) \, .
\ee
A $U(3)_L \otimes U(3)_R$ description \cite{HerreraSiklody:1996pm,HerreraSiklody:1997kd,Kaiser:2000gs}
of the pseudoscalar particles, including the singlet $\eta_{\raisebox{-1pt}{$\scriptstyle 1$}}$ field, allows one to understand many properties of the $\eta$ and $\eta'$ mesons in a quite systematic way.

\section{Massive fields and low-energy constants}
\label{sec:MassiveFields}

The main limitation of the EFT approach is the proliferation of unknown LECs. At LO, the symmetry constraints severely restrict the allowed operators, making possible to derive many phenomenological implications in terms of a small number of dynamical parameters. However, higher-order terms in the chiral expansion are much more sensitive to the non-trivial aspects of the underlying QCD dynamics. All LECs are in principle calculable from QCD, but, unfortunately, we are not able at present to perform such a first-principles computation. Although the functional integration over the quark fields in  \eqn{eq:generatingfunctional} can be explicitly done, we do not know how to perform analytically the remaining integral over the gluon field configurations. Numerical simulations in a discretized space-time lattice offer a promising tool to address the problem, but the current techniques are still not good enough to achieve a complete matching between QCD and its low-energy effective theory. On the other side, a perturbative evaluation of the gluonic contribution would obviously fail in reproducing the correct dynamics of $\chi$SB.

A more phenomenological approach consists in analysing the massive states of the hadronic QCD spectrum that, owing to confinement, is a dual asymptotic representation of the quark and gluon degrees of freedom. The QCD resonances encode the most prominent features of the non-perturbative strong dynamics, and it seems rather natural to expect that the lowest-mass states, such as the $\rho$ mesons, should have an important impact on the physics of the pseudoscalar bosons. In particular, the exchange of those resonances should generate sizeable contributions to the chiral couplings. Below the $\rho$ mass scale, the singularity associated with the pole of a resonance propagator is replaced by the corresponding momentum expansion:
\bel{eq:RhoPropagator}
\frac{1}{s-M_R^2}\, =\, -\frac{1}{M_R^2}\;\sum_{n=0} \left(\frac{s}{M_R^2}\right)^n
\qquad\qquad (s\ll M_R^2)\, .
\ee
Therefore, the exchange of virtual resonances generates
derivative Nambu--Goldstone couplings proportional to powers of $1/M_R^2$.

\subsection{Resonance chiral theory}

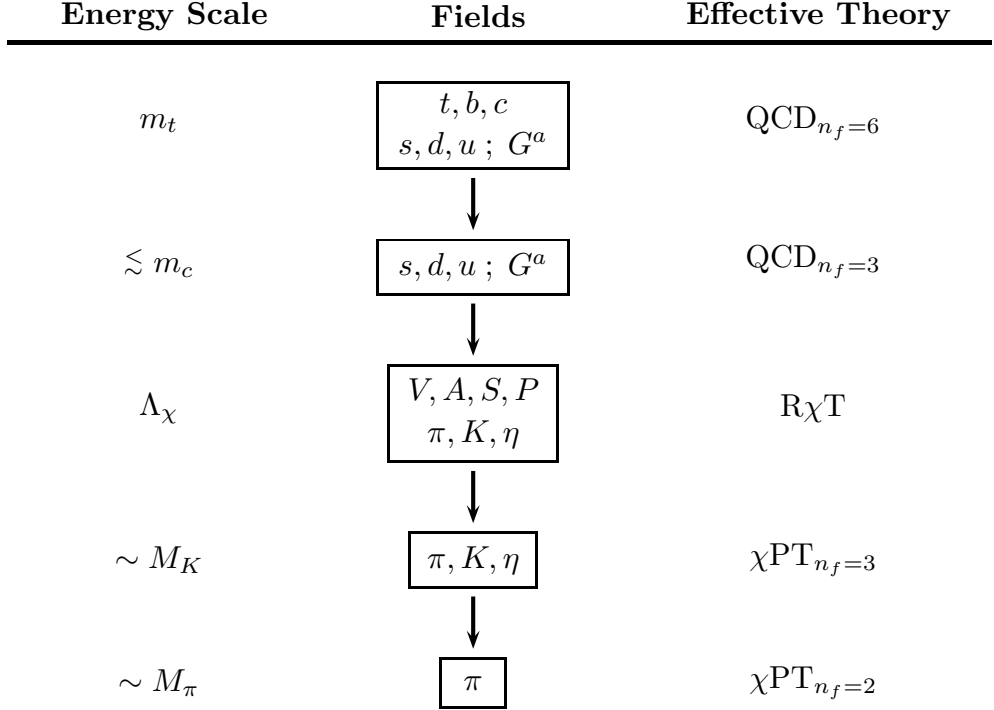
\begin{figure}[t]
\begin{center}
{\setlength{\unitlength}{.58mm}
\begin{picture}(190,140)      
\put(0,0){\makebox(190,140){}} 
\put(5,130){\makebox(50,10){ {\bf Energy\ Scale}}}
\put(70,130){\makebox(40,10){ {\bf Fields}}}
\put(125,130){\makebox(60,10){ {\bf Effective\ Theory}}}
\put(1,130){{\linethickness{1.2pt}\line(1,0){188}}} {
\put(15,106){\makebox(30,17){{$m_t$}}}
\put(72,106){{\linethickness{.6pt}\framebox(36,16){{$\ba
      t, b, c \\[1pt] s, d, u\; ;\; G^a \ea $}}}}
     }
\put(135,106){\makebox(40,17){{QCD${}_{n_f=6}$}}} {
\put(15,79){\makebox(30,18){{$\lsim m_c$}}}
\put(72,82){{\linethickness{.6pt}\framebox(36,10){{$s, d, u\;
;\; G^a $}}}}
             }
\put(135,79){\makebox(40,18){{QCD${}_{n_f=3}$}}} {
\put(15,51){\makebox(30,18){{$\Lambda_\chi$}}}
\put(74,50){{\linethickness{.6pt}\framebox(32,18){{$\ba V, A, S, P  \\[1pt]
            \pi, K,\eta  \ea $}}}}
         }
\put(135,51){\makebox(40,18){{R$\chi$T}}} {
\put(15,22){\makebox(30,18){{$\sim M_K$}}}
\put(78,26){{\linethickness{.6pt}\framebox(24,10){{$\pi, K,\eta
$}}}}
             }
\put(135,22){\makebox(40,18){{$\chi\mathrm{PT}_{n_f=3}$}}} {
\put(15,0){\makebox(30,15){{$\sim M_\pi$}}}
\put(84,3){{\linethickness{.6pt}\framebox(12,9){{$\pi$}}}}
             }
\put(135,0){\makebox(40,15){{$\chi\mathrm{PT}_{n_f=2}$}}}

{\psset{xunit=.58mm,yunit=.58mm}
\psline[linewidth=1.3pt]{->}(90,104)(90,94)
\psline[linewidth=1.3pt]{->}(90,80)(90,70)
\psline[linewidth=1.3pt]{->}(90,48)(90,38)
\psline[linewidth=1.3pt]{->}(90,24)(90,14)}
\end{picture}
} 
\end{center}
\caption{Chain of effective field theories, from $m_t$ to the pion mass scale.}
\label{fig:EFTchain}
\end{figure}

A systematic analysis of the role of resonances in the $\chi$PT Lagrangian was first performed at $\cO(p^4)$ in Refs.~\cite{Ecker:1989yg,Ecker:1988te}, and extended later to the $\cO(p^6)$ LECS \cite{Cirigliano:2006hb}. One writes a general chiral-invariant Lagrangian $\cL(U,V,A,S,P)$, describing the couplings of meson resonances of the type $V(1^{--})$, $A(1^{++})$, $S(0^{++})$ and $P(0^{-+})$ to the Nambu--Goldstone bosons, at LO in derivatives. The coupling constants of this Lagrangian are phenomenologically extracted from physics at the resonance mass scale. One has then an effective chiral theory defined in the intermediate energy region, with the generating functional \eqn{eq:generatingfunctional} given by the path-integral formula
\be
\exp{\{i Z\}} \, = \,
\int \, \cD U\, \cD V \,\cD A \,\cD S \,\cD P
\, \exp{\left\{ i \int d^4x \;\cL(U,V,A,S,P) \right\}}\, .
\label{eq:pi_relation}
\ee
This resonance chiral theory (R$\chi$T) constitutes an interpolating representation between the short-distance QCD description and the low-energy $\chi$PT framework, which can be schematically visualized through the chain of effective field theories
displayed in Fig.~\ref{fig:EFTchain}. The functional integration of the heavy fields  leads to a low-energy theory with only Nambu--Goldstone bosons. At LO, this integration can be explicitly performed through a perturbative expansion around the classical solution for the resonance fields.  Expanding the resulting non-local action in powers of momenta, one gets then the local $\chi$PT Lagrangian with its LECs predicted in terms of the couplings and masses of the R$\chi$T.

The massive states of the hadronic spectrum have definite transformation properties under the vacuum symmetry group $H\equiv SU(3)_V$. In order to couple them to the Nambu--Goldstone modes, in a chiral-invariant way, we can take advantage of the compensating transformation $h(\vec\phi,g)$, in eqn~\eqn{eq:h_def}, which appears under the action of $G$ on the canonical coset representative $\xi_R^{\phantom{\dagger}}(\vec{\phi\,}) = \xi_L^\dagger (\vec{\phi\,})\equiv u(\vec{\phi\,})$:
\be\label{eq:h_def_2}
u(\vec{\phi\,})\,\toG\, g_R^{\phantom{\dagger}}\,u(\vec{\phi\,})\, h^\dagger(\vec\phi,g)\, = \,
h(\vec\phi,g)\,u(\vec{\phi\,})\, g_L^\dagger \, . 
\ee
A chiral transformation of the quark fields $(g_L^{\phantom{\dagger}},g_R^{\phantom{\dagger}})\in G$ induces a corresponding transformation $h(\vec\phi,g)\in H$, acting on the hadronic states. 

In practice, we shall only be interested in resonances transforming as octets or singlets under $SU(3)_V$. Denoting  the resonance multiplets generically by                    
$R_8= \vec{\lambda}\vec{R}/\sqrt{2}$  (octet) and $R_1$ (singlet), the            
non-linear realization of $G$ is given by                                      
\be
R_8\,\toG\, h(\vec\phi,g) \; R_8 \; h^\dagger (\vec\phi,g)\, ,                                                                            
\qquad\qquad\qquad
R_1\,\toG\, R_1 \, .                                               
\label{eq:R_transformation}
\ee
Since the action of $G$ on the octet field $R_8$ is local, we must define a covariant derivative                                                                      
\be                                                                    
\nabla_\mu R_8 \, =\, \partial_\mu R_8 + [\Gamma_\mu,R_8] \, ,
\label{eq:d_covariant}                         
\ee
with the connection                                                                         
\be                                                        
\Gamma_\mu \, =\,                                                                
\frac{1}{2} \left\{ u^\dagger (\partial_\mu - ir_\mu) u +                   
  u(\partial_\mu - i\ell_\mu) u^\dagger \right\}                               
\label{eq:connection}                                                                             
\ee
ensuring the proper transformation                                              
\be                                                                            
\nabla_\mu R_8 \,\toG\, h(\vec\phi,g)\; \nabla_\mu R_8 \;\, h^\dagger(\vec\phi,g)\, .
\label{eq:dc_transf}                                             
\ee

It is useful to define the covariant quantities 
\be\label{eq:octet_objects}
u_\mu \equiv  i\, u^\dagger (D_\mu U) u^\dagger  =  u_\mu^\dagger\, ,                                                               
\qquad\quad                                                     
\chi_\pm^{\phantom{\dagger}}  \equiv  u^\dagger \chi u^\dagger \pm u \chi^\dagger u\, ,            
\qquad\quad
f^{\mu\nu}_\pm   =   u F_L^{\mu\nu} u^\dagger \pm                    
u^\dagger   F_R^{\mu\nu} u \, ,
\ee
which transform as $SU(3)_V$ octets: $X\toG h(\vec\phi,g)\, X\, h^\dagger(\vec\phi,g)$. Remembering that $U=u^2$, it is a simple exercise to rewrite all $\chi$PT operators in terms of these variables. For instance,
$\langle D_\mu U^\dagger D^\mu U \rangle = \langle u^\mu u_\mu\rangle$ and
$\langle U^\dagger\chi +\chi^\dagger U\rangle = \langle\chi_+^{\phantom{\dagger}}\rangle$. The advantage of the quantities \eqn{eq:octet_objects} is that they can be easily combined with the resonance fields to build chiral-invariant structures.

In the large-$N_C$ limit, the octet and singlet resonances become degenerate in the chiral limit. We can then collect them in a nonet multiplet 
\bel{eq:nonet}
R\, \equiv\,  R_8 + \frac{1}{\sqrt{3}}\, R_0\, I_3
\, =\, \frac{1}{\sqrt{2}}\,\vec{\lambda}\,\vec{R}  + \frac{1}{\sqrt{3}}\, R_0\, I_3\, ,
\ee
with a common mass $M_R$.
To determine the resonance-exchange contributions to the $\cO(p^4)$ $\chi$PT Lagrangian, we only need the LO couplings to the Nambu--Goldstone modes that are linear in the resonance fields. The relevant resonance Lagrangian can be written as \cite{Ecker:1988te,Pich:2002xy}
\be
\cL_{\mathrm{R}\chi\mathrm{T}}\,\dot=\, \sum_R\,\cL_R 
\, .                                                                            
\label{eq:res_Lagrangian}                                                                            
\ee

The spin-0 pieces take the form ($R=S,P$)
\be\label{eq:L_S_P}
\cL_R \, =\, {1\over 2} \, \langle \nabla^\mu R\, \nabla_\mu R - M^2_R\, R^2\rangle
+ \langle R\,\chi_R^{\phantom{\dagger}}\rangle \, .
\ee
Imposing $P$ and $C$ invariance, the corresponding resonance interactions
are governed by the $\cO(p^2)$ chiral structures
\be\label{eq:chiSP}
\chi_{S}^{\phantom{\dagger}} \, = \,  c_d \, u_\mu u^\mu + c_m \, \chi_+^{\phantom{\dagger}}\, ,
\qquad\qquad\qquad
\chi_P^{\phantom{\dagger}}\, =\, d_m \, \chi_-^{\phantom{\dagger}}\, .
\ee

At low energies, the solutions of the resonance equations of motion,
\bel{eq:EoM_S_P}
(\nabla^2 + M_R^2)\, R \, =\, \chi_R^{\phantom{\dagger}} \, ,
\ee
can be expanded in terms of the local chiral operators that only contain light fields:
\be
R \, =\, \frac{1}{M_R^2}\;\chi_R^{\phantom{\dagger}}
\, +\, \cO\!\left(\frac{p^4}{M_R^4}\right)\, .
\ee
Substituting these expressions back into $\cL_R$, in eqn~\eqn{eq:L_S_P}, one obtains the corresponding LO contributions to the $\cO(p^4)$ $\chi$PT Lagrangian: 
%
\be 
\Delta\cL_4^R\, =\, \sum_{R=S,P}\,\frac{1}{2M_R^2}\;
\langle\chi_R^{\phantom{\dagger}}\chi_R^{\phantom{\dagger}}\rangle \, ,
\ee
Rewriting this result in the standard basis of $\chi$PT operators in eqn~\eqn{eq:l4}, one finally gets the spin-0 resonance-exchange contributions to the $\cO(p^4)$ LECs \cite{Ecker:1988te,Pich:2002xy}:
\bel{eq:s_exchange}
L_3^S\, =\, \frac{c_d^2}{2 M_S^2} \, , 
\qquad\qquad
L_5^S \, =\,  \frac{c_d c_m}{M_S^2} \, , 
\qquad\qquad 
L_8^{S+P}\, =\, \frac{c_m^2}{2 M_S^2}  - \frac{d_m^2}{2 M_P^2}\, .
\ee
Thus, scalar exchange contributes to $L_3$, $L_5$ and $L_8$, while the exchange of pseudoscalar resonances only shows up in $L_8$.

We must also take into account the presence of the $\eta_{\raisebox{-1pt}{$\scriptstyle 1$}}$ state, which is the lightest pseudoscalar resonance in the hadronic spectrum. Owing to the $U(1)_A$ anomaly, this singlet state has a much larger mass than the octet of Nambu--Goldstone bosons, and it is integrated out together with the other massive resonances. Its LO coupling can be easily extracted from the $U(3)_L\otimes U(3)_R$ chiral Lagrangian, which incorporates the matrix $\widetilde  U(\vec{\phi\,} )$ that collects the pseudoscalar nonet:
\bel{eq:Leta1}
\cL_2^{U(3)}\, \dot=\, \frac{F^2}{4}\; \langle \widetilde U^\dagger\chi +\chi^\dagger\widetilde U\rangle\quad\longrightarrow\quad
-i\,\frac{F}{\sqrt{24}}\;\eta_{\raisebox{-1pt}{$\scriptstyle 1$}}\;\langle\chi_-^{\phantom{\dagger}}\rangle\, .
\ee
The exchange of an $\eta_{\raisebox{-1pt}{$\scriptstyle 1$}}$ meson generates then the $\chi$PT operator $\langle\chi_-^{\phantom{\dagger}}\rangle^2$, with a coupling\footnote{As displayed in eqn~\eqn{eq:s_exchange}, the exchange of a complete nonet of pseudoscalars does not contribute to $L_7$. The singlet and octet contributions exactly cancel at large $N_C$.\label{foot:L7}}
\bel{eq:L7}
L_7^{\eta_{\raisebox{-1pt}{$\scriptscriptstyle 1$}}}\, =\, - \frac{F^2}{48 M_{\eta_{\raisebox{-1pt}{$\scriptscriptstyle 1$}}}^2}\, .
\ee

\begin{figure}[t]
\begin{center}
\includegraphics[width=8.5cm]{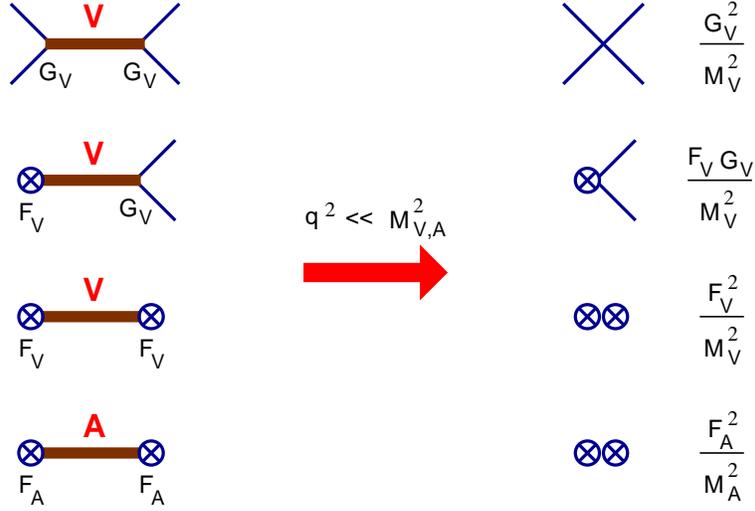}
\end{center}
\caption{Contributions to the $\chi$PT LECs from $V$ and $A$ resonance exchange. The cross denotes the insertion of an external vector or axial-vector current.}
\label{fig:VectorExchange}
\end{figure}

For technical reasons, the vector and axial-vector mesons are more conveniently described in terms of antisymmetric tensor fields $V_{\mu\nu}$ and $A_{\mu\nu}$  \cite{Ecker:1988te,Gasser:1983yg}, respectively, instead of the more familiar Proca field formalism.\footnote{
The antisymmetric formulation of spin-1 fields avoids mixings with the Nambu--Goldstone modes and has better ultraviolet properties. The two descriptions are related by a change of variables in the corresponding path integral~\cite{Bijnens:1995ii,Kampf:2006yf} and give the same physical predictions, once a proper short-distance behaviour is required \cite{Ecker:1989yg,Pich:2016lew}.}
Their corresponding Lagrangians read ($R=V,A$)
\be\label{eq:L_V_A} 
\cL_R \, = \, 
    - {1\over 2}\, \langle \nabla^\lambda R_{\lambda\mu}                      
\nabla_\nu R^{\nu\mu} -{M^2_R\over 2} \, R_{\mu\nu} R^{\mu\nu}\rangle
\, +\, \langle R_{\mu\nu}\chi_R^{\mu\nu}\rangle\,  ,
\ee
with the $\cO(p^2)$ chiral structures
\be\label{eq:chiVA}
\chi_V^{\mu\nu}\, =\,\frac{F_V}{2\sqrt{2}} \; f_+^{\mu\nu} +                                   
  \frac{i\, G_V}{\sqrt{2}} \;  u^\mu u^\nu\, ,
\qquad\qquad\qquad
\chi_A^{\mu\nu}\, =\,\frac{F_A}{2\sqrt{2}} \; f_-^{\mu\nu}\, .
\ee
Proceeding in the same way as with the spin-0 resonances, one easily gets the vector and axial-vector contributions to the $\cO(p^4)$ $\chi$PT LECS \cite{Ecker:1988te}:
\beqn
L_1^V &\, = &\, \frac{G_V^2}{8 M_V^2}\, , 
\qquad\qquad 
L_2^V\, =\, 2 L_1^V\, ,
\qquad\qquad 
L_3^V\, =\, -6 L_1^V\, , 
\no\\  
L_9^V &\, = &\, \frac{F_V G_V}{2 M_V^2}\, , 
\qquad\qquad\qquad
L_{10}^{V+A}\, =\, - \frac{F_V^2}{4 M_V^2} + \frac{F_A^2}{4 M_A^2} \, . 
\label{eq:vmd_results}
\eeqn
The dynamical origin of these results is graphically displayed in Fig.~\ref{fig:VectorExchange}. Therefore, vector-meson exchange generates contributions to $L_1$, $L_2$, $L_3$, $L_9$ and $L_{10}$ \cite{Donoghue:1988ed,Ecker:1988te}, while $A$ exchange only contributes to $L_{10}$ \cite{Ecker:1988te}.

The estimated resonance-exchange contributions to the $\chi$PT LECS bring a dynamical understanding of the phenomenological values of these constants shown in Table~\ref{tab:Lcouplings}. The couplings $L_4$ and $L_6$, which do not receive any resonance contribution, are much smaller than the other $\cO(p^4)$ LECs, and consistent with zero. $L_1$ is correctly predicted to be positive and the relation $L_2 = 2 L_1$ is approximately well satisfied. $L_7$ is negative, as predicted in \eqn{eq:L7}. The absolute magnitude of this parameter can be estimated from the quark-mass expansion of $M_\eta^2$ and $M_{\eta'}^2$, which fixes $M_{\eta_{\raisebox{-1pt}{$\scriptscriptstyle 1$}}} = 804$~MeV \cite{Ecker:1988te}. Taking $F\approx F_\pi = 92.2$~MeV, one predicts $L_7=-0.3\cdot 10^{-3}$ in perfect agreement with the value given in Table~\ref{tab:Lcouplings}.

To fix the vector-meson parameters, we will take $M_V=M_\rho = 775$~MeV and $|F_V|= 154$ MeV, from $\Gamma(\rho^0\to e^+e^-)$. The electromagnetic pion radius determines $F_V G_V>0$, correctly predicting a positive value for $L_9$. From this observable one gets $|G_V| = 53$ MeV, but this is equivalent to fit $L_9$. A similar value $|G_V|\sim 69$~MeV, but with a much larger uncertainty, can be extracted from $\Gamma(\rho^0\to 2\pi)$ \cite{Ecker:1988te}. The axial parameters can be determined using the old Weinberg sum rules \cite{Weinberg:1967kj}:
$F_A^2 = F_V^2 - F_\pi^2 = (123 \, {\rm MeV})^2$ and $M_A^2 = M_V^2
F_V^2/ F_A^2 = (968 \, {\rm MeV})^2$ (see the next subsection).

The resulting numerical values of the $L_i$ couplings \cite{Ecker:1988te} 
are summarized in the fifth column of Table~\ref{tab:Lcouplings}. Their comparison  with the phenomenologically determined values of $L_i^r(M_\rho)$ clearly establish a chiral version of vector (and axial-vector) meson dominance: whenever they can contribute at all, $V$ and $A$ exchange seem to completely dominate the relevant coupling constants.
Since the information on the scalar sector is quite poor, the values of $L_5$ and $L_8$ have actually been used to determine $c_d/M_S$ and $c_m/M_S$ (neglecting completely the contribution to $L_8$ from the much heavier pseudoscalar nonet). Therefore, these results cannot be considered as evidence for scalar dominance, although they provide a quite convincing demonstration of its consistency.

\subsection{Short-Distance Constraints}
\label{subsec:SDconstraints}

Since the R$\chi$T is an effective interpolation between QCD and $\chi$PT, the short-distance properties of the underlying QCD dynamics provide useful constraints on its parameters \cite{Ecker:1989yg}:

\begin{enumerate}

\item {\it Vector Form Factor.} \
At tree-level, the matrix element of the vector current between two Nambu--Goldstone states, is characterised by the form factor
\bel{eq:VFF-RChT}
F_V(t)\, =\, 1\, + \, 
\frac{F_{V}\, G_{V}}{F^2}\; \frac{t}{M_{V}^2-t} \, .
\ee
Since $F_V(t)$ should vanish at infinite momentum transfer $t$, the resonance couplings should satisfy
\bel{eq:SD1}
F_{V}\, G_{V}\, =\, F^2\, .
\ee

\item {\it Axial Form Factor.} \
The matrix element of the axial current between one Nambu--Goldstone boson and
one photon is parametrized by the axial form factor $G_A(t)$. The resonance
Lagrangian \eqn{eq:L_V_A} implies
\bel{eq:AFF}
G_A(t)\, =\, 
\frac{2\, F_{V}\, G_{V}- F_{V}^2}{M_{V}^2}\, +\,
\frac{F_{A}^2}{M_{A}^2-t} \, ,
\ee
which vanishes at $t\to\infty$ provided that
\bel{eq:SD2}
2\, F_{V}\, G_{V}\, =\, F_{V}^2\, .
\ee

\item {\it Weinberg Sum Rules.} \
In the chiral limit, the two-point function built from a left-handed and a right-handed vector quark currents,
\bel{eq:LRcorrelator}
2 i\int d^4x\;\e^{iqx}\;\langle 0 | T[J_{L,12}^\mu(x) J_{R,12}^{\nu\dagger}(0)] | 0\rangle\, =\, (-g^{\mu\nu} q^2 + q^\mu q^\nu)\; \Pi_{LR}(q^2)\, ,
\ee
defines the correlator
\bel{eq:WSR}
\Pi_{LR}(t)\, =\, {F^2\over t} \, +\,
{F_{V}^2\over M_{V}^2-t} \, -\,
{F_{A}^2\over M_{A}^2-t} \, .
\ee
Since gluonic interactions preserve chirality, $\Pi_{LR}(t)$ is identically zero in QCD perturbation theory. At large momenta, its operator product expansion can only get non-zero contributions from operators that break chiral symmetry, which 
have dimensions $d\ge 6$ when $m_q=0$. This implies that $\Pi_{LR}(t)$ vanishes faster than $1/t^2$ at $t\to\infty$,  \cite{Bernard:1975cd,Floratos:1978jb,Pascual:1981jr}. Imposing this condition on \eqn{eq:WSR}, one gets the relations \cite{Weinberg:1967kj}
\bel{eq:SD3}
F_{V}^2 - F_{A}^2\, =\, F^2 \, ,
\qquad\qquad\qquad
M_{V}^2 F_{V}^2 - M_{A}^2 F_{A}^2 = 0 \, .
\ee
They imply that $F_V> F_A$ and $M_V < M_A$. Moreover,
\bel{eq:FV_A}
F_V^2\, =\, \frac{F^2 M_A^2}{M_A^2-M_V^2}\, ,
\qquad\qquad\qquad
F_A^2\, =\, \frac{F^2 M_V^2}{M_A^2-M_V^2}\, .
\ee

\item {\it Scalar Form Factor.} \
The matrix element of the scalar quark current between one kaon and one pion
contains the form factor \cite{Jamin:2000wn,Jamin:2001zq}
\bel{eq:SFF}
F^S_{K\pi}(t)\, =\, 1\, + \, 
\frac{4\, c_{m}}{F^2}\left\{ c_{d} +
\left( c_{m}-c_{d}\right)\,
\frac{M_K^2-M_\pi^2}{M_{S}^2}\right\}
 \frac{t}{M_{S}^2-t} \, .
\ee
Requiring $F^S_{K\pi}(t)$ to vanish at $t\to\infty$, one gets the constraints 
\bel{eq:SD4}
4\, c_{d}\, c_{m}\, =\, F^2 \, ,
\qquad\qquad\qquad
c_{m}\, =\, c_{d}\, .
\ee
%

\item {\it $SS-PP$ Sum Rules.} \
The two-point correlation functions of two scalar or two pseudoscalar
currents would be equal if chirality was absolutely preserved. Their
difference is easily computed in the R$\chi$T:
\bel{eq:SSR}
\Pi_{SS-PP}(t)\, =\, 16\, B_0^2\,\left\{ 
\frac{c_{m}^2}{M_{S}^2-t} \, -\,
\frac{d_{m}^2}{M_{P}^2-t} \, +\, {F^2\over 8\, t}\right\}\, .
\ee
For massless quarks, $\Pi_{SS-PP}(t)$ vanishes as $1/t^2$ when
$t\to\infty$, with a coefficient proportional to $\alpha_s\,\langle\bar q\Gamma q\,\bar q\Gamma q\rangle$ \cite{Jamin:1994vr,Shifman:1978by,Shifman:1978bx}.
The vacuum four-quark condensate provides a non-perturbative breaking
of chiral symmetry. In the large-$N_C$ limit, it factorizes as
$\alpha_s\,\langle\bar q q\rangle^2 \sim \alpha_s\, B_0^2$.
Imposing this behaviour on \eqn{eq:SSR}, one gets \cite{Golterman:1999au}
\bel{eq:SD5}
c_{m}^2 - d_{m}^2 \, =\, \frac{F^2}{8}\,  ,
\qquad\qquad\qquad
c_{m}^2 M_{S}^2 - d_{m}^2 M_{P}^2\, =\, 
\frac{3\,\pi\alpha_s}{4}\; F^4\, .
\ee

\end{enumerate}

The relations~\eqn{eq:SD1}, \eqn{eq:SD2} and \eqn{eq:FV_A}
determine the vector and axial-vector couplings in terms of $M_V$
and $F$ \cite{Ecker:1989yg}:
\bel{eq:VA_coup}
F_V = 2\, G_V = \sqrt{2}\, F_A = \sqrt{2}\, F \, ,
\qquad\qquad\qquad
M_A = \sqrt{2}\, M_V \, .
\ee
The scalar \cite{Jamin:2000wn,Jamin:2001zq} and pseudoscalar parameters are obtained from the analogous constraints \eqn{eq:SD4} and \eqn{eq:SD5} \cite{Pich:2002xy}:
\bel{eq:SP_coup}
c_m = c_d = \sqrt{2}\, d_m = F/2 \, ,
\qquad\qquad\qquad
M_P = \sqrt{2}\, M_S \, \left(1 - \delta\right)^{1/2}\, .
\ee
The last relation involves a small correction \
$\delta \approx 3\,\pi\alpha_s F^2/M_S^2 \sim 0.08\,\alpha_s$ \
that can be neglected together with the tiny contributions from the
light quark masses.

Inserting these values into \eqn{eq:s_exchange} and \eqn{eq:vmd_results},
one obtains quite strong predictions for the $\cO(p^4)$ LECs in terms
of only $M_V$, $M_S$ and $F$:
\bel{eq:Li_SRA_1}
2\, L_1 = L_2 = \frac{1}{4}\, L_9 = -\frac{1}{3}\, L_{10}
= \frac{F^2}{8\, M_V^2}\, ,
\no\ee
\bel{eq:Li_SRA_2}
L_3 = -\frac{3\, F^2}{8\, M_V^2} + \frac{F^2}{8\, M_S^2}\, ,
\qquad\qquad
L_5 =\frac{F^2}{4\, M_S^2}\, ,
\qquad\qquad
L_8 = \frac{3\, F^2}{32\, M_S^2}\, .
\ee
The last column in Table~\ref{tab:Lcouplings} shows the numerical results obtained with $M_V = 0.775$~GeV,
$M_S = 1.4$~GeV and $F=92.2$~MeV. Also shown is the $L_7$
prediction in \eqn{eq:L7}, taking
$M_{\eta_{\raisebox{-1pt}{$\scriptstyle 1$}}} = 0.804$~GeV. The excellent agreement with the measured values demonstrates that the lightest resonance multiplets give indeed the dominant contributions to the $\chi$PT LECs.

\section{The limit of a very large number of QCD colours}
\label{sec:LargeNC}

The phenomenological success of resonance exchange can be better understood in the $N_C\to\infty$ limit of QCD \cite{tHooft:1973alw,tHooft:1974pnl,Witten:1979kh}. The $N_C$ dependence of the $\beta$ function determines that the strong coupling scales as $\alpha_s\sim\cO(1/N_C)$. Moreover, the fact that there are $N_C^2-1\approx N_C^2$ gluons, while quarks only have $N_C$ colours, implies that the gluon dynamics becomes dominant at large values of $N_C$. 

\begin{figure}[h]
\begin{center}
\includegraphics[width=12cm]{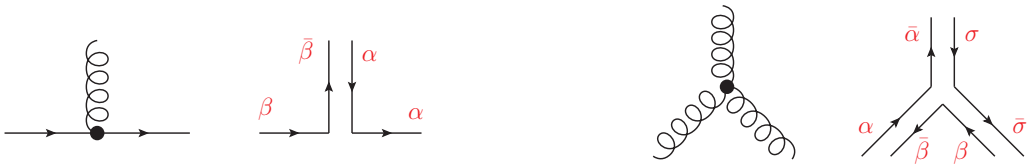}
\end{center}
\caption{Double-line notation, representing the gluonic colour flow.}
\label{fig:DlineNotation}
\end{figure}

The counting of colour factors in Feynman diagrams is most easily done considering the gluon fields as $N_C\times N_C$ matrices in colour space, $(G_\mu)^\alpha_{\phantom{\alpha}\beta} = G_\mu^a\, (T^a)^\alpha_{\phantom{\alpha}\beta}$, so that the colour flow becomes explicit as in $\bar q_\alpha (G_\mu)^\alpha_{\phantom{\alpha}\beta}\, q^\beta$. This can be represented diagrammatically with a double-line, indicating the gluon colour-anticolour, as illustrated in Fig.~\ref{fig:DlineNotation}.

\begin{figure}[t]
\begin{center}
\mbox{}\vskip .3cm
\begin{minipage}[c]{6cm}\centering
\includegraphics[width=6cm]{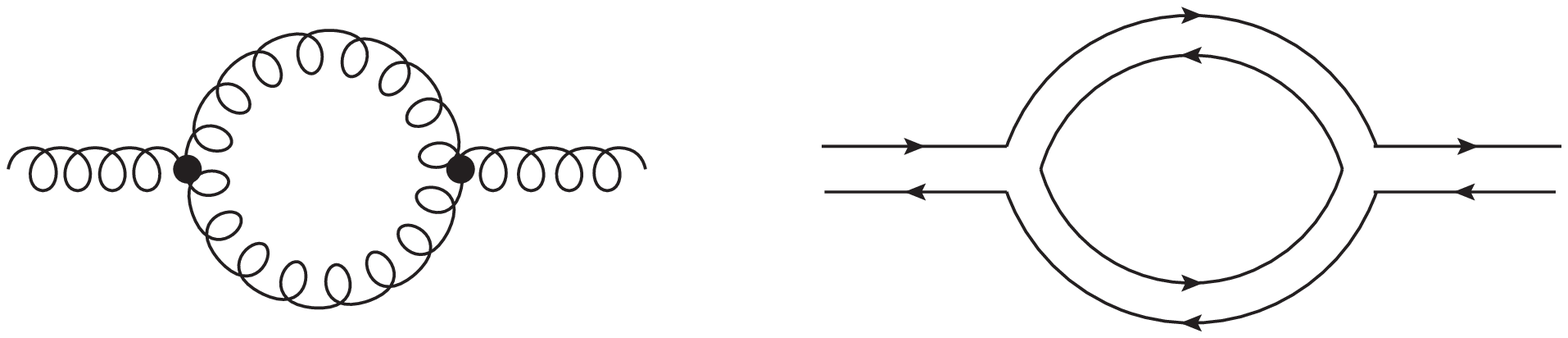}
\vskip .25cm
$(1/\sqrt{N_C})^2\; N_C = 1$
\end{minipage}
\hskip .75cm
\begin{minipage}[c]{6cm}\centering
\includegraphics[width=6cm]{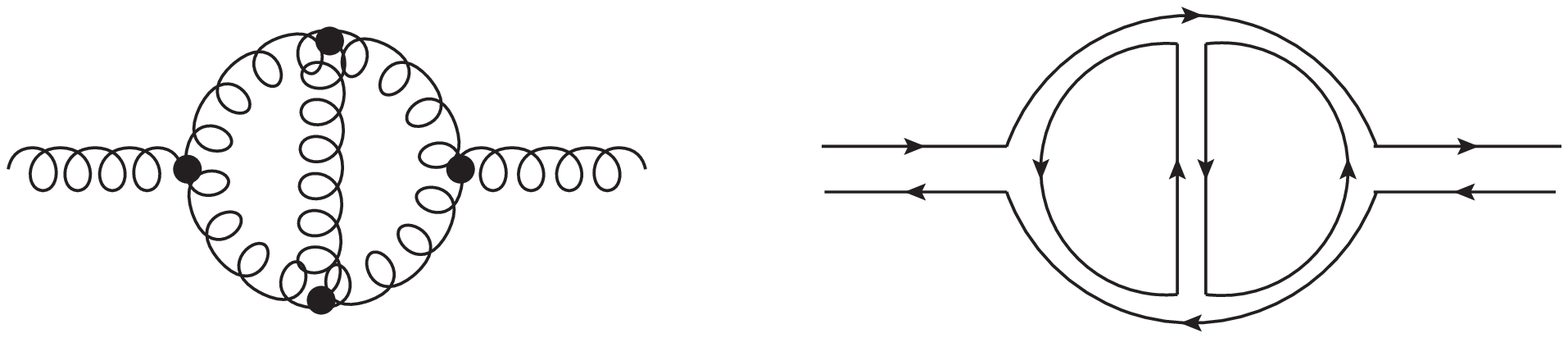}
\vskip .25cm
$(1/\sqrt{N_C})^4\; N_C^2 = 1$
\end{minipage}
\vskip .7cm
\begin{minipage}[c]{6cm}\centering
\includegraphics[width=6cm]{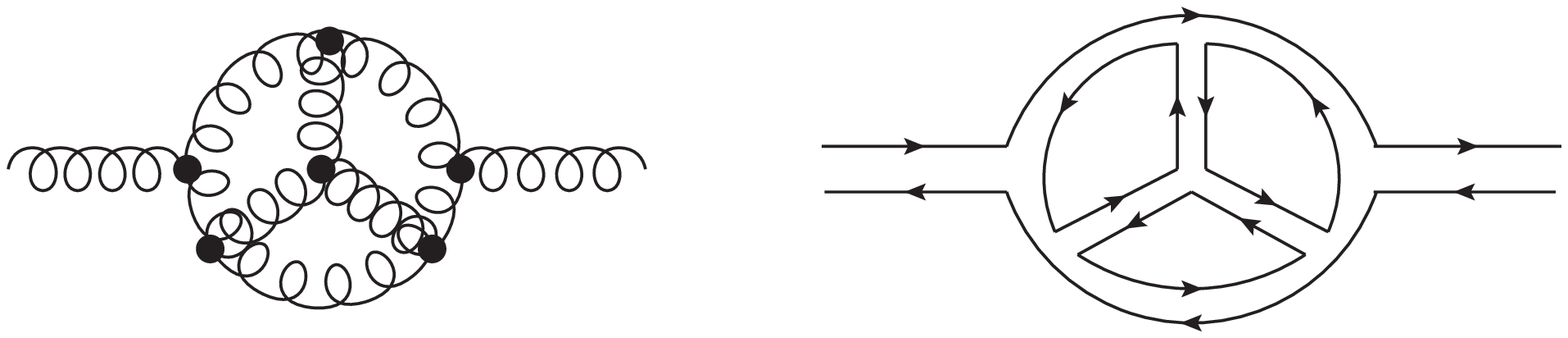}
\vskip .25cm
$(1/\sqrt{N_C})^6\; N_C^3 = 1$
\end{minipage}
\hskip .75cm
\begin{minipage}[c]{6cm}\centering
\includegraphics[width=6cm]{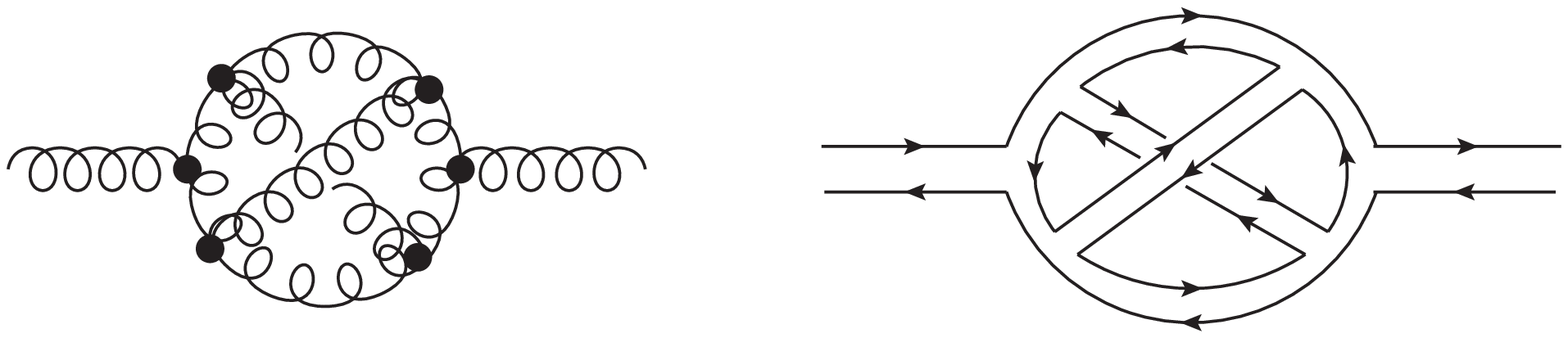}
\vskip .25cm
$(1/\sqrt{N_C})^6\; N_C = 1/N_C^2$
\end{minipage}
\end{center}
\caption{Large-$N_C$ counting of different gluon topologies.}
\label{fig:GluonNcCounting}
\end{figure}

\begin{figure}[t]
\begin{center}
\begin{minipage}[c]{6cm}\centering
\includegraphics[width=6cm]{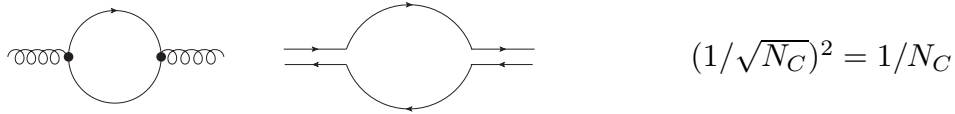}
\end{minipage}
\hskip 1.5cm
\begin{minipage}[c]{3cm}\centering
$(1/\sqrt{N_C})^2 = 1/N_C$
\end{minipage}
\end{center}
\caption{$N_C$ suppression of quark loops.}
\label{fig:QuarkNcCounting}
\end{figure}

\begin{figure}[t]
\begin{center}
\begin{minipage}[c]{1.75cm}\centering
\mbox{}\vskip .4cm
\includegraphics[height=1.5cm]{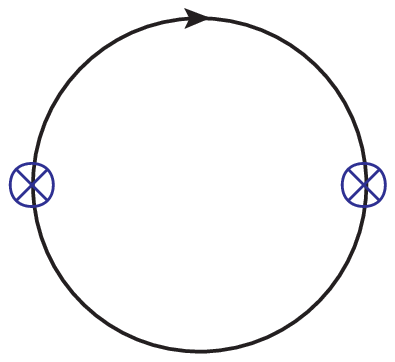}
\vskip .25cm
$N_C$
\end{minipage}
\hskip .8cm
\begin{minipage}[c]{4.5cm}\centering
\mbox{}\vskip .4cm
\includegraphics[height=1.5cm]{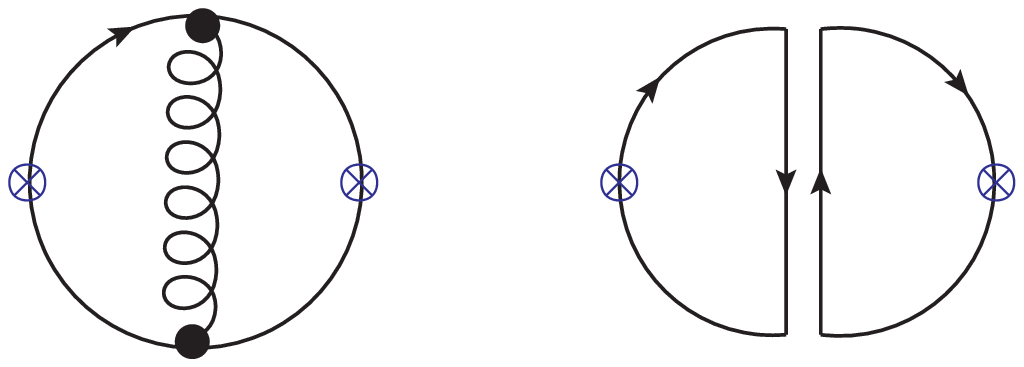}
\vskip .25cm
$(1/\sqrt{N_C})^2\; N_C^2 = N_C$
\end{minipage}
\hskip .8cm
\begin{minipage}[c]{4.5cm}\centering
\includegraphics[height=1.8cm]{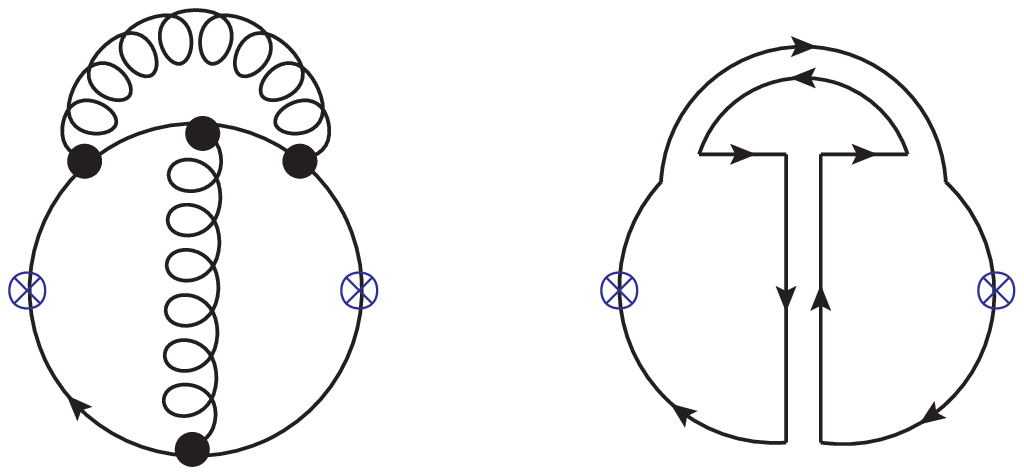}
\vskip .25cm
$(1/\sqrt{N_C})^4\; N_C = 1/N_C$
\end{minipage}
\vskip .7cm
\begin{minipage}[c]{6.cm}\centering
\includegraphics[width=4.75cm]{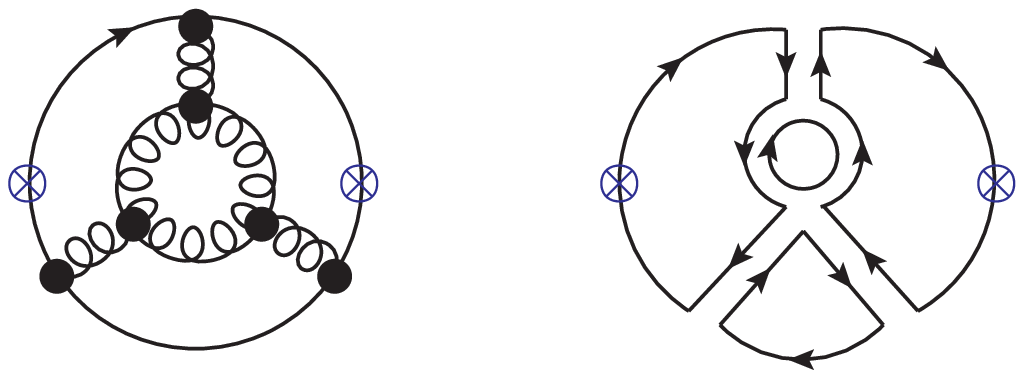}
\vskip .25cm
$(1/\sqrt{N_C})^6\; N_C^4 = N_C$
\end{minipage}
\hskip .75cm
\begin{minipage}[c]{6.cm}\centering
\includegraphics[width=4.75cm]{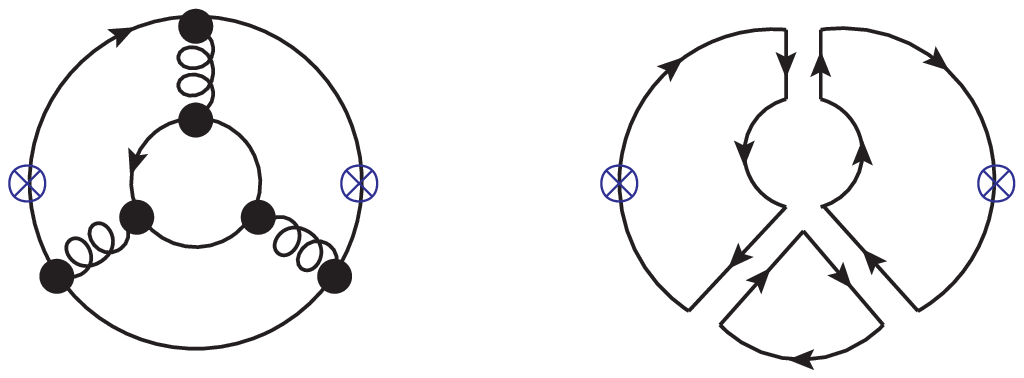}
\vskip .25cm
$(1/\sqrt{N_C})^6\; N_C^3 = 1$
\end{minipage}
\end{center}
\caption{Large-$N_C$ counting of quark correlation functions. The crosses indicate the insertion of a quark current $\bar q\Gamma q$.}
\label{fig:QuarkcorrNcCounting}
\end{figure}

Figures~\ref{fig:GluonNcCounting}, \ref{fig:QuarkNcCounting} and
\ref{fig:QuarkcorrNcCounting} display a selected set of topologies, with their associated colour factors. The combinatorics of Feynman diagrams at large $N_C$ results in simple counting rules, which characterize the $1/N_C$ expansion:
\begin{enumerate}
\item Dominance of planar diagrams with an arbitrary number of gluon exchanges
(Fig.~\ref{fig:GluonNcCounting}), and a single quark loop at the edge in the case of quark  correlation functions (Fig.~\ref{fig:QuarkcorrNcCounting}).
\item Non-planar diagrams are suppressed by factors of $1/N_C^2$ (last topology in Fig.~\ref{fig:GluonNcCounting} and third one in Fig.~\ref{fig:QuarkcorrNcCounting}).
\item Internal quark loops are suppressed by factors of $1/N_C$ (Fig.~\ref{fig:QuarkNcCounting} and last topology in Fig.~\ref{fig:QuarkcorrNcCounting}).
\end{enumerate}
%

The summation of the leading planar diagrams is a very formidable task
that we are still unable to perform. Nevertheless, making the very
plausible assumption that colour confinement persists at $N_C\to\infty$,
a very successful qualitative picture of the meson world emerges.

\begin{figure}[t]
\begin{center}
\includegraphics[width=10cm]{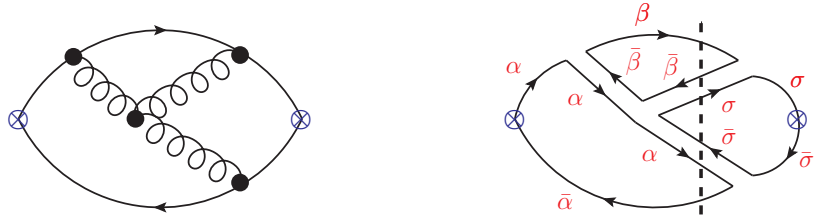}
\end{center}
\caption{Planar 3-gluon correction to a 2-point function. The absorptive cut indicated by the dashed line corresponds to the colour-singlet configuration $\bar q_\alpha\, G^\alpha_{\protect\phantom{\alpha}\sigma}\,
G^\sigma_{\protect\phantom{\sigma}\beta}\, q^\beta$.
}
\label{fig:ColourCut}
\end{figure}

Let us consider a generic $n$-point function of local quark bilinears\
$J = \bar q\,\Gamma q$:
\bel{eq:n-point}
\langle 0 | T\left[ J_1(x_1) \cdots J_n(x_n)\right] | 0 \rangle \,\sim\, \cO(N_C)\, .
\ee
A simple diagrammatic analysis shows that, at large $N_C$, the only singularities correspond to one-meson poles \cite{Witten:1979kh}. This is illustrated in Fig.~\ref{fig:ColourCut} with the simplest case of a two-point function. The dashed vertical line indicates an absorptive cut, {\it i.e.}, an on-shell intermediate state. Clearly, being a planar diagram with quarks only at the edges, the cut can only contain a single $q \bar q$ pair. Moreover, the colour flow clearly shows that the intermediate on-shell quarks and gluons form a single colour-singlet state $\bar q_\alpha G^\alpha_{\protect\phantom{\alpha}\sigma}
G^\sigma_{\protect\phantom{\sigma}\beta} q^\beta$; no smaller combination of them is separately colour singlet. Therefore, in a confining theory, the intermediate state is a perturbative approximation to a single meson. Analysing other diagrammatic configurations, one can easily check that this is a generic feature of planar topologies. Therefore, at $N_C\to\infty$,
%
%
the two-point function has the following spectral decomposition in momentum space:
\bel{eq:2-point}
\langle 0 | J_1(k) \, J_2(-k) | 0 \rangle\, =\, \sum_n \frac{F_n^2}{k^2-M_n^2} \, .
\ee
From this expression, one can derive the following interesting implications:
\begin{itemize}
\item[i)] Since the left-hand side is of $\cO(N_C)$,
$F_n = \langle 0|J|n\rangle \sim \cO(\sqrt{N_C}\,)$ \ and \
$M_n\sim \cO(1)$.

\item[ii)] There are an infinite number of meson states because, in QCD, the correlation function
$\langle 0 | J_1(k) \, J_2(-k) | 0 \rangle$
behaves logarithmically for large $k^2$ (the sum on the right-hand side would behave as $1/k^2$ with a finite number of states).

\item[iii)] The one-particle poles in the sum are on the real axis. Therefore, the meson states are free, stable and non-interacting at $N_C\to\infty$.
\end{itemize}

\begin{figure}[t]
\begin{center}
\includegraphics[width=7cm]{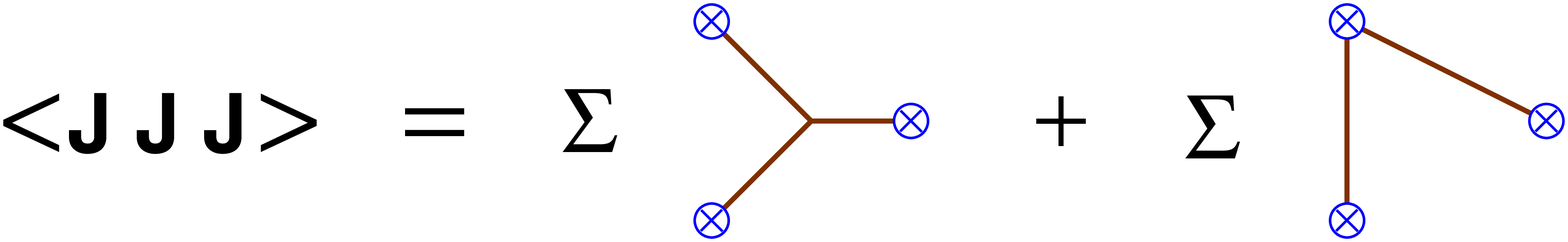}
\vskip .5cm
\includegraphics[width=13cm]{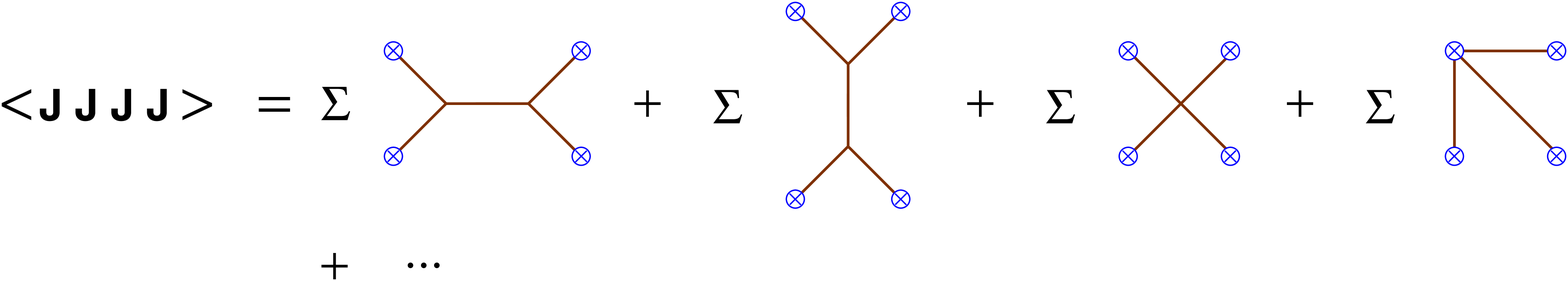}
\end{center}
\caption{Three-point and four-point correlation functions at large $N_C$.}
\label{fig:3point}
\end{figure}

Analysing in a similar way n-point functions, with $n>2$, confirms that the only singularities of the leading planar diagrams, in any kinematical channel, are one-particle single poles \cite{Witten:1979kh}. Therefore, at $N_C\to\infty$, the corresponding amplitudes are given by sums of tree-level diagrams with exchanges of free meson propagators, as indicated in Fig.~\ref{fig:3point} for the 3-point and 4-point correlators. There may be simultaneous poles in several kinematical variables $p_1^2,\, p_2^2,\,\ldots\, p_n^2$. For instance, the 3-point function contains contributions with single-poles in three variables, plus topologies with two simultaneous poles. The coefficients of these pole contributions should be non-singular functions of momenta; {\it i.e.}, polynomials that can be interpreted as local interaction vertices. Thus, the 3-pole terms contain an interaction vertex among three mesons, while the 2-pole diagrams include a current coupled to two mesons.
Similarly, the 4-point function contains topologies with four (five) simultaneous single-poles and interaction vertices among four (three) mesons, plus 3-pole contributions with some currents coupled to two or three mesons.

Since these correlation functions are of $\cO(N_C)$,
the local interaction vertices among 3 and 4 mesons, shown in the figure, scale as $V_3\sim\cO(N_C^{-1/2})$ and $V_4\sim\cO(N_C^{-1})$, respectively. Moreover, the 2-meson current matrix element behaves as $\langle 0|J|M_1 M_2\rangle\sim \cO(1)$, while $\langle 0|J|M_1 M_2 M_3\rangle\sim \cO(N_C^{-1/2})$.  In general, the $N_C$ dependence of an effective local interaction vertex among $m$ mesons is $V_m\sim\cO(N_C^{1-m/2})$, and a current matrix element with $m$ mesons scales as $\langle 0|J|M_1\cdots M_m\rangle\sim \cO(N_C^{1-m/2})$. Each additional meson coupled to the current $J$ or to an interaction vertex brings then a suppression factor $1/\sqrt{N_C}$.

Including gauge-invariant gluon operators, 
such as $J_G = \mathrm{Tr}\left(G_{\mu\nu}G^{\mu\nu}\right)$,
the diagrammatic analysis can be easily extended to glue states \cite{Witten:1979kh}. Since
$\langle 0| T\left(J_{G_1} \cdots J_{G_n}\right)| 0\rangle \sim \cO(N_C^2)$,
one derives the large-$N_C$ counting rules: \
$\langle 0|J_G|G_1\cdots G_m\rangle\sim \cO(N_C^{2-m})$ \
and \ $V[G_1,\cdots,G_m]\sim \cO(N_C^{2-m})$.
Therefore, at $N_C\to\infty$, glueballs are also free, stable, 
non-interacting and infinite in number. The mixed correlators involving both quark and gluon operators satisfy
$\langle 0| T\left(J_1 \cdots J_n J_{G_1} \cdots J_{G_m}| 0\right)\rangle 
\sim \cO(N_C)$, which implies 
$V[M_1,\cdots,M_p;G_1,\cdots,G_q]\sim \cO(N_C^{1-q-p/2})$.
Therefore, glueballs and mesons decouple at large $N_C$, their mixing
being suppressed by a factor $1/\sqrt{N_C}$.

Many known phenomenological features of the hadronic world are easily understood at LO in the $1/N_C$ expansion: suppression of the $\bar q q$ sea (exotics),
quark model spectroscopy, Zweig's rule, $SU(3)$ meson nonets,
narrow resonances, multiparticle decays dominated by resonant
two-body final states, etc.
In some cases, the large-$N_C$ limit is in fact the only
known theoretical explanation that is sufficiently general.
Clearly, the expansion in powers of $1/N_C$ appears to be
a sensible physical approximation at $N_C = 3$.
Notice that the QED coupling has a similar size \
$e = \sqrt{4\pi\alpha}\approx 0.3$.

The large-$N_C$ limit provides a weak coupling regime to perform quantitative QCD studies. At LO in $1/N_C$, the scattering amplitudes are given by sums of tree diagrams with physical hadrons exchanged. Crossing and unitarity imply that this
sum is the tree approximation to some local effective Lagrangian. 
Higher-order corrections correspond to hadronic loop diagrams.

\subsection{$\boldsymbol{\chi}$PT at large $\mathbf{N_C}$}

The large-$N_C$ counting rules are obviously well satisfied within $\chi$PT.
The $U(\vec{\phi\,})$ matrix that parametrizes the Nambu--Goldstone modes contains the scale $F\sim\cO(\sqrt{N_C})$, which compensates the canonical dimension of the fields $\vec{\phi}$. Therefore, each additional pseudoscalar field brings a suppression factor $1/\sqrt{N_C}$. The same scale $F$ governs the single-pseudoscalar coupling to the axial current, which is then of $\cO(\sqrt{N_C})$.

The generating functional \eqn{eq:generatingfunctional} involves classical sources that are coupled to the QCD quark bilinears. Since correlation functions of quark currents are of $\cO(N_C)$, the chiral Lagrangian should also scale as $\cO(N_C)$, at large $N_C$. Moreover, the LO terms in $1/N_C$ must involve a single flavour trace because each additional quark loop brings a suppression factor $1/N_C$. The $\cO(p^2)$ Lagrangian \eqn{eq:lowestorder} contains indeed a single flavour trace and an overall factor $F^2\sim\cO(N_C)$. The coupling $B_0$ must then be of $\cO(1)$, which is corroborated by \eqn{eq:b0}. While $\cL_{\mathrm{eff}}\sim\cO(N_C)$, chiral loops have a suppression factor $(4\pi F)^{-2}\sim \cO(1/N_C)$ for each loop. Thus, the $1/N_C$ expansion is equivalent to a semiclassical expansion.

The $\cO(p^4)$ Lagrangian \eqn{eq:l4} contains operators with a single flavour trace that are of $\cO(N_C)$, and chiral structures with two traces that should be of $\cO(1)$ because of their associated $1/N_C$ suppression. Actually, this is not fully correct due to the algebraic relation \eqn{eq:CayleyHamilton}, which has been used to rewrite the LO operator $\langle D_\mu U^\dagger D_\nu U D^\mu U^\dagger D^\nu U \rangle$ as $\frac{1}{2}\, O^{L_1} + O^{L_2} -2\, O^{L_3}$. Taking this into account, the large-$N_C$ scaling of the $\cO(p^4)$ LECs is given by
\beqn\label{eq:Nc_LECs}
\cO(N_C):&\quad & L_1\, ,\, L_2 \, ,\, L_3 \, ,\, L_5 \, ,\, L_8 \, ,\, L_9 \, ,\, L_{10} \, , 
\no\\
\cO(1):&\quad & 2 L_1-L_2 \, ,\, L_4 \, ,\, L_6 \, ,\, L_7 \, .
\eeqn
This hierarchy of couplings appears clearly manifested in Table~\ref{tab:Lcouplings}, where the phenomenological (and lattice) determinations of $L_4$ and $L_6$ are compatible with zero, and the relation $2L_1-L_2=0$ is well satisfied. The R$\chi$T determinations of the LECs are also in perfect agreement with the large-$N_C$ counting rules because they originate from tree-level resonance exchanges that are of LO in $1/N_C$.

A very subtle point arises concerning the coupling $L_7$. This LEC gets contributions from the exchange of singlet and octet pseudoscalars that, owing to their nonet-symmetry multiplet structure, exactly cancel each other at $N_C\to\infty$, in agreement with \eqn{eq:Nc_LECs}. However, the $U(1)_A$ anomaly, which is an $\cO(1/N_C)$ effect, generates a heavy mass for the $\eta_{\raisebox{-1pt}{$\scriptstyle 1$}}$ that decouples this state from the octet of Nambu--Goldstone pseudoscalars. Therefore, when the $\eta_{\raisebox{-1pt}{$\scriptstyle 1$}}$ field is integrated out from the low-energy theory, the chiral coupling $L_7$ receives the $\eta_{\raisebox{-1pt}{$\scriptstyle 1$}}$-exchange contribution \eqn{eq:L7}. Since $M_{\eta_{\raisebox{-1pt}{$\scriptscriptstyle 1$}}}^2\sim\cO(1/N_C,\cM)$, the coupling $L_7$ could then be naively considered to be of $\cO(N_C^2)$ \cite{Gasser:1984gg}. However, taking the limit of a heavy $\eta_{\raisebox{-1pt}{$\scriptstyle 1$}}$ mass, while keeping $m_s$ small, amounts to consider $N_C$ small and the large-$N_C$ counting is no-longer consistent \cite{Peris:1994dh}.

\subsection{R$\boldsymbol{\chi}$T estimates of LECs}

The R$\chi$T Lagrangian provides an explicit implementation of the large-$N_C$ limit of QCD, which, however, has been truncated to the lowest-mass resonance states. The true hadronic realization of the QCD dynamics at $N_C\to\infty$ corresponds to a (tree-level) local effective Lagrangian with an infinite number of  massive resonances. Thus, for each possible choice of quantum numbers, one must include an infinite tower of states with increasing masses. This
can be easily implemented in the R$\chi$T Lagrangian and taken into account in the determination of the chiral LECs. For the $\cO(p^4)$ couplings, the resulting predictions just reproduce the expressions in eqn~\eqn{eq:s_exchange} and \eqn{eq:vmd_results}, adding to each term the corresponding sum over the tower of states with the given quantum numbers  \cite{Pich:2002xy}:
\begin{displaymath}
2\, L_1\, =\, L_2 \, =\, \sum_i\; {G_{V_i}^2\over 4\, M_{V_i}^2}\, , 
\qquad\qquad\quad
L_3 \, =\, \sum_i\;\left\{ -{3\, G_{V_i}^2\over 4\, M_{V_i}^2} +
{c_{d_i}^2\over 2\, M_{S_i}^2}\right\} \, ,
\end{displaymath}
\begin{displaymath}
L_5 \, =\, \sum_i\; {c_{d_i}\, c_{m_i}\over M_{S_i}^2} \, ,
\qquad\quad
L_7\, =\, - \frac{F^2}{48 M_{\eta_1}^2}\, ,
\qquad\quad
L_8 \, =\, \sum_i\;\left\{ {c_{m_i}^2\over 2\, M_{S_i}^2} -
{d_{m_i}^2\over 2\, M_{P_i}^2}\right\} \, ,
\end{displaymath}
\bel{eq:LargeNcLecs}
L_9 \, =\, \sum_i\; {F_{V_i}\, G_{V_i}\over 2\, M_{V_i}^2}\, ,
\qquad\qquad\quad
L_{10} \, =\, \sum_i\;\left\{ {F_{A_i}^2\over 4\, M_{A_i}^2}
 - {F_{V_i}^2\over 4\, M_{V_i}^2}\right\}  \, .
\ee
Owing to the explicit $1/M_{R_i}^2$ suppression, the largest contributions originate from the exchange of the lightest resonances that we considered in Section~\ref{sec:MassiveFields}.

The short-distance conditions discussed in Section~\ref{subsec:SDconstraints} must also incorporate the towers of massive states. Since there is an infinite number of resonance couplings involved, one would need to consider an infinite number of constraints through the study of all possible QCD Green functions. Obviously, this is not a feasible task. However, truncating the sums to a few states, one can easily analyse the sensitivity of the results to the inclusion of higher-mass contributions \cite{Bijnens:2003rc,Golterman:2001pj,Knecht:1997ts,Knecht:1998sp}.
With a given set of meson states, the R$\chi$T Lagrangian provides an effective dynamical description that interpolates between the high-energy QCD behaviour and the very low-energy $\chi$PT realization. The short-distance conditions, rather than determining the physical values of the resonance parameters, just fix these couplings so that the interpolation behaves properly. When adding more states, the resonance parameters get readjusted to ensure the best possible interpolation with the available mass spectrum. This explains the amazing success of the simplest determination of LECs with just the lowest-mass resonances \cite{Ecker:1989yg,Ecker:1988te,Pich:2002xy}.

The effects induced by more exotic resonance exchanges with $J^{PC} = 1^{+-}$ and $2^{++}$ have been also investigated \cite{Ecker:2007us}. The short-distance constraints eliminate any possible contribution to the $\cO(p^4)$ LECs from $1^{+-}$ resonances, and only allow a tiny $2^{++}$ correction to $L_3$, $L_3^T = 0.16\cdot 10^{-3}$, which is negligible compared to the sum of vector and scalar contributions \cite{Ecker:2007us}. This small tensor correction had been previously obtained in the $SU(2)$ theory \cite{Ananthanarayan:1998hj,Toublan:1995bk}.

In order to determine the $\cO(p^6)$ (and higher-order) LECs, one must also consider local operators with several massive resonances, as dictated by the large-$N_C$ rules. The relevant R$\chi$T couplings can be constrained by studying appropriate sets of 3-point (or higher) functions \cite{Cirigliano:2005xn,Cirigliano:2004ue,Knecht:2001xc,Moussallam:1994xp,Moussallam:1997xx,RuizFemenia:2003hm}. 
A very detailed analysis of the ensuing $\cO(p^6)$ predictions can be found in Ref~\cite{Cirigliano:2006hb}.

Although the large-$N_C$ limit provides a very successful description of the low-energy dynamics, we are still lacking a systematic procedure to incorporate contributions of NLO in the $1/N_C$ counting. Some relevant subleading effects can be easily pinned down, such  as the resonance widths which regulate the poles of the meson propagators \cite{GomezDumm:2000fz,Guerrero:1997ku,Pich:2001pj,SanzCillero:2002bs}, or the role of final state interactions in the physical amplitudes \cite{Dobado:1989qm,GomezDumm:2000fz,GomezNicola:2001as,Guerrero:1997ku,Hannah:1997ux,Jamin:2000wn,Jamin:2001zq,Ledwig:2014cla,Nieves:2011gb,Oller:1998zr,Pallante:1999qf,Pallante:2000hk,Pallante:2001he,Pich:2001pj,SanzCillero:2002bs,Truong:1988zp}.

More recently, methods to determine the LECs of $\chi$PT at the NLO in $1/N_C$ have been developed \cite{Cata:2001nz,Pich:2008jm,Pich:2010sm,Rosell:2004mn,Rosell:2006dt}. This is a very relevant goal because the dependence of the LECs with the chiral renormalization scale is a subleading effect in the $1/N_C$ counting, of the same order than the loop contributions. Since the LO resonance-saturation estimates are performed at $N_C\to\infty$, they cannot control the $\mu$ dependence of the LECs. According to Table~\ref{tab:Lcouplings}, the large-$N_C$ predictions seem to work at $\mu\sim M_\rho$, which appears to be physically reasonable. However, a NLO analysis is mandatory in order to get the right $\mu$ dependence of the LECs.

At NLO one needs to include quantum loops involving virtual resonance propagators \cite{Kampf:2009jh,Rosell:2005ai,Rosell:2004mn}. This constitutes a major technical challenge because their  ultraviolet  divergences require higher-dimensional counterterms, which could generate a problematic behaviour at large momenta. Therefore, a careful investigation of short-distance QCD constraints, at the NLO in $1/N_C$, becomes necessary to enforce a proper ultraviolet behaviour of R$\chi$T and determine the needed renormalized couplings  \cite{Portoles:2006nr,Rosell:2005ai,Rosell:2009yb,Rosell:2004mn,SanzCillero:2007ib,SanzCillero:2009ap,Xiao:2007pu}. Using analyticity and unitarity, it is possible to avoid the technicalities associated with the renormalization procedure, reducing the calculation of one-loop Green functions to tree-level diagrams plus dispersion relations \cite{Pich:2008jm,Pich:2010sm,Rosell:2006dt}.  This allows us to understand the underlying physics in a much more transparent way.

\begin{figure}[t]
\begin{center}
\begin{minipage}[c]{6.9cm}\centering
\includegraphics[width=6.9cm,clip]{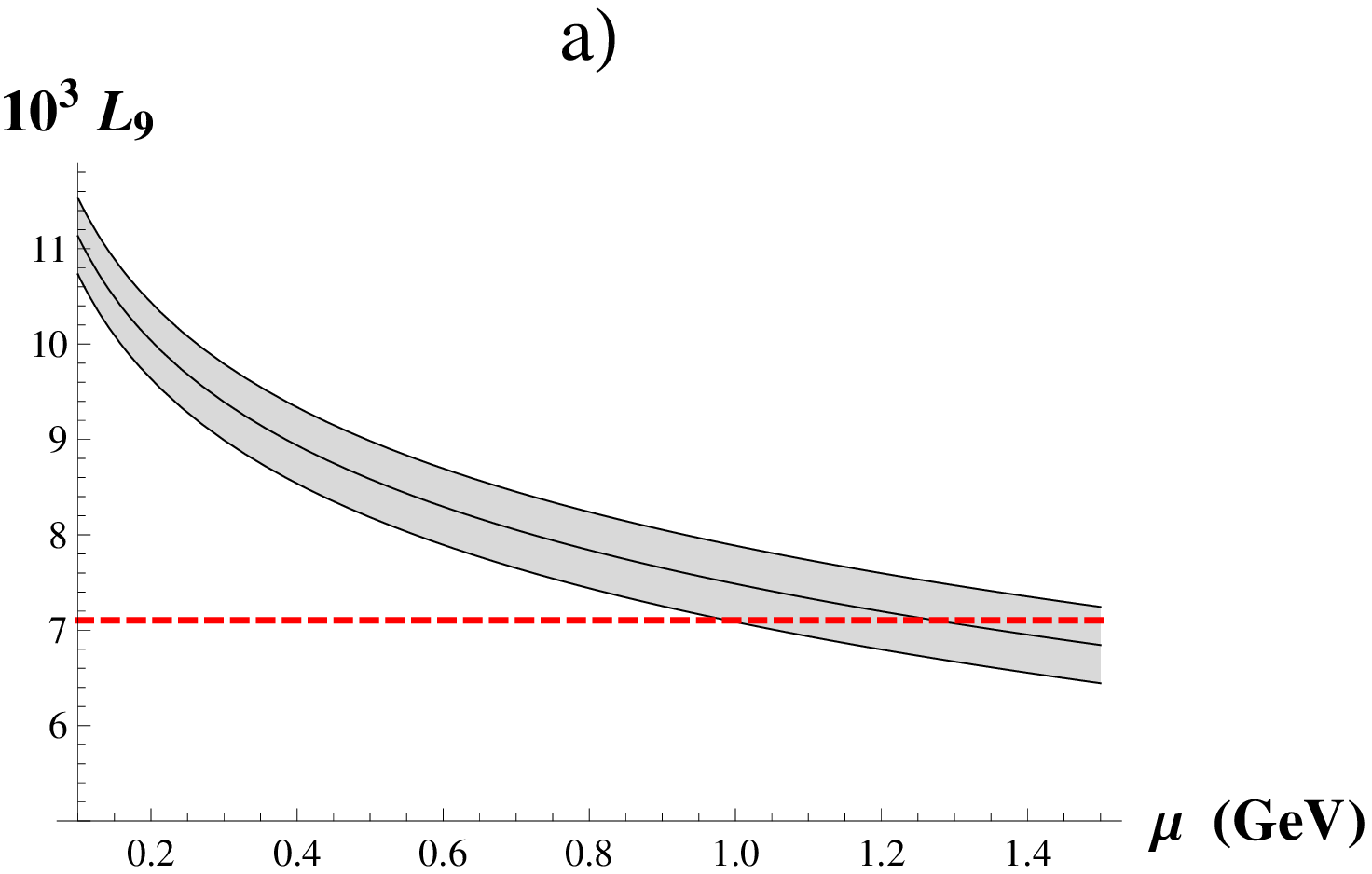}
\end{minipage}
\hskip .5cm
\begin{minipage}[c]{5.5cm}\centering
\includegraphics[width=5.5cm,clip]{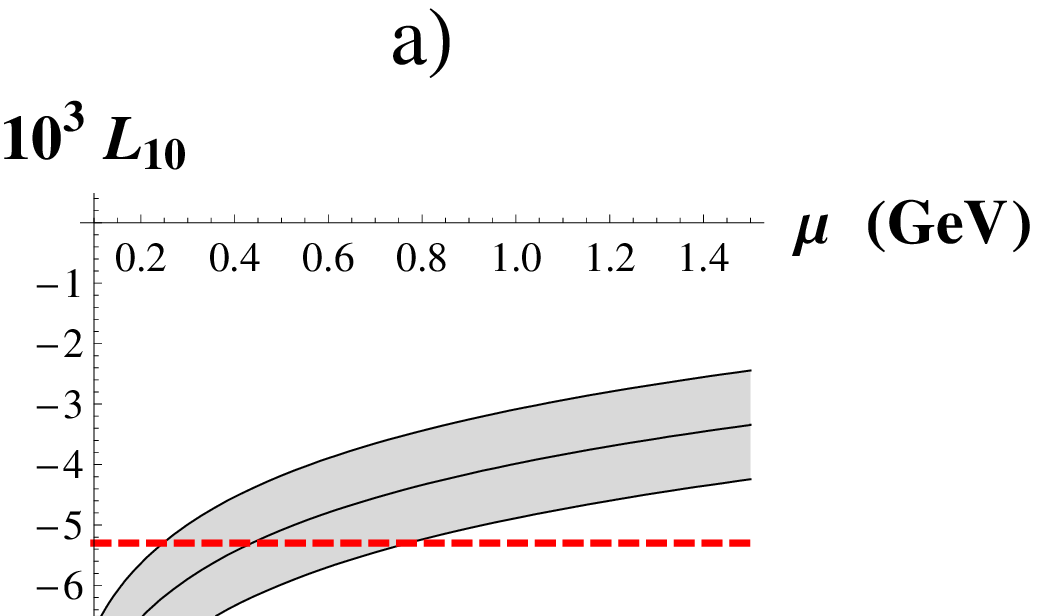}
\end{minipage}
\end{center}
\caption{R$\chi$T determinations of $L_9^r(\mu)$ and $L_{10}^r(\mu)$, at NLO in $1/N_C$ \cite{Pich:2008jm,Pich:2010sm}. The horizontal dashed lines show the LO predictions.}
\label{fig:NLO-LECs}
\end{figure}

Three $\cO(p^4)$ (and three $\cO(p^6)$) LECs have been already determined at NLO in $1/N_C$, keeping full control of their $\mu$ dependence, with the results \cite{Pich:2008jm,Pich:2010sm,Rosell:2006dt}: 
\begin{displaymath}
L_8^r(M_\rho)\, =\, (0.6\pm 0.4)\cdot 10^{-3}\, ,
\qquad \qquad\qquad
L_9^r(M_\rho)\, =\, (7.9\pm 0.4)\cdot 10^{-3}\, ,
\end{displaymath}
\be
L_{10}^r(M_\rho)\, =\, -(4.4\pm 0.9)\cdot 10^{-3}\, .
\ee
These numerical values are in good agreement with the phenomenological determinations in Table~\ref{tab:Lcouplings}. The result for $L_8^r(M_\rho)$ also agrees with the more precise lattice determination quoted in the table. The NLO R$\chi$T predictions for $L_9^r(\mu)$ and $L_{10}^r(\mu)$ are shown in Fig.~\ref{fig:NLO-LECs}, as function of the renormalization scale $\mu$ (gray bands).
%
%
The comparison with their LO determinations (dashed red lines) corroborates the numerical success of the $N_C\to\infty$ estimates. At scales $\mu\sim M_\rho$, the differences between the LO and NLO estimates are well within the expected numerical uncertainty of $\cO(1/N_C)$.

\section{Electroweak symmetry breaking}
\label{sec:EWSB}

The discovery of the Higgs boson represents a major achievement in fundamental physics and has established the SM as the correct description of the electroweak interactions at the experimentally explored energy scales. There remain, however, many open questions that the SM is unable to address, such as the nature of the mysterious dark matter that seems to be abundantly present in our Universe, the huge asymmetry between matter and antimatter, and the dynamical origin of the measured hierarchies of fermion masses and mixings, which span many orders of magnitude. Although we are convinced that new physics is needed in order to properly understand these facts, so far, all LHC searches for new phenomena beyond the SM have given negative results. The absence of new massive states at the available energies suggest the presence of an energy gap above the electroweak scale, separating it from the scale where the new dynamics becomes relevant. 
EFT provides then an adequate framework to parametrize our ignorance about the underlying high-energy dynamics is a model-independent way.

While the measured Higgs properties comply with the SM expectations, the current experimental uncertainties are large enough to accommodate other alternative scenarios of electroweak symmetry breaking (EWSB) \cite{Pich:2015lkh}. We still ignore whether the discovered Higgs corresponds indeed to the SM scalar, making part of an $SU(2)_L$ doublet together with the three electroweak Nambu--Goldstone fields, or it is a different degree of freedom, decoupled somehow from the Nambu--Goldstone modes. Additional bosons could also exist, within an extended scalar sector with much richer phenomenological implications. Another possibility would be a dynamical EWSB, similarly to what happens in QCD.

To formulate an electroweak EFT, one needs to specify its field content (the light degrees of freedom) and symmetry properties. Moreover, a clear power counting must be established in order to organise the effective Lagrangian. The simplest possibility is to build an EFT with the SM gauge symmetries and particle spectrum, assuming that the Higgs boson belongs indeed to the SM scalar doublet. The EWSB is then linearly realised, and the effective Lagrangian is ordered according to the canonical dimensions of all possible gauge-invariant operators:
\bel{eq:SMEFT}
\cL_{\mathrm{SMEFT}} \, =\, \cL_{\mathrm{SM}}\, +\,
\sum_{d>4}\sum_i\,\frac{c_i^{(d)}}{\Lambda_{\mathrm{NP}}^{d-4}}\; O_i^{(d)}\, .
\ee
The SM is just the LO approximation with dimension $d=4$, while operators with $d>4$ are suppressed by a factor $1/\Lambda_{\mathrm{NP}}^{d-4},$ with $\Lambda_{\mathrm{NP}}^{\phantom{d}}$ representing the new physics scale. The dimensionless couplings $c_i^{(d)}$ contain information on the underlying dynamics. For instance, the exchange of a heavy massive vector boson $X_\mu$, coupled to a quark vector current, would generate a $d=6$ four-fermion operator in the effective Lagrangian:
\bel{eq:X-exchange}
\cL_{\mathrm{NP}}\,\dot=\, g_X^{\phantom{\dagger}}\,\left(\bar q_L^{\phantom{\dagger}}\gamma^\mu q_L^{\phantom{\dagger}}\right)\, X_\mu
\quad\longrightarrow\quad
- \frac{g_X^{2\phantom{\dagger}}}{M_X^2}\;\left(\bar q_L^{\phantom{\dagger}}\gamma^\mu q_L^{\phantom{\dagger}}\right)\,\left(\bar q_L^{\phantom{\dagger}}\gamma_\mu q_L^{\phantom{\dagger}}\right)\, .
\ee

There is only one operator with $d=5$ (up to Hermitian conjugation and flavour assignments), which violates lepton number by two units \cite{Weinberg:1979sa}.
Assuming the separate conservation of the baryon and lepton numbers, and a single SM fermion family, the NLO piece of $\cL_{\mathrm{SMEFT}}$ contains 59 independent operators with $d=6$  \cite{Buchmuller:1985jz,Grzadkowski:2010es}. When the three fermion families are included, this number blows up to 1350 CP-even plus 1149 CP-odd operators \cite{Alonso:2013hga}. In addition, there are five $d=6$ operators that violate $B$ and $L$ \cite{Abbott:1980zj,Wilczek:1979hc}.

A large number of recent works have studied different aspects of this SM effective field theory (SMEFT), working out the phenomenological constraints on the $d=6$ Lagrangian. 
The full anomalous dimension matrix of the $d=6$ operators is already known  \cite{Alonso:2013hga}, making possible to perform analyses at the one-loop level.
An overview of the current status can be found in Ref.~\cite{Brivio:2017vri}. Here, we prefer to discuss a more general EFT framework, which does not make any assumption about the nature of the Higgs field. The EWSB is realised non-linearly in terms of its Nambu--Goldstone modes, and the light Higgs is incorporated into the effective Lagrangian as a generic $SU(2)_L$ singlet field with unconstrained couplings to the EWSB sector
\cite{Buchalla:2012qq,Buchalla:2013rka,Pich:2015kwa,Pich:2016lew}.

\subsection{Custodial Symmetry}
\label{subsec:custodial}

A massless gauge boson has only two polarizations, while a massive spin-1 particle should have three. In order to generate the missing longitudinal polarizations of the $W^\pm$ and $Z$ bosons, without breaking gauge invariance, three additional degrees of freedom must be added to the massless $SU(2)_L\otimes U(1)_Y$ gauge theory. The SM incorporates instead a complex scalar doublet $\Phi(x)$ containing four real fields and, therefore, one massive neutral scalar, the Higgs boson, remains in the spectrum after the EWSB.
It is convenient to collect the four scalar fields in the $2\times 2$ matrix~\cite{Appelquist:1980vg}
\be
\Sigma \,\equiv\, \left( \Phi^c, \Phi\right) \, =\,
\left( \bat \Phi^{0*} & \Phi^+  \\ -\Phi^- &  \Phi^0 \ea\right)\, ,
\label{eq:sigma_matrix}
\ee
with $\Phi^c = i \sigma_2\Phi^*$ the charge-conjugate of the scalar doublet $\Phi$.
The SM scalar Lagrangian can then be written in the form \cite{Pich:1995bw,Pich:1998xt}
\bel{eq:l_sm}
\cL(\Phi)\, =\, \frac{1}{2}\, \langle\, \left(D^\mu\Sigma\right)^\dagger D_\mu\Sigma\,\rangle
- \frac{\lambda}{16} \left(\langle\,\Sigma^\dagger\Sigma\,\rangle
- v^2\right)^2 ,
\ee
where
$D_\mu\Sigma \equiv \partial_\mu\Sigma
+ i g \,\frac{\vec{\sigma}}{2}\vec{W}_\mu \,\Sigma - i g' \,\Sigma \,\frac{\sigma_3}{2} B_\mu $
is the usual gauge-covariant derivative. 

The Lagrangian $\cL(\Phi)$ has a global
$G\equiv SU(2)_L\otimes SU(2)_R$ symmetry,
\be
\Sigma \quad \longrightarrow \quad g_L^{\phantom{\dagger}} \,\Sigma\, g_R^\dagger\, ,
\qquad\qquad\qquad\qquad
g_{L,R}^{\phantom{\dagger}}  \in SU(2)_{L,R} \, ,
\label{eq:sigma_transf}
\ee
while the vacuum choice $\langle 0|\Phi^0|0\rangle = v$ is only preserved when $g_L^{\phantom{\dagger}}=g_R^{\phantom{\dagger}}$, {\it i.e.}, by the custodial symmetry group $H\equiv SU(2)_{L+R}$ \cite{Sikivie:1980hm}. Thus, the scalar sector of the SM has the same pattern of chiral symmetry breaking \eqn{eq:scsb} than QCD with $n_f=2$ flavours. In fact, $\cL(\Phi)$ is formally identical to the linear sigma-model Lagrangian \eqn{eq:sigma3}, up to a trivial normalization of the $\Sigma$ field and the presence of the electroweak covariant derivative. In the SM, $SU(2)_L$ is promoted to a local gauge symmetry, but only the $U(1)_Y$ subgroup of $SU(2)_R$ is gauged. Therefore, the $U(1)_Y$ interaction in the covariant derivative breaks the $SU(2)_R$ symmetry.

The polar decomposition
\be\label{eq:polarEW}
\Sigma(x) \, = \, \frac{1}{\sqrt{2}}
\left[ v + h(x) \right] \; U(\vec{\phi\,})
\ee
separates the Higgs field $h(x)$, which is a singlet under $G$ transformations, from the three Nambu--Goldstone excitations $\vec{\phi}(x)$, parametrized through the $2\times 2$ matrix 
%
\bel{eq:Goldstones}
U(\vec{\phi\,}) \, =\,  \exp{\left\{ i \vec{\sigma} \, \vec{\phi} / v \right\} }
\quad \longrightarrow \quad g_L^{\phantom{\dagger}} \, U(\vec{\phi\,})\, g_R^\dagger\, .
\ee
One can rewrite $\cL(\Phi)$ in the form~\cite{Appelquist:1980vg,Longhitano:1980iz,Longhitano:1980tm}:
\be
\cL(\Phi)\, =\, \frac{v^2}{4}\,
\langle\, D_\mu U^\dagger D^\mu U \,\rangle \, +\,
\cO\left( h/v \right) \, ,
\label{eq:sm_goldstones}
\ee
which makes explicit the universal Nambu--Goldstone Lagrangian \eqn{eq:l2} associated with the symmetry breaking \eqn{eq:scsb}. Up to terms containing the scalar Higgs, the only difference is the presence of the electroweak gauge fields through the covariant derivative 
$D_\mu U \equiv \partial_\mu U
+ i g \,\frac{\vec{\sigma}}{2}\vec{W}_\mu \, U - i g'\, U \,\frac{\sigma_3}{2} B_\mu$.
Thus, the same Lagrangian that describes the low-energy pion interactions in two-flavour QCD governs the SM Nambu--Goldstone dynamics, with the trivial change $F\to v$ \cite{Pich:1998xt}.
%
The electroweak precision data have confirmed that the pattern of symmetry breaking implemented in the SM is the correct one, and have determined the fundamental scale
$v = (\sqrt{2}\, G_F)^{-1/2} = 246\;\mathrm{GeV}$~\cite{Pich:2012sx}.

In the unitary gauge, $U(\vec{\phi\,})=1$, the Nambu--Goldstone fields are rotated away and the Lagrangian~\eqn{eq:sm_goldstones} reduces to a quadratic mass term for the electroweak gauge bosons, giving the SM prediction for the $W^\pm$ and $Z$ masses:
\bel{eq:W_Z_masses}
m_W\, =\, m_Z \,\cos{\theta_W}\, =\,\frac{1}{2}\, v g\, , 
\ee
with
$Z^\mu \equiv \cos{\theta_W} W_3^\mu - \sin{\theta_W} B^\mu$ and
$\tan{\theta_W} = g'/ g$. Thus, these masses are generated by the Nambu--Goldstone modes, not by the Higgs field.\footnote{The QCD pions generate, in addition, a tiny correction $\delta m_W = \delta m_Z \,\cos{\theta_W} = F_\pi g/2$, through their coupling to the SM gauge bosons encoded in the $\chi$PT Lagrangian \eqn{eq:lowestorder}.} 
The successful mass relation \eqn{eq:W_Z_masses} is a direct consequence of the pattern of EWSB, providing a clear confirmation of its phenomenological validity. The particular dynamical structure of the SM scalar Lagrangian can only be tested through the Higgs properties. The measured Higgs mass determines the quartic coupling, 
\bel{eq:QuarticHiggs}
\lambda\, =\, \frac{m_h^2}{2 v^2}\, =\, 0.13\, , 
\ee
while its gauge couplings are consistent with the SM prediction within the present experimental uncertainties \cite{Pich:2015lkh}.

\subsection{Equivalence theorem}
\label{subsec:EquivalenceTh}

The elastic scattering amplitudes of the electroweak Nambu--Goldstone bosons can be obtained from the analogous results for the QCD pions. Thus, they obey the (weak) isospin and crossing relation \eqn{eq:PionScattering-1L}, where $A(s,t,u)$ is given in~\eqn{eq:PionScattering-1L-b} with the change $F\to v$. Actually, eqn~\eqn{eq:PionScattering-1L-b} includes also the one-loop correction with the corresponding $\cO(p^4)$ LECs $\ell_{1,2}$ that get different names in the electroweak EFT (see later).

\begin{figure}[t]
\begin{center}
\includegraphics[width=13cm]{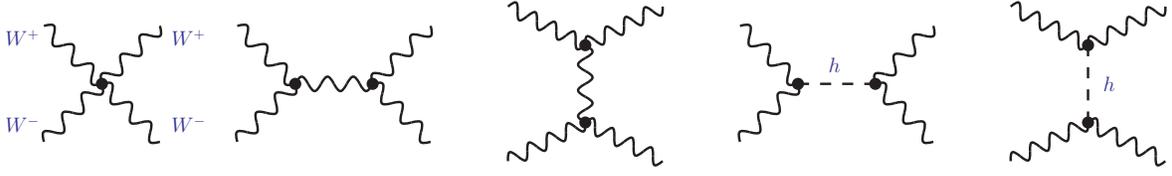}
\end{center}
\caption{SM Feynman diagrams contributing to the elastic $W^+W^-$ scattering at LO.}
\label{fig:WWtoWW}
\end{figure}

The electroweak Nambu--Goldstone modes correspond to the
longitudinal polarizations of the gauge bosons. Therefore, the $\chi$PT results directly give the scattering amplitudes of longitudinally-polarized gauge bosons, up to corrections proportional to their non-zero masses. In the absence of the Higgs field, one gets the LO result \cite{Cornwall:1974km,Lee:1977eg,Vayonakis:1976vz}:
\be\label{eq:EquivalenceTh}
T(W^+_L W^-_L\to W^+_L W^-_L) \, = \, T(\phi^+\phi^-\to \phi^+\phi^-)\, +\, \cO\left(\frac{m_W}{\sqrt{s}}\right)
\, = \, \frac{s+t}{v^2}\, +\, \cO\left(\frac{m_W}{\sqrt{s}}\right)\, .
\ee
At high energies the amplitude grows as $s/v^2,$ which implies a tree-level violation of unitarity. This is the expected behaviour from the derivative coupling in the Lagrangian \eqn{eq:sm_goldstones}. Making a direct calculation with longitudinal spin-1 bosons, one could naively expect a harder energy dependence, $T(W^+_L W^-_L\to W^+_L W^-_L)\sim g^2 E^4/m_W^4$, because each longitudinal polarization vector grows as
$\epsilon^\mu_L(\vec{k\,})\sim k^\mu/m_W$ and there are four gauge-boson external legs.\footnote{In the reference frame $k^\mu = (k^0,0,0,|\vec{k}|)$,
$\epsilon^\mu_L(\vec{k\,}) =\frac{1}{m_W}\, (|\vec{k}|,0,0,k^0) = \frac{k^\mu}{m_W} + \cO(m_W/|\vec{k}|)$.}  
Thus, there is a strong gauge cancellation among the first three diagrams in Fig.~\ref{fig:WWtoWW}, which eliminates those contributions with the highest powers of energy.

The SM scalar doublet $\Phi$ gives rise to a renormalizable Lagrangian, with good short-distance properties, that obviously satisfies unitarity. The right high-energy behaviour of the scattering amplitude is recovered through the additional contributions from Higgs boson exchange in Fig.~\ref{fig:WWtoWW}, which cancel the unphysical growing:
\beqn
T(W^+_L W^-_L\to W^+_L W^-_L)_{\mbox{\tiny SM}}&\, =&\,
\frac{1}{v^2}\,\left\{ s + t - \frac{s^2}{s-m_h^2}
- \frac{t^2}{t-m_h^2}\right\}
\, +\, \cO\left(\frac{m_W}{\sqrt{s}}\right)
\no\\ &\, =&\,
-\frac{m_h^2}{v^2}\,\left\{ \frac{s}{s-m_h^2}+ \frac{t}{t-m_h^2}\right\}
\, +\, \cO\left(\frac{m_W}{\sqrt{s}}\right)\, .\qquad\qquad
\eeqn
The SM gauge structure implies an exact cancellation of all terms that grow with energy, changing an $E^4$ behaviour into $E^0$. Any small modification of the couplings in Fig.~\ref{fig:WWtoWW} would spoil this cancellation, generating a phenomenologically unacceptable result. Therefore,  large deviations of the gauge couplings from their SM predictions should not be expected. Moreover, small changes in the couplings would require the presence of new contributions to the scattering amplitude in order to maintain the cancellations.

At very large energies, $s\gg m_W^2$, $T(W^+_L W^-_L\to W^+_L W^-_L)_{\mbox{\tiny SM}}\approx -2 m_h^2/v^2$, which has an S-wave component
\bel{eq:Swave}
a_0\,\equiv\,\frac{1}{32\pi}\,\int_{-1}^1 d\cos{\theta}\;\; 
T(W^+_L W^-_L\to W^+_L W^-_L)_{\mbox{\tiny SM}}\,\approx\,
-\frac{m_h^2}{8\pi v^2}\, .
\ee
The elastic unitarity constraint $|a_0|\le 1$ puts then an upper bound on the Higgs mass. Taking into account the inelastic channels into $ZZ$ and $hh$, the bound becomes stronger by a factor $\sqrt{2/3}$, leading to the final result  
$m_h \le \sqrt{8\pi v}\,\sqrt{2/3}\approx 1$~TeV \cite{Lee:1977eg}. The measured Higgs mass, $m_h = (125.09\pm 0.24)$~GeV \cite{Aad:2015zhl}, is well below this upper limit. Notice that the bound has been extracted from a perturbative LO calculation. A much heavier Higgs, above 1~TeV, would have indicated the need for large higher-order contributions and a strong-coupling regime with $\lambda > 1$.

QCD provides a good illustration of the role of unitarity at strong coupling.
Although the $\chi$PT result for the $\pi\pi$ elastic scattering amplitude grows with energy, QCD is a renormalizable theory that satisfies unitarity.
The chiral prediction~\eqn{eq:PionScattering-1L-b} is only valid at low energies, below the resonance mass scale. Once higher-mass states are taken into account, as done in R$\chi$T, the $\pi\pi$ scattering amplitude recovers its good unitarity properties, provided the proper QCD short-distance behaviour is imposed. The P-wave isovector amplitude ($J=I=1$), for instance, gets unitarized by $\rho$ exchange. The unitarisation of the S-wave isoscalar amplitude ($J=I=0$), with vacuum quantum numbers, proceeds in a much more subtle way because pion loop corrections are very large in this channel; their resummation generates a unitarized amplitude with a pole in the complex plane that corresponds to the controversial $\sigma$ or $f_0(500)$ meson, a broad resonance structure that is absent at $N_C\to\infty$ \cite{Ledwig:2014cla,Pelaez:2015qba}.

\section{Electroweak effective theory}
\label{sec:EWET}

In order to formulate the electroweak effective theory (EWET) we must consider the most general low-energy Lagrangian that satisfies the SM symmetries and only contains the known light spectrum: the SM gauge bosons and fermions, the electroweak Nambu--Goldstone modes and the Higgs field $h$ \cite{Buchalla:2012qq,Buchalla:2013rka,Pich:2015kwa,Pich:2016lew}. In addition to the SM gauge symmetries, our main assumption will be the pattern of EWSB:
\bel{eq:ewsb}
G \equiv SU(2)_L\otimes SU(2)_R \;
\longrightarrow\; H \equiv SU(2)_{L+R}\, .
\ee
As we have done before with $\chi$PT, we will organise the Lagrangian as an expansion in powers of derivatives and symmetry breakings over the EWSB scale (and/or any new-physics heavy mass scale). The purely Nambu--Goldstone terms are of course formally identical to those present in $\chi$PT with $n_f=2$. The EWET contains, however, a richer variety of ingredients, since we must include the SM gauge symmetries, a fermion sector and a quite different type of symmetry-breaking effects:
\bel{eq:L_EWET}
\cL_{\mathrm{EWET}}\, =\, \cL_2^{\mathrm{EW}} + \cL_4^{\mathrm{EW}} + \cdots
\, =\, \cL_{\mathrm{SM}}^{(0)} +  \Delta \cL_2 + \cL_4^{\mathrm{EW}} + \cdots
\ee
The dots denote the infinite tower of operators with higher chiral dimensions. The LO term $\cL_2^{\mathrm{EW}}$ contains the renormalizable massless (unbroken) SM Lagrangian,
\be
\cL_{\rm SM}^{(0)} \, =\, \cL_{\rm YM} +  i \, \sum_{f} \bar{f}\gamma^\mu D_\mu f\, ,
\ee
where $D_\mu$ is the covariant derivative of the $SU(3)_C\otimes SU(2)_L \otimes U(1)_Y $ gauge group, $\cL_{\rm YM}$ the corresponding Yang-Mills Lagrangian and the sum runs over all SM fermions $f$. The additional LO piece $\Delta \cL_2$ incorporates the EWSB contributions.

We will parametrize the Nambu--Goldstone fields with the coset coordinates \eqn{eq:h_def} and the choice of canonical coset representative $\xi_L^{\phantom{\dagger}}(\vec{\phi\,}) = \xi_R^\dagger (\vec{\phi\,}) \equiv u(\vec{\phi\,})$.
This convention is opposite to \eqn{eq:h_def_2}, usually adopted in $\chi$PT, but it looks more natural to describe the SM gauge group.\footnote{The two conventions are just related by a permutation of left and right, or equivalently, $u(\vec{\phi\,}) \leftrightarrow u^\dagger (\vec{\phi\,})$. Thus, analogous terms in both theories only differ by the change $F \leftrightarrow -v$.}
Therefore,
\be
u(\vec{\phi\,})\,\toG\, g_L^{\phantom{\dagger}}\, u(\vec{\phi\,})\, g_h^\dagger
\, = \,
g_h^{\phantom{\dagger}} \, u(\vec{\phi\,})\, g_R^\dagger \, , 
\label{eq:h_def_ew}
\ee
with $g_h^{\phantom{\dagger}}\equiv h(\vec\phi,g)\in H$ the compensating transformation needed to get back to the chosen coset representative.

The gauge fields are incorporated in the same way as the left and right sources of $\chi$PT, although here we must add the Lagrangian $\cL_{\rm YM}$ because we need to quantize them. Thus, we formally introduce the $SU(2)_L$ and $SU(2)_R$ matrix fields, $\hat{W}_\mu$ and $\hat{B}_\mu$ respectively, transforming as
\bel{eq:FakeTransform}
\hat{W}^\mu\quad\longrightarrow\quad g_L^{\phantom{\dagger}}\, \hat{W}^\mu g_L^\dagger + i\, g_L^{\phantom{\dagger}}\, \partial^\mu g_L^\dagger\, ,
\qquad\qquad\quad
\hat{B}^\mu\quad\longrightarrow\quad g_R^{\phantom{\dagger}}\, \hat{B}^\mu g_R^\dagger + i\, g_R^{\phantom{\dagger}}\, \partial^\mu g_R^\dagger\, ,
\ee
the covariant derivative ($U = u^2\,\to\, g_L^{\phantom{\dagger}} U g_R^\dagger $)
\bel{eq:DU}
D_\mu U \, =\, \partial_\mu U -  i\, \hat{W}_\mu  U + i\, U \hat{B}_\mu
\quad\longrightarrow\quad g_L^{\phantom{\dagger}}\, D_\mu U \, g_R^\dagger
\, ,
\ee
and the field-strength tensors
\beqn
\hat{W}_{\mu\nu}  &\, =\, & \partial_\mu \hat{W}_\nu - \partial_\nu \hat{W}_\mu
- i\, [\hat{W}_\mu,\hat{W}_\nu]
\quad \longrightarrow\quad g_L^{\phantom{\dagger}}\, \hat{W}_{\mu\nu} \, g_L^\dagger
\, ,
\no\\
\hat{B}_{\mu\nu}  &\, =\, & \partial_\mu \hat{B}_\nu - \partial_\nu \hat{B}_\mu
- i\, [\hat{B}_\mu,\hat{B}_\nu]
\quad\longrightarrow\quad
g_R^{\phantom{\dagger}}\, \hat{B}_{\mu\nu} \, g_R^\dagger
\, .
\eeqn
The SM gauge fields are recovered through the identification \cite{Pich:2012jv}
\bel{eq:SMgauge}
\hat{W}^\mu \, =\, -g\;\frac{\vec{\sigma}}{2}\, \vec{W}^\mu \, ,
\qquad\qquad\qquad\quad
\hat{B}^\mu\, =\, -g'\;\frac{\sigma_3}{2}\, B^\mu\, ,
\ee
which explicitly breaks the $SU(2)_R$ symmetry group while preserving the $SU(2)_L\otimes U(1)_Y$ gauge symmetry.

We also define the covariant quantities
\be\label{eq.cov-bosonic-tensors}
u_\mu  \, =\, 
i\, u\, (D_\mu U)^\dagger u \, =\, -i\, u^\dagger D_\mu U\, u^\dagger
\, =\, u_\mu^\dagger\, ,
\qquad\qquad\quad
f_\pm^{\mu\nu} \, = \,
u^\dagger \hat{W}^{\mu\nu}  u \pm u\, \hat{B}^{\mu\nu} u^\dagger\, ,
\ee
which transform as $( u_\mu , f_\pm^{\mu\nu} )\,\to\, g_h^{\phantom{\dagger}}\, ( u_\mu , f_\pm^{\mu\nu} )\, g_h^\dagger $.

The bosonic part of $\Delta\cL_2$ can then be written as \cite{Pich:2016lew}
\bel{eq:L2bosonic}
\Delta\cL_2^{\mathrm{Bosonic}}\, =\, \frac{1}{2}\,
\partial_\mu h\,\partial^\mu h
\, -\,\frac{1}{2}\, m_h^2\, h^2 \, -\, V(h/v)
\, +\,
\frac{v^2}{4}\,\cF_u(h/v)\;\langle u_\mu u^\mu\rangle\, ,
\ee
where $\langle u_\mu u^\mu\rangle$ is the usual LO Nambu--Goldstone operator and
\be\label{eq:Fhu_V}
  V(h/v)\, = \, v^4\;\sum_{n=3} c^{(V)}_n \left(\frac{h}{v}\right)^n\, ,
\qquad\qquad
\cF_u(h/v)\, = \, 1\, +\, \sum_{n=1} c^{(u)}_n \left(\frac{h}{v}\right)^n\, .
\ee
Each chiral-invariant structure can be multiplied with an arbitrary function
of $h/v$ because the Higgs field is a singlet under $SU(2)_L\otimes SU(2)_R$\cite{Grinstein:2007iv}. The electroweak scale $v$ is used to compensate the powers of both the Higgs and the Nambu--Goldstone fields, since they are expected to have a similar underlying origin and, therefore, the coefficients $c_n^{(u)}$ could be conjectured to be of $\cO(1)$. 
The scalar Lagrangian of the SM corresponds to
 $c^{(V)}_3 = \frac{1}{2}\, m_h^2/v^2$, $c^{(V)}_4 = \frac{1}{8}\, m_h^2/v^2$, $c^{(V)}_{n>4} = 0$, $c^{(u)}_1 = 2$, $c^{(u)}_2 = 1$ and $c^{(u)}_{n>2} = 0$.

In principle, the quadratic derivative term of the Higgs should also be multiplied with an arbitrary function $\cF_h(h/v)$. However, this function can be reabsorbed into a redefinition of the field $h$ \cite{Giudice:2007fh}.

\subsection{Fermionic fields}

The SM fermion multiplets can be organised into $SU(2)_L$ and $SU(2)_R$ doublets,
\be
\psi_L\, =\, \left( \begin{array}{c} t_L \\ b_L \end{array}\right)\, ,
\qquad\qquad
\psi_R\,  =\,\left( \begin{array}{c} t_R \\ b_R \end{array}\right) \, ,
\ee
and analogous definitions for the other quark and lepton flavours, extending the symmetry group to $\cG=SU(2)_L\otimes SU(2)_R\otimes U(1)_{X}$
with $X=(\mathrm{B}-\mathrm{L})/2$, being $\mathrm{B}$ and $\mathrm{L}$ the baryon and lepton numbers, respectively~\cite{Hirn:2005fr}.
They transform under $\cG$ as
\be 
\psi_L  \quad\longrightarrow\quad g_X^{\phantom{\dagger}}\, g_L^{\phantom{\dagger}}\;  \psi_L   \, ,
\qquad\qquad\qquad
\psi_R  \quad\longrightarrow\quad g_X^{\phantom{\dagger}}\, g_R^{\phantom{\dagger}}\;  \psi_R   \, ,
\label{eq.psi-transformation}
\ee 
with $g_X^{\phantom{\dagger}} \in U(1)_X$.

One must introduce the $U(1)_X$ field $\hat{X}_\mu$, transforming like
\bel{eq:Xtransform}
\hat{X}^\mu\quad\longrightarrow\quad \hat{X}^\mu
+ i\, g_X^{\phantom{\dagger}}\, \partial^\mu g_X^\dagger\, ,
\ee
and its corresponding field strength tensor
\be
\hat{X}_{\mu\nu}\, =\,\partial_\mu\hat{X}_\nu -\partial_\nu \hat{X}_\mu
\ee
that is a singlet under $\cG$.
The fermion covariant derivatives take then the form \cite{Pich:2016lew}
\bel{eq.dpsi}
D_\mu^L\psi_L\, =\,\left(\partial_\mu - i\,\hat{W}_\mu - i\, x_{\psi}\,
\hat{X}_\mu \right) \psi_L\, ,
\qquad\quad
D_\mu^R\psi_R\, =\,\left(\partial_\mu - i\,\hat{B}_\mu - i\, x_{\psi}\,
\hat{X}_\mu
\right) \psi_R\, ,
\ee
where $x_{\psi}$ is the corresponding $U(1)_X$ charge of the fermion field.

To recover the SM gauge interactions, the auxiliary fields must be frozen to the values given in~(\ref{eq:SMgauge}) and
\bel{eq:Xmu-fixing}
\hat{X}_\mu  \, =\, \,-\, g'\, B_\mu \, ,
\ee
which introduces an explicit breaking of the symmetry group $\cG$
to the SM subgroup $SU(2)_L\otimes U(1)_Y$, with
$Y = T_{3R} +\frac{1}{2}\, (\mathrm{B}-\mathrm{L})$~\cite{Senjanovic:1975rk}, {\it i.e.},
\bel{eq:Q_LR}
Q\, =\, T_{3L} + T_{3R} + \frac{1}{2}\, \left( \mathrm{B}-\mathrm{L}\right)\, .
\ee
Since bosons have $\mathrm{B}=\mathrm{L}=0$, the enlargement of the symmetry group does not modify the bosonic sector.

In order to easily build chiral-invariant structures, it is convenient to introduce covariant fermion doublets \cite{Pich:2016lew}
\bel{eq:CovFerm}
\xi_{L} \,\equiv\, 
u^\dagger \, \psi_L \, ,
\qquad\quad
\xi_{R} \,\equiv\, 
u \,\psi_R \, ,
\qquad\quad
d_\mu^{L} \xi_{L}\,\equiv\, u^\dagger \, D_\mu^L\psi_L\, ,
\qquad\quad
d_\mu^{R} \xi_{R}\,\equiv\, u \, D_\mu^R\psi_R\, ,
\ee
transforming under $\cG$ as
\be 
\left(\xi_L , \xi_R, d_\mu^{L} \xi_{L}, d_\mu^{R} \xi_{R}\right) \quad\longrightarrow\quad g_X^{\phantom{\dagger}}\, g_h^{\phantom{\dagger}} \;
\left(\xi_L , \xi_R, d_\mu^{L} \xi_{L}, d_\mu^{R} \xi_{R}\right)\, .
\ee
The combined field $\xi\equiv \xi_L + \xi_R$ and $d_\mu\xi\equiv d_\mu^L\xi_L + d_\mu^R\xi_R$ transform obviously in the same way.

The kinetic fermion Lagrangian can be easily written in covariant form:
\be
\cL_{\rm Fermionic}^{(0)}   \, =\, \sum_\xi
i\,\bar\xi \gamma^\mu d_\mu \xi \, =\, \sum_\psi \left(
i\,\overline\psi_L \gamma^\mu D_\mu^L\psi_L \, +\,
i\,\overline\psi_R \gamma^\mu D_\mu^R\psi_R \right)\, .
\ee
Similarly, the usual fermion bilinears are compactly expressed as:
\bel{eq:bilinears}
J_\Gamma\,\equiv\,
\bar{\xi}\,\Gamma\, \xi \, =\, \left\{ \bat
\overline{\psi}_L\Gamma\psi_L \,+\, \overline{\psi}_R\Gamma\psi_R\, ,
& \qquad\quad\mbox{\small  $\Gamma=\gamma^\mu\, (V) , \, \gamma^\mu \gamma_5\, (A)$} ,
\\[10pt]
\overline{\psi}_L \Gamma\, U(\vec{\phi\,})\,\psi_R
\,+\, \overline{\psi}_R \Gamma \, U^\dagger (\vec{\phi\,}) \psi_L\, ,
& \qquad\quad\mbox{\small $\Gamma=1\, (S) ,\, i\gamma_5\, (P) ,\, \sigma^{\mu\nu}\, (T)$}  .
\ea\right.
\ee
The Nambu--Goldstone coordinates disappear whenever the left and right sectors decouple; they only remain in those structures that mix the left and right chiralities, such as the scalar (S), pseudoscalar (P) and tensor (T) bilinears. While the spinorial indices get closed within $J_\Gamma$, the fermion bilinear is an $SU(2)$ tensor, with indices $ J^{mn}_\Gamma\, =\, \bar{\xi}_n\,\Gamma\, \xi_m$, that transforms covariantly: $J_\Gamma\to g_h^{\phantom{\dagger}} \,J_\Gamma\, g_h^\dagger $. 

Fermion masses constitute an explicit breaking of chiral symmetry, which can be incorporated in the effective Lagrangian with a right-handed spurion field $\cY_R$, transforming as
\bel{eq:Yspurion}
\cY_R\quad\longrightarrow\quad g_R^{\phantom{\dagger}}\,\cY_R\, g_R^\dagger\, ,
\qquad\qquad\qquad
\cY\, \equiv\, u\,\cY_R\, u^\dagger
\quad\longrightarrow\quad g_h^{\phantom{\dagger}}\,\cY\, g_h^\dagger
\, ,
\ee
that allows us to add the invariant term
\bel{eq:Yukawas}
\Delta\cL_2^{\mathrm{Fermionic}}\, =\, -v\; \bar\xi_L\, \cY\, \xi_R \, +\, \mathrm{h.c.}
\, =\, -v\;\bar\psi_L\, U(\vec{\phi\,})\,\cY_R\,\psi_R\, +\, \mathrm{h.c.}
\ee
The Yukawa interaction is recovered when the spurion field is frozen to the value \cite{Appelquist:1984rr,Bagan:1998vu}
\bel{eq:SM_Yspurion}
\cY\, =\, \hat{Y}_t(h/v)\,\cP_+ + \hat Y_b(h/v)\,\cP_-\, ,
\qquad\qquad\qquad
\cP_\pm \,\equiv\,\frac{1}{2}\,\left( I_2\pm\sigma_3\right)\, ,
\ee
where \cite{Buchalla:2013rka}
\bel{eq:SM_Yspurion_h}
\hat{Y}_{t,b}(h/v)\, =\, \sum_{n=0}\, \hat{Y}_{t,b}^{(n)}\, \left(\frac{h}{v}\right)^n\, .
\ee
In the SM, the $SU(2)$ doublet structure of the Higgs implies $\hat{Y}_{t,b}^{(0)} = \hat{Y}_{t,b}^{(1)} = m_{t,b}/v$, while $\hat{Y}_{t,b}^{(n\ge 2)}=0$.

To account for the flavour dynamics, one promotes the fermion doublets $\xi_{L,R}$ to flavour vectors $\xi^A_{L,R}$ with family index $A$. The spurion field $\cY$ becomes then a $3\times 3$ flavour matrix, with up-type and down-type components $\hat{Y}_u(h/v)$ and $\hat{Y}_d(h/v)$ that parametrize the custodial and flavour symmetry breaking \cite{Espriu:2000fq}. The corresponding expansion coefficient matrices $\hat{Y}_{u,d}^{(n)}(h/v)$, multiplying the $h^n$ term, could be different for each power $n$ because chiral symmetry does not constrain them. Additional dynamical inputs are needed to determine their flavour structure.

\subsection{Power counting}

The structure of the LO Lagrangian determines the power-counting rules of the EWET, in a quite straightforward way \cite{Buchalla:2013eza,Pich:2016lew}. The Higgs and the Nambu--Goldstone modes do not carry any chiral dimension $\hat d$, while their canonical field dimensions are compensated by the electroweak scale $v$. The external gauge sources $\hat W_\mu$, $\hat B_\mu$ and $\hat X_\mu$ are of $\cO(p)$, as they appear in the covariant derivatives, and their corresponding field-strength tensors are quantities of $\cO(p^2)$. Since any on-shell particle satisfies $p^2 = m^2$, all EWET fields have masses of $\cO(p)$. This implies that the gauge couplings are also of $\cO(p)$ because $m_W = g v/2$ and $m_Z = \sqrt{g^2+g'^2}\, v/2$, while $\vec{W}_\mu , B_\mu \sim \cO(p^0)$. All terms in $\cL_{\mathrm{YM}}$ and $\Delta\cL_2^{\mathrm{Bosonic}}$ are then of $\cO(p^2)$, provided the Higgs potential is also assigned this chiral dimension, which is consistent with the SM Higgs self-interactions being proportional to $m_h^2\sim \cO(p^2)$. Thus,
\begin{displaymath}
u(\vec{\phi\,}) , h ,  \cF_u(h/v) , \vec{W}_\mu , B_\mu \,  \sim \, \cO(p^0)\, ,
\qquad\qquad
\hat{W}_{\mu\nu} , \hat{B}_{\mu\nu} , \hat{X}_{\mu\nu} , f_{\pm\, \mu\nu} , c_n^{(V)} \, \sim \, \cO (p^2)\, ,
\end{displaymath}
\be\label{eq:bosonic_PC}
D_\mu U  , u_\mu , \partial_\mu ,  \hat{W}_\mu ,  \hat{B}_\mu , \hat{X}_\mu  , m_h , m_W ,  m_Z ,  g , g'\, \sim \, \cO (p)\, .
\ee

The assignment of chiral dimensions is slightly more subtle in the fermion sector. The chiral fermion fields must scale as $\xi_{L,R}\sim\cO(p^{1/2})$, so that the fermionic component of $\cL_{\mathrm{SM}}^{(0)}$ is also of $\cO(p^2)$. The spurion $\cY$ is a quantity of $\cO(p)$, since the Yukawa interactions must be consistent with the chiral counting of fermion masses, and the fermion mass terms are then also of $\cO(p^2)$. Therefore,
\be\label{eq:fermionic_PC}
\xi  , \bar\xi , \psi ,  \bar\psi \, \sim \,  \cO (p^{1/2})\, ,
\qquad\qquad\qquad
m_\psi , \cY \, \sim \, \cO (p)\, .
\ee

The chiral dimensions reflect the infrared behaviour at low momenta and lead to a consistent power counting to organise the EWET. In particular, the chiral low-energy expansion preserves gauge invariance order by order, because the kinetic, cubic and quartic gauge terms have all $\hat d = 2$  \cite{Hirn:2005fr}. With a straightforward dimensional analysis \cite{Manohar:1983md}, one can easily generalise the Weinberg's power-counting theorem \eqn{eq:WeinbergPC} \cite{Weinberg:1978kz}. An arbitrary Feynman diagram $\Gamma$ scales with momenta as \cite{Buchalla:2012qq,Buchalla:2013rka,Buchalla:2013eza,Hirn:2005fr,Pich:2016lew}
\bel{eq:Gamma_scaling}
\Gamma \,\sim\, p^{\hat{d}_\Gamma}\, ,
\qquad\qquad\qquad
\hat{d}_\Gamma\, =\, 2 + 2L + \sum_{\hat{d}} (\hat{d} -2)\, N_{\hat{d}} \, ,
\ee
with $L$ the number of loops and $N_{\hat{d}}$ the number of vertices with a given value of $\hat{d}$. 

As in $\chi$PT, quantum loops increase the chiral dimension by two units and their divergences get renormalized by higher-order operators.  The loop corrections are suppressed by the usual geometrical factor $1/(4\pi)^2$, giving rise to a series expansion in powers of momenta over the electroweak chiral scale $\Lambda_{\mathrm{EWET}} = 4\pi v\sim 3~\mathrm{TeV}$. Short-distance contributions from new physics generate EWET operators, suppressed by the corresponding new-physics scale $\Lambda_{\mathrm{NP}}$. One has then a combined expansion in powers of $p/\Lambda_{\mathrm{EWET}}$ and $p/\Lambda_{\mathrm{NP}}$. 

Local four-fermion operators, originating in short-distance exchanges of heavier states, carry a corresponding factor $g_{\mathrm{NP}}^2/\Lambda_{\mathrm{NP}}^2$, and similar suppressions apply to operators with a higher number of fermion pairs. Assuming that the SM fermions are weakly coupled, {\it i.e.}, $g_{\mathrm{NP}}^{\phantom{2}}\sim\cO( p)$, one must then assign an additional $\cO( p)$ suppression to the fermion bilinears  \cite{Buchalla:2013eza,Pich:2016lew}:
\bel{eq:4-fermion-counting}
(\bar\eta\,\Gamma\,\zeta)^n \quad\sim\quad\cO\left(  p^{2n}\right)\, .
\ee
This has already been taken into account in the Yukawa interactions through the spurion $\cY$. Obviously, the power-counting rule \eqn{eq:4-fermion-counting} does not apply to the kinetic term and, moreover, it is consistent with the loop expansion. 

When the gauge sources are frozen to the values \eqn{eq:SMgauge} and \eqn{eq:Xmu-fixing}, they generate an explicit breaking of custodial symmetry that gets transferred to higher orders by quantum corrections. Something similar happens in $\chi$PT with the explicit breaking of chiral symmetry through electromagnetic interactions \cite{Ecker:1988te,Ecker:2000zr,Urech:1994hd}. This can be easily incorporated into the EWET with the right-handed spurion \cite{Pich:2016lew}
\bel{eq:T_R}
\cT_R\quad\longrightarrow\quad g_R^{\phantom{\dagger}}\, \cT_R\, g_R^\dagger\, ,
\qquad\qquad\qquad
\cT\, =\, u\, \cT_R\, u^\dagger \quad\longrightarrow\quad g_h^{\phantom{\dagger}} \cT g_h^\dagger\, .
\ee
Chiral-invariant operators with an even number of $\cT$ fields account for the custodial symmetry-breaking structures induced through quantum loops with $B_\mu$ propagators, provided one makes the identification
\bel{eq:T_R-value}
\cT_R    \, =\, -g'\;\frac{\sigma_3}{2}\, .
\ee
Being proportional to the coupling $g'$, this spurion has chiral dimension $\hat d =1$ \cite{Pich:2016lew}:
\bel{eq:T_R-counting}
\cT_R\;\sim\; \cT\;\sim\; \cO( p)\, .
\ee

\subsection{NLO Lagrangian}

Assuming invariance under CP transformations, the most general $\cO( p^4)$ bosonic Lagrangian contains eleven P-even ($\cO_i$) and three P-odd ($\widetilde \cO_i$) operators \cite{Pich:2015kwa,Pich:2016lew}:
\bel{eq:L4}
\cL_4^{\mathrm{Bosonic}}\, =\, \sum_{i=1}^{11} \cF_i(h/v)\; \cO_i
\, +\, \sum_{i=1}^{3}\widetilde\cF_i(h/v)\; \widetilde \cO_i \, .
\ee
For simplicity, we will only consider a single fermion field. The CP-invariant fermionic Lagrangian of $\cO( p^4)$ involves then seven P-even ($\cO_i^{\psi^2}$) plus three P-odd ($\widetilde \cO_i^{\psi^2}$) operators with one fermion bilinear, and ten P-even ($\cO_i^{\psi^4}$) plus two P-odd ($\widetilde \cO_i^{\psi^4}$) four-fermion operators \cite{Pich:2016lew}:
\beqn\label{eq:L4-fermionic}
\cL_4^{\mathrm{Fermionic}}&\, =&\, \sum_{i=1}^{7}
\cF_i^{\psi^2}\! (h/v)\; \cO_i^{\psi^2}
\, +\, \sum_{i=1}^{3}
\widetilde\cF_i^{\psi^2}\! (h/v)\; \widetilde \cO_i^{\psi^2}
\no\\ & + &\,
 \sum_{i=1}^{10}\cF_i^{\psi^4}\! (h/v)\; \cO_i^{\psi^4}
\, +\, \sum_{i=1}^{2}\widetilde\cF_i^{\psi^4}\! (h/v)\; \widetilde \cO_i^{\psi^4}\, .
\eeqn
The basis of independent P-even and P-odd operators is listed in Tables~\ref{tab:Peven-Op4} and \ref{tab:Podd-Op4}, respectively.\footnote{
A much larger number of operators appears in former EWET analyses, owing to a slightly different chiral counting that handles the breakings of custodial symmetry less efficiently~\cite{Alonso:2012px,Buchalla:2012qq,Buchalla:2013rka}.}
The number of chiral structures has been reduced through field redefinitions, partial integration, equations of motion and algebraic identities.\footnote{
When QCD interactions are taken into account, the covariant derivatives incorporate the gluon field $\hat G_\mu = g_s G_\mu^a T^a$ and two additional P-even operators must be included:
$\mathrm{Tr}_C ( \hat G_{\mu\nu} \hat G^{\mu\nu} )$ and 
$\mathrm{Tr}_C ( \hat G_{\mu\nu} \langle J^{8\,\mu\nu}_T\rangle )$, where $\hat G_{\mu\nu} = \partial_\mu \hat G_\nu -  \partial_\nu \hat G_\mu - i\, [ \hat G_\mu , \hat G_\nu ]$, $\mathrm{Tr}_C (A)$ indicates the colour trace of $A$
 and $J^{8\,\mu\nu}_T$ is the colour-octet tensorial quark bilinear \cite{Krause:2018cwe}.} 
The number of fermionic operators increases dramatically when the quark and lepton flavours are taken into account.

\begin{table}[t!]
\tableparts
{\caption{
CP-invariant, P-even operators of the $\cO(p^4)$ EWET \cite{Pich:2015kwa,Pich:2016lew}.}
\label{tab:Peven-Op4}}
{\renewcommand{\arraystretch}{1.4}
\begin{tabular}{cccc}
\hline
$i$ & $\cO_i$ &  $\cO^{\psi^2}_i$ & $\cO^{\psi^4}_i$ \\
\hline
$1$  & $\frac{1}{4}\,\bra {f}_+^{\mu\nu} {f}_{+\, \mu\nu}
- {f}_-^{\mu\nu} {f}_{-\, \mu\nu}\ket$
&  $\bra J_S \ket \bra u_\mu u^\mu \ket$ & $\bra J_{S} J_{S} \ket $
\\
$2$  & $ \frac{1}{2}\, \bra {f}_+^{\mu\nu} {f}_{+\, \mu\nu}
+ {f}_-^{\mu\nu} {f}_{-\, \mu\nu}\ket$
& $i \,  \bra J_T^{\mu\nu} \left[ u_\mu, u_\nu \right] \ket$ & 
$\bra J_{P} J_{P} \ket $
\\
$3$  & $\frac{i}{2}\,  \bra {f}_+^{\mu\nu} [u_\mu, u_\nu] \ket$
& $\bra J_T^{\mu \nu} f_{+ \,\mu\nu} \ket $ & 
$\bra J_{S} \ket \bra  J_{S} \ket $
\\
$4$  & $\bra u_\mu u_\nu\ket \, \bra u^\mu u^\nu\ket $ & 
$\hat{X}_{\mu\nu} \bra J_T^{\mu \nu} \ket $ 
& $\bra J_{P} \ket \bra  J_{P} \ket $
\\
$5$  & $\bra u_\mu u^\mu\ket^2$ & 
$\frac{1}{v} \, (\partial_\mu h)\,\bra u^\mu J_P \ket $ 
&  $\bra J_V^\mu J_{V,\mu}^{\phantom{\mu}}\ket $
\\
$6$ & $\frac{1}{v^2}\, (\partial_\mu h)(\partial^\mu h)\,\bra u_\nu u^\nu \ket$
& $\bra J_A^\mu \ket \bra u_\mu \mathcal{T} \ket $ & 
 $\bra J_A^\mu J_{A,\mu}^{\phantom{\mu}}\ket $
\\
$7$  & 
$\frac{1}{v^2} \, (\partial_\mu h)(\partial_\nu h)\,\bra u^\mu u^\nu \ket$
& $\frac{1}{v^2}\, (\partial_\mu h) (\partial^\mu h)\, \bra J_S\ket $
& $\bra J_V^\mu\ket \bra J_{V,\mu}^{\phantom{\mu}}\ket $
\\
$8$ & 
$\frac{1}{v^4}\, (\partial_\mu h)(\partial^\mu h)(\partial_\nu h)(\partial^\nu h)$
& --- & $\bra J_A^\mu\ket \bra J_{A,\mu}^{\phantom{\mu}}\ket $
\\
$9$ & $\frac{1}{v}\, (\partial_\mu h)\,\bra f_-^{\mu\nu}u_\nu \ket$ &  ---
& $\bra J^{\mu\nu}_{T} J_{T ,\mu\nu}^{\phantom{\mu}} \ket $
\\
$10$ & $\langle \cT u_\mu\rangle^2$  & --- &
$\bra J^{\mu\nu}_{T} \ket \bra J_{T ,\mu\nu}^{\phantom{\mu}} \ket $
\\ 
$11$ & $ \hat{X}_{\mu\nu} \hat{X}^{\mu\nu}$ & --- & ---
\\
\hline
\end{tabular}
}
\vskip .6cm 
\tableparts
{\caption{
CP-invariant, P-odd operators of the $\cO(p^4)$ EWET \cite{Pich:2016lew}.}
\label{tab:Podd-Op4}}
{\renewcommand{\arraystretch}{1.4}
\begin{tabular}{cccc}
\hline
$i$ & $\widetilde{\cO}_i$ & $\widetilde{\cO}^{\psi^2}_i$
& $\widetilde{\cO}^{\psi^4}_i$ 
\\ \hline
$1$ &  $\frac{i}{2}\, \bra {f}_-^{\mu\nu} [u_\mu, u_\nu] \ket$
& $\bra J_T^{\mu \nu} f_{- \,\mu\nu} \ket $ &
$\bra J_V^\mu J_{A,\mu}^{\phantom{\mu}}\ket $
\\
$2$  & $\bra {f}_+^{\mu\nu} {f}_{-\, \mu\nu} \ket $
& $\frac{1}{v}\, (\partial_\mu h) \, \bra u_\nu J^{\mu\nu}_T \ket $
 &  $\bra J_V^\mu\ket \bra J_{A,\mu}^{\phantom{\mu}}\ket $
\\
$3$  &  $\frac{1}{v}\, (\partial_\mu h)\,\bra f_+^{\mu\nu}u_\nu \ket$
& $\bra J_V^\mu \ket \bra u_\mu \mathcal{T} \ket $ & ---
\\
\hline
\end{tabular}
}
\end{table}

All coefficients $\cF_i^{(\psi^{2,4})}(z)$ and $\widetilde\cF_i^{(\psi^{2,4})}(z)$ are functions of $z=h/v$; {\it i.e.},
\bel{eq:mFi}
\cF_i^{(\psi^{2,4})}(z) \, =\, \sum_{n=0} \cF_{i,n}^{(\psi^{2,4})}\, z^n\, , 
\qquad\qquad\qquad
\widetilde\cF_i^{(\psi^{2,4})}(z)\, =\,\sum_{n=0} \widetilde\cF_{i,n}^{(\psi^{2,4})}\, z^n\, .
\ee
When the gauge sources are frozen to the values~\eqn{eq:SMgauge} and \eqn{eq:Xmu-fixing}, the Higgsless term 
$\cF_{2,0}\,\cO_2  + \cF_{11,0} \, \cO_{11} +\widetilde{\cF}_{2,0}\,\widetilde{\cO}_2$
%
becomes a linear combination of the $W_\mu$ and $B_\mu$ Yang-Mills Lagrangians. Therefore, it could be eliminated through a redefinition of the corresponding gauge couplings.

To compute the NLO amplitudes, one must also include one-loop corrections with the LO Lagrangian $\cL_2^{\mathrm{EW}}$. The divergent contributions can be computed with functional methods, integrating all one-loop fluctuations of the fields in the corresponding path integral. The divergences generated by the scalar sector (Higgs and Nambu--Goldstone fluctuations) were first computed in Ref.~\cite{Guo:2015isa}, while the full one-loop renormalization has been accomplished recently \cite{Alonso:2017tdy,Buchalla:2017jlu}.

According to the EWET power counting, the one-loop contributions are of $\cO(p^4)$. Nevertheless, they can give rise to local structures already present in the LO Lagrangian, multiplied by $\cO(p^2)$ factors such as $m_h^2$, $c_n^{(V)}$ or the product of two gauge couplings. In fact, the fluctuations of gauge bosons and fermions belong to the renormalizable sector and do not generate new counterterms \cite{Buchalla:2017jlu}. They must be reabsorbed into $\cO(p^4)$ redefinitions of the LO Lagrangian parameters and fields. Genuine $\cO(p^4)$ structures originate in the scalar fluctuations, owing to the non-trivial geometry of the scalar field manifold~\cite{Alonso:2015fsp,Alonso:2016oah,Alonso:2017tdy}, and the mixed loops between the renormalizable and non-renormalizable sectors.

In order to avoid lengthy formulae, we only detail here the $\cO(p^4)$ divergences, originating from Higgs and Nambu--Goldstone fluctuations, that renormalize the couplings of the bosonic $\cO_i$ operators:
\bel{eq:cFi-renorm-EWET}
\cF_i(h/v)\, =\, \cF_i^{\, r}(\mu, h/v) + \Gamma_{\cO_i}(h/v)\;\Delta_{\overline{\mathrm{MS}}}
\; =\; \sum_{n=0} \left[\cF_{i,n}^{\, r}(\mu) + \gamma_{i,n}^\cO\; \Delta_{\overline{\mathrm{MS}}} \right] \;
\left(\frac{h}{v}\right)^n\, ,
\ee
with
\bel{eq:MSbar}
\Delta_{\overline{\mathrm{MS}}} \, =\, \frac{\mu^{D-4}}{32\pi^2}\,\left\{ \frac{2}{D-4} -\log{(4\pi)} + \gamma_E\right\}\, ,
\ee
the usual $\overline{\mathrm{MS}}$ subtraction. The computed coefficient functions $\Gamma_{\cO_i}(z)$ are given in Table~\ref{tab:Gamma-Oi}, in terms of $\cF_u(z)$ and~\cite{Guo:2015isa}
\be 
\cK(z)\,\equiv \, \frac{\cF_u^{\, '}(z)}{\cF_u^{1/2}(z)}\, ,
\qquad\qquad\qquad
\Omega(z)\,\equiv \, 2\,\frac{\cF_u^{\, ''}(z)}{\cF_u(z)} - \left( \frac{\cF_u^{\, '}(z)}{\cF_u(z)} \right)^2\, ,
\ee
where $\cF_u^{\, '}$ and $\cF_u^{\, ''}$ indicate the first and second derivative of $\cF_u(z)$ with respect to the variable $z=h/v$. The last column of the table shows explicitly the zero-order factors $\gamma_{i,0}^\cO$ in the expansion of $\Gamma_{\cO_i}(h/v)$ in powers of the Higgs field, which only depend on the LO couplings $a \equiv \frac{1}{2}\, c_1^{(u)}$ and $b \equiv c_2^{(u)}$.
The coefficients  $\Gamma_{\cO_i}(z)$ and $\gamma_{i,n}^\cO$ govern the running with the renormalization scale of the corresponding $\cF_i^{\, r}(\mu,z)$ functions and $\cF^{\, r}_{i,n}(\mu)$ factors, respectively, as dictated in \eqn{eq:running} for the $\chi$PT LECs.

\begin{table}[t!]
\tableparts
{\caption{Coefficient functions $\Gamma_{\cO_i}(h/v)$, 
generated by one-loop scalar fluctuations, and their zero-order expansion factors $\gamma_{i,0}^\cO$~\cite{Guo:2015isa}.} 
\label{tab:Gamma-Oi}}
{\renewcommand{\arraystretch}{1.4}
\begin{tabular}{ccc}
\hline
$i$ & $\Gamma_{\cO_i}(h/v)$ &  $\gamma_{i,0}^\cO$ \\
\hline
1 & $\frac{1}{24}\, (\cK^2 -4)$ & $-\frac{1}{6}\, (1 -a^2)$
\\
2 & $-\frac{1}{48}\, (\cK^2 +4)$ & $-\frac{1}{12}\, (1 +a^2)$
\\
3 & $\frac{1}{24}\, (\cK^2 -4)$ & $-\frac{1}{6}\, (1 -a^2)$
\\
4 & $\frac{1}{96}\, (\cK^2 -4)^2$ & $\frac{1}{6}\, (1 -a^2)^2$
\\
5 & $\quad\frac{1}{192}\, (\cK^2 -4)^2 + \frac{1}{128}\,\cF_u^2\,\Omega^2\quad $ 
& $\frac{1}{8}\, (a^2 -b)^2 + \frac{1}{12}\, (1 -a^2)^2$
\\
6 & $\frac{1}{16}\, \Omega\, (\cK^2 -4) - \frac{1}{96}\,\cF_u\,\Omega^2$ 
& $-\frac{1}{6}\, (a^2 -b) \, (7 a^2-b-6)$
\\
7 & $\frac{1}{24}\,\cF_u\,\Omega^2$ & $\frac{2}{3}\, (a^2 -b)^2$
\\
8 & $\frac{3}{32}\,\Omega^2$ & $\frac{3}{2}\, (a^2 -b)^2$
\\
9 & $\frac{1}{24}\,\cF_u^{\, '}\,\Omega$ & $-\frac{1}{3}\, a\, (a^2 -b)$
\\ \hline
\end{tabular}}
\end{table}

Taking $\cF_u(z)=1$ ($a=b=0$), the Higgs decouples from the Nambu--Goldstone modes in $\Delta\cL_2^{\mathrm{Bosonic}}$, and one recovers the known renormalization factors of the Higgsless electroweak chiral Lagrangian \cite{Appelquist:1980vg,Herrero:1993nc,Herrero:1994iu,Longhitano:1980iz,Longhitano:1980tm}. The renormalization factors $\gamma_{4,n}^\cO$ and $\gamma_{5,n}^\cO$ reproduce in this case their $\chi$PT counterparts $\frac{1}{4}\,\gamma_2$ and $\frac{1}{4}\,\gamma_1$ in \eqn{eq:gammas-su2}.
In the SM, $\cK = 2$ and $\Omega = 0$ ($a=b=1$). Thus, all $\cO(p^4)$ divergences disappear, as it should,  except $\Gamma_{\cO_2}(z) = \gamma_{2,0}^\cO = -\frac{1}{6}$ that is independent of the Higgs field and gets reabsorbed through a renormalization of the gauge couplings in the Yang-Mills Lagrangian.

Since $\frac{1}{4}\,\Gamma_{\cO_1}(z) + \frac{1}{2}\,\Gamma_{\cO_2}(z) = -\frac{1}{12}$  is independent of the Higgs field, the chiral structures $h^n\,\langle f_+^{\mu\nu} f_{+\,\mu\nu}^{\phantom{+}}\rangle$ are not renormalized when $n\not= 0$.
Therefore, the interaction vertices $h^n \gamma\gamma$ and $h^n Z\gamma$, with $n\ge 1$, are renormalization-group invariant~\cite{Azatov:2013ura,Delgado:2014jda,Guo:2015isa}.

\subsection{Scattering of longitudinal gauge bosons at NLO}

The one-loop corrections to the elastic scattering of two electroweak Nambu--Goldstone modes (or, equivalently, longitudinal gauge bosons) 
can be partly taken from the $\chi$PT expression \eqn{eq:PionScattering-1L-b}, which only includes the virtual contributions from massless $\vec\phi$ propagators. One must add the corrections induced by the Higgs boson, and replace the $\cO(p^4)$ $\chi$PT LECs by their corresponding EWET counterparts: $\ell_1\to 4\,\cF_{5,0}$, $\ell_2\to 4\,\cF_{4,0}$. In the limit $g=g'=0$, custodial symmetry becomes exact and
the scattering amplitudes follow the weak isospin decomposition \eqn{eq:PionScattering-1L}, with the $\cO(p^4)$ function \cite{Delgado:2013hxa,Espriu:2013fia}
\beqn\label{eq:WWscattering-NLO}
\lefteqn{A(s,t,u) \, = \, \frac{s}{v^2} \; (1-a^2)\, +\,
\frac{4}{v^4}\,\left[ \cF^r_{4,0}(\mu)\, (t^2+u^2) + 
2\, \cF^r_{5,0}(\mu)\, s^2 \right]} &&
\no\\[3pt]
&& \hskip .3cm\mbox{} + \frac{1}{96\pi^2 v^4}\,\left\{ 
\frac{2}{3}\, (14\, a^4 -10\, a^2-18\, a^2 b+ 9\, b^2 + 5)\; s^2 
+ \frac{13}{3}\, (1-a^2)^2\; (t^2+u^2)
\right.\no\\[3pt] && \hskip 1.85cm\left.\mbox{}
 + \frac{1}{2}\, (1-a)^2\,\left[
 (s^2-3 t^2-u^2)\,\log{\left(\frac{-t}{\mu^2}\right)} +
(s^2- t^2-3 u^2)\,\log{\left(\frac{-u}{\mu^2}\right)}\right]
\right.\no\\[3pt] && \hskip 1.85cm\left.\mbox{}
-3\, (2\, a^4 -2\, a^2 -2\, a^2 b + b^2 + 1)\; s^2\,\log{\left(\frac{-s}{\mu^2}\right)} 
\right\}\, ,
\eeqn
where $a \equiv \frac{1}{2}\, c_1^{(u)}$ and $b \equiv c_2^{(u)}$ are the relevant LO couplings of the Higgs to the $\vec\phi$ fields in eqn~\eqn{eq:L2bosonic}. All boson masses have been neglected, and the $\cO(p^4)$ couplings have been renormalized in the $\overline{\mathrm{MS}}$ scheme, as indicated in \eqn{eq:cFi-renorm-EWET} and \eqn{eq:MSbar}. Taking into account the different renormalization schemes, this result agrees with its corresponding $\chi$PT expression~\eqn{eq:PionScattering-1L-b} when $a=b=0$. In the SM, $a=b=1$ and $\cF_{4,0} = \cF_{5,0}=0$, which implies $A(s,t,u) = 0$. Owing to unitarity, the SM amplitude is of $\cO(m_h^2/s)$.

With the same approximations ($g=g'=m_h=0$) and neglecting in addition the scalar potential, {\it i.e.}, assuming the Higgs self-interactions to be of $\cO(m_h^2)$, the elastic scattering amplitude of two Higgs bosons takes the form
\cite{Delgado:2013hxa}
\beqn\label{eq:hh2hh-NLO}
\lefteqn{\cA(hh\to hh) \, = \, 
\frac{2}{v^4}\,\cF^r_{8,0}(\mu)\, (s^2 +t^2+u^2)} &&
\no\\[3pt]
& & \hskip .3cm\mbox{} + \frac{3\, (a^2-b)^2}{32\pi^2 v^4}\left\{
2\, (s^2 +t^2+u^2) - s^2\, \log{\left(\frac{-s}{\mu^2}\right)}
- t^2\, \log{\left(\frac{-t}{\mu^2}\right)}- u^2\, \log{\left(\frac{-u}{\mu^2}\right)}\right\} ,
\no\\ &&
\eeqn 
which also vanishes in the SM limit ($a=b=1$, $\cF_{8,0}=0$).

The analogous computation of the $\phi^i\phi^j\to hh$ scattering amplitudes gives~\cite{Delgado:2013hxa}:
\beqn\label{eq:WW2hh-NLO}
\lefteqn{\cA(\phi^i\phi^j\to hh) \, = \, \delta_{ij}\,\left\{
\frac{a^2-b}{v^2} \; s\, +\,
\frac{1}{v^4}\,\left[ 2\,\cF^r_{6,0}(\mu)\, s^2 + \cF^r_{7,0}(\mu)\, (t^2+u^2) \right]\right. } &&
\no\\[3pt]
& & \hskip .65cm\mbox{} + \frac{a^2-b}{192\pi^2 v^4}\, \left( 
   (a^2-b)\;  t^2\,
 \left[\frac{26}{3}-3\,\log{\left(\frac{-t}{\mu^2}\right)}- \log{\left(\frac{-u}{\mu^2}\right)}\right]
\right.\no\\[3pt] & & \hskip 2.45cm\mbox{}
 +  (a^2-b)\; u^2\,
\left[\frac{26}{3}-\log{\left(\frac{-t}{\mu^2}\right)}- 3\,\log{\left(\frac{-u}{\mu^2}\right)}\right]
\no\\[3pt] & & \hskip 2.45cm\mbox{}
+  (a^2-b)\; s^2\, \left[ \log{\left(\frac{-t}{\mu^2}\right)} + \log{\left(\frac{-u}{\mu^2}\right)}  \right]
+ 12\, (a^2-1)\; s^2\, \log{\left(\frac{-s}{\mu^2}\right)}
\no\\[3pt] & & \left.\left.\hskip 2.45cm\mbox{}
+ \frac{1}{3}\, (72-88\, a^2 + 16\, b )\; s^2 
\right)\right\} .
\eeqn
Time-reversal invariance implies that the same expression applies for $\cA(hh\to \phi^i\phi^j)$.

With the renormalization factors given in Table~\ref{tab:Gamma-Oi}, one can easily check that the scattering amplitudes \eqn{eq:WWscattering-NLO}, \eqn{eq:hh2hh-NLO} and \eqn{eq:WW2hh-NLO} are independent of the renormalization scale $\mu$.
Similar expressions have been derived for other interesting scattering amplitudes such as 
$\cA(\gamma\gamma\to\phi^i\phi^j)$  \cite{Delgado:2014jda}, $\cA(\gamma\gamma\to hh)$ \cite{Delgado:2016rtd},
$\cA(\phi^i\phi^j\to t\bar t)$  \cite{Castillo:2016erh} and $\cA(hh\to t\bar t)$ \cite{Castillo:2016erh}.

\section{Fingerprints of heavy scales}
\label{sec:fingerprints}

The couplings of the EWET encode all the dynamical information about the underlying ultraviolet dynamics that is accessible at the electroweak scale. Different scenarios of new physics above the energy gap would imply different patterns of LECs, with characteristic correlations that could be uncovered through high-precision measurements. Given a generic strongly-coupled theory of EWSB with heavy states above the gap, we would like to identify the imprints that its lightest excitations leave on the effective Lagrangian couplings. To simplify the discussion, we will concentrate on the bosonic CP-conserving operators $\cO_i$ \cite{Pich:2015kwa}. A more general analysis, including the fermion sector and gluonic interactions has been given in Refs.~\cite{Krause:2018cwe,Pich:2016lew}.

Let us consider an effective Lagrangian containing the SM fields coupled to the lightest scalar, pseudoscalar, vector and axial-vector colour-singlet resonance multiplets $S$, $P$, $V^{\mu\nu}$ and $A^{\mu\nu}$, transforming as $SU(2)_{L+R}$ triplets ($R\to g_h^{\phantom{\dagger}} R\, g_h^{\dagger}$), and the corresponding singlet states $S_1$, $P_1$, $V^{\mu\nu}_1$ and $A^{\mu\nu}_1$ ($R_1\to R_1$). 
In order to analyse their implications on the $\cO(p^4)$ EWET couplings, we only need to keep those structures with the lowest number of resonances and derivatives. At LO, the most general bosonic interaction with at most one resonance field, that is invariant under the symmetry group $G$, has the form:
\bel{eq:HeavyFields}
\cL_{\mathrm{EWR}\chi\mathrm{T}}^{\phantom{*}}\, \dot=\,\langle V_{\mu\nu}\,\chi_V^{\mu\nu} \rangle + \langle A_{\mu\nu}\,\chi_A^{\mu\nu} \rangle
+ \langle P\,\chi_P \rangle +  S_1\, \chi_{S_1} \, ,
\ee
with
\begin{displaymath}
\chi_V^{\mu\nu}\, =\,\frac{F_V}{2\sqrt{2}}\, f^{\mu\nu}_+
+ \frac{i\, G_V}{2\sqrt{2}}\, [u^\mu, u^\nu]\, ,
\qquad\qquad
\chi_A^{\mu\nu}\, =\,\frac{F_A}{2\sqrt{2}}\, f^{\mu\nu}_-
+ \sqrt{2}\, \lambda_1^{hA}\,  \partial^\mu h \, u^\nu\, ,
\end{displaymath}
\bel{eq:EWchis}
\chi_P\, =\, \frac{d_P}{v}\, (\partial_\mu h)\, u^\mu\, ,
\qquad\qquad\qquad
\chi_{S_1}\, =\, \frac{c_d}{\sqrt{2}}\,\langle u_\mu u^\mu \rangle + \lambda_{hS_1} v\, h^2\,.
\ee
The spin-1 fields are described with antisymmetric tensors and all couplings must be actually understood as functions of $h/v$. At this chiral order, the singlet vector, axial-vector and pseudoscalar fields, and the scalar triplet cannot couple to the Nambu--Goldstone modes and gauge bosons.

Integrating out the heavy fields, one easily recovers the bosonic $\cO(p^4)$ EWET structures. In addition, the exchange of the scalar singlet field $S_1$ generates a correction to the LO EWET Lagrangian,
\bel{eq:S1-exchange}
\Delta\cL_{S_1}^{(2)}\, =\,\frac{\lambda_{hS_1}}{2 M_{S_1}^2}\; v\, h^2\, \left\{ \lambda_{hS_1}\, v\, h^2 + \sqrt{2}\,  c_d\,  \bra u_\mu u^\mu\ket\right\}\, ,
\ee
that is suppressed by two powers of the heavy mass scale $M_{S_1}$. The function $\lambda_{hS_1}(h/v)$ should be assigned a chiral dimension 2 to get a consistent power counting, so that the two terms in \eqn{eq:S1-exchange} have the same chiral order $\cO(p^4)$. The term proportional to $\lambda_{hS_1} c_d$ contributes to $\cF_u(h/v)$ in \eqn{eq:L2bosonic}, while the $(\lambda_{hS_1})^2$ term corrects the Higgs potential $V(h/v)$. 

The predicted values of the $\cO(p^4)$ couplings $\cF_i(h/v)$, in terms of resonance parameters, are detailed in Table~\ref{tab:EW-LECs}.
The exchange of triplet vector states contributes to $\cF_{1-5}$, while axial-vectors triplets leave their imprints on $\cF_{1,2,6,7,9}$. The spin-0 resonances have a much more reduced impact on the LECs: a pseudoscalar triplet only manifests in $\cF_7$ and the singlet scalar contributes only to $\cF_5$.
Possible low-energy contributions from more exotic $J^{PC}=1^{+-}$ heavy states haven been analysed in Ref.~\cite{Cata:2014fna}.

\begin{table}[tb]
\tableparts
{\caption{Resonance contributions to the bosonic $\cO(p^4)$ LECs. The right column includes short-distance constraints \cite{Pich:2015kwa}.}
\label{tab:EW-LECs}}
{\renewcommand{\arraystretch}{1.6}
\begin{tabular}{cccc}
\hline
$\cF_1\quad =$ & $\frac{F_A^2}{4M_A^2}- \frac{F_V^2}{4M_V^2}$ & = &
$-\frac{v^2}{4}\,\left(\frac{1}{M_V^2}+\frac{1}{M_A^2}\right)$
\\
$\cF_2\quad =$ & $-\frac{F_A^2}{8M_A^2}- \frac{F_V^2}{8M_V^2}$ &=&
$-\frac{v^2 (M_V^4+M_A^4)}{8 M_V^2M_A^2 (M_A^2-M_V^2)}$
\\
$\cF_3\quad =$ & $-  \frac{F_VG_V}{2M_V^2}$ &=& $-\frac{v^2}{2M_V^2}$
\\
$\cF_4\quad =$ & $\frac{G_V^2}{4 M_V^2}$ &=&
$\frac{(M_A^2-M_V^2) v^2}{4 M_V^2 M_A^2}$
\\
$\cF_5\quad =$ & $\frac{c_{d}^2}{4M_{S_1}^2} -\frac{G_V^2}{4M_V^2}$ &=&
$\frac{c_{d}^2}{4M_{S_1}^2} -\frac{(M_A^2-M_V^2) v^2}{4 M_V^2 M_A^2}$
\\
$\cF_6\quad =$ & $-\frac{(\lambda_1^{hA})^2v^2}{M_A^2}$ &=&
$-\frac{M_V^2 (M_A^2-M_V^2) v^2}{M_A^6}$
\\
$\cF_7\quad =$ & $\frac{d_P^2}{2 M_P^2}+ \frac{(\lambda_1^{hA})^2v^2}{M_A^2}$ &=&
$\frac{d_P^2}{2 M_P^2}+\frac{M_V^2 (M_A^2-M_V^2) v^2}{M_A^6}$
\\
$\cF_8\quad =$ & $0$ && 
\\
$\cF_9\quad =$ & $- \frac{F_A \lambda_1^{hA} v}{M_A^2}$ &=& $-\frac{M_V^2 v^2}{M_A^4}$
\\[2pt] \hline
\end{tabular}
}
\end{table}

The predicted vector and axial-vector contributions are independent of the (antisymmetric) formalism adopted to describe the spin-1 resonances. The same results are obtained using Proca fields or a hidden-gauge formalism~\cite{Bando:1987br,Casalbuoni:1988xm}, once a proper utraviolet behaviour is required (physical Green functions should not grow at large momenta) \cite{Pich:2016lew}.

\subsection{Short-distance behaviour}

The previous resonance-exchange predictions can be made more precise with mild assumptions about the ultraviolet properties of the underlying fundamental theory. The procedure is similar to the one adopted before in R$\chi$T. One imposes the expected fall-off at large momenta of specific Green functions, and obtains constraints on the resonance parameters that are valid in broad classes of dynamical theories.

The functional derivatives of the action with respect to the external sources $\hat W^\mu$ and $\hat B^\mu$ define the corresponding left and right (vector and axial) currents. In complete analogy with QCD, the matrix element of the vector current between two Nambu--Goldstone bosons is characterized by a vector form factor that has the same functional form of eqn~\eqn{eq:VFF-RChT}, with $F_V$ and $G_V$ the vector couplings in \eqn{eq:EWchis} and the electroweak scale $v$ replacing the pion decay constant $F$. Requiring that this form factor should vanish at infinite momentum transfer implies the constraint 
\bel{eq:FVGV}
F_V\, G_V\, =\, v^2\, .
\ee
Enforcing a similar condition on the matrix element of the axial current between the Higgs and one Nambu--Goldstone particle, one obtains \cite{Pich:2012dv,Pich:2013fea}:
\bel{eq:FAkappa}
F_A \,\lambda_1^{hA}\, =\, \frac{1}{2}\, c_1^{(u)}\, v\,\equiv\, a \, v\, ,
\ee
where $c_1^{(u)}$ is the $h\,\partial_\mu\vec\phi\,\partial^\mu\vec\phi$ coupling in \eqn{eq:L2bosonic}, which for $a \equiv \frac{1}{2}\, c_1^{(u)} = 1$ reproduces the gauge coupling of the SM Higgs.

The two-point correlation function of one left-handed and one right-handed currents is an order parameter of EWSB, formally given by the same expressions as in QCD, eqn~\eqn{eq:LRcorrelator} and \eqn{eq:WSR}, changing $F$ by $v$. In asymptotically-free gauge theories, $\Pi_{LR}(t)$ vanishes as $1/t^3$, at large momenta \cite{Bernard:1975cd}, which implies the two Weinberg Sum Rules \cite{Weinberg:1967kj}:
\bel{eq:WSRs}
F_{V}^2 - F_{A}^2  = v^2\, ,
\qquad\qquad\qquad\quad
F_{V}^2  \,M_{V}^2 - F_{A}^2 \, M_{A}^2  = 0\, .
\ee
These conditions require $M_V < M_A$ and determine $F_V$ and $F_A$ in terms of the resonance masses and the electroweak scale, as in \eqn{eq:FV_A}.
At the one-loop level, and together with \eqn{eq:FVGV} and \eqn{eq:FAkappa}, the super-convergence properties of $\Pi_{LR}(t)$ also impose a relation between the Higgs gauge coupling and the resonance masses \cite{Pich:2012dv,Pich:2013fea}:
\bel{eq:kappa}
a\, =\, M_V^2/M_A^2\, <\, 1\, .
\ee
Thus, the coupling of the Higgs to the electroweak gauge bosons is predicted to be smaller than the SM value.

With the identities \eqn{eq:FVGV}, \eqn{eq:FAkappa}, \eqn{eq:WSRs} and \eqn{eq:kappa}, all vector and axial-vector contributions to the $\cO(p^4)$ LECs can be written in terms of $M_V$, $M_A$ and $v$. The resulting expressions are shown in the last column of Table~\ref{tab:EW-LECs}, and are valid in dynamical scenarios where the two Weinberg sum rules are fulfilled, as happens in asymptotically-free theories. Softer conditions can be obtained imposing only the first sum rule \cite{Pich:2012dv,Pich:2013fea}, {\it i.e.}, requiring $\Pi_{LR}(t)\sim 1/t^2$ at large momentum transfer,
which is also valid in gauge theories with non-trivial UV fixed points~\cite{Peskin:1990zt}.

The expressions derived for the LECs in terms of the vector and axial-vector couplings are generic relations which include the functional dependence on $h/v$ hidden in the couplings. This is however no-longer true for the improved predictions incorporating short-distance constraints, where only constant couplings have been considered. Thus, the expressions on the right-hand side of Table~\ref{tab:EW-LECs} refer to $\cF_{i,0}$, the $\cO(h^0)$ terms in the expansion of the corresponding LECs in powers of $h/v$.

\begin{figure*}
\begin{center}
\begin{minipage}[c]{6cm}
\includegraphics[width=6cm]{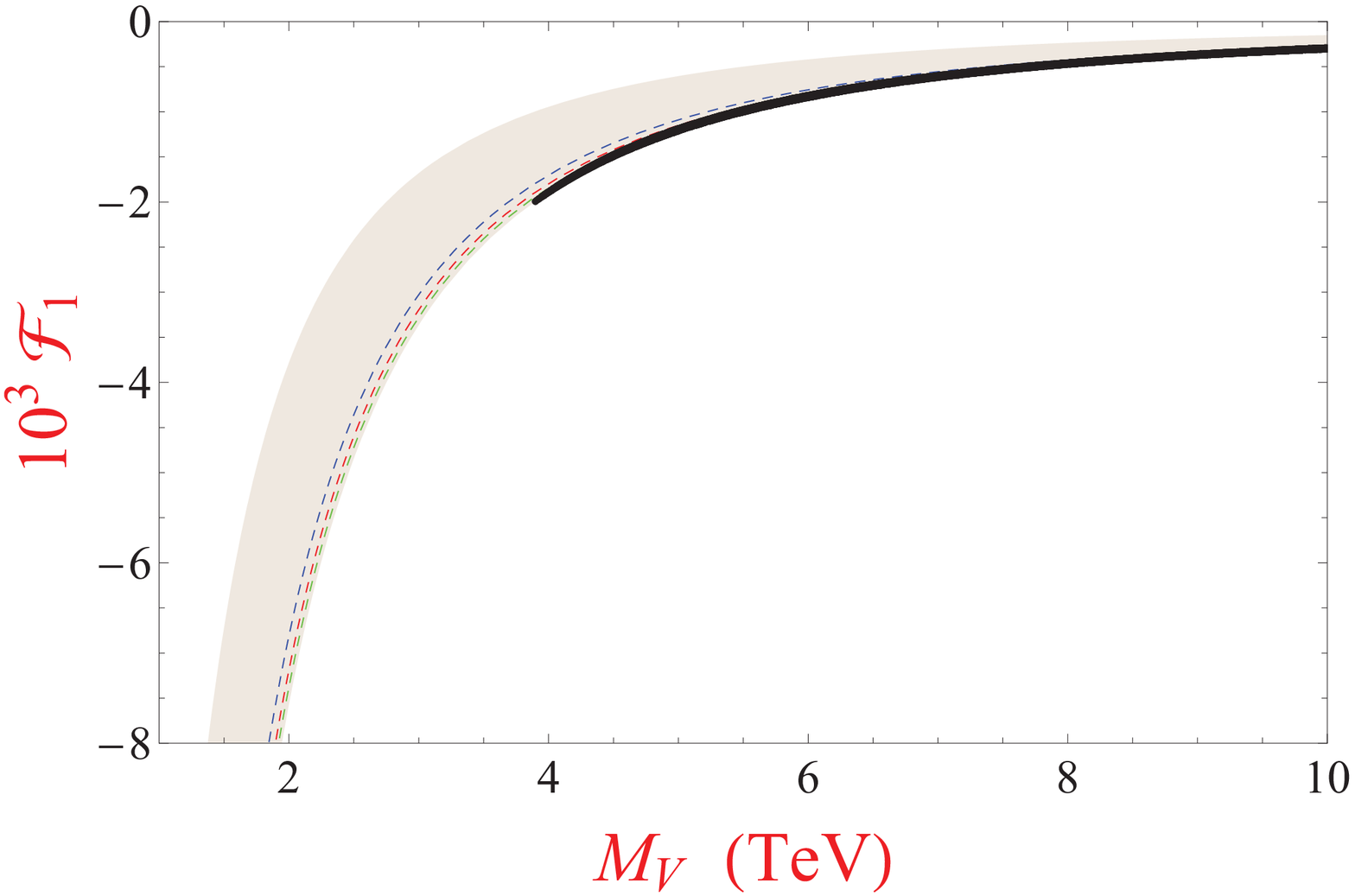}
\end{minipage}
\hskip .5cm
\begin{minipage}[c]{6cm}
\includegraphics[width=6cm]{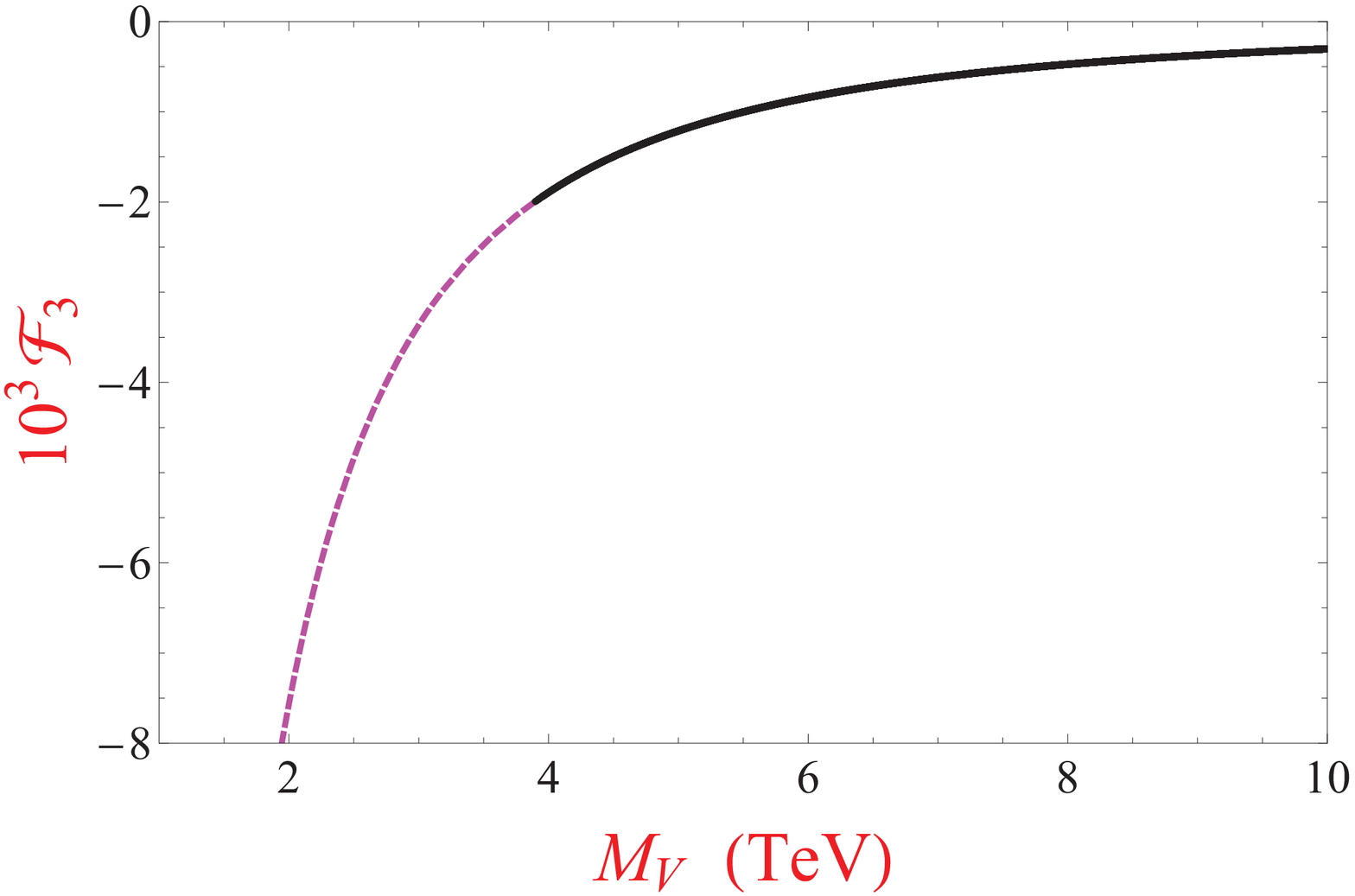}
\end{minipage}
\\[8pt]
\begin{minipage}[c]{6cm}
\includegraphics[width=6cm]{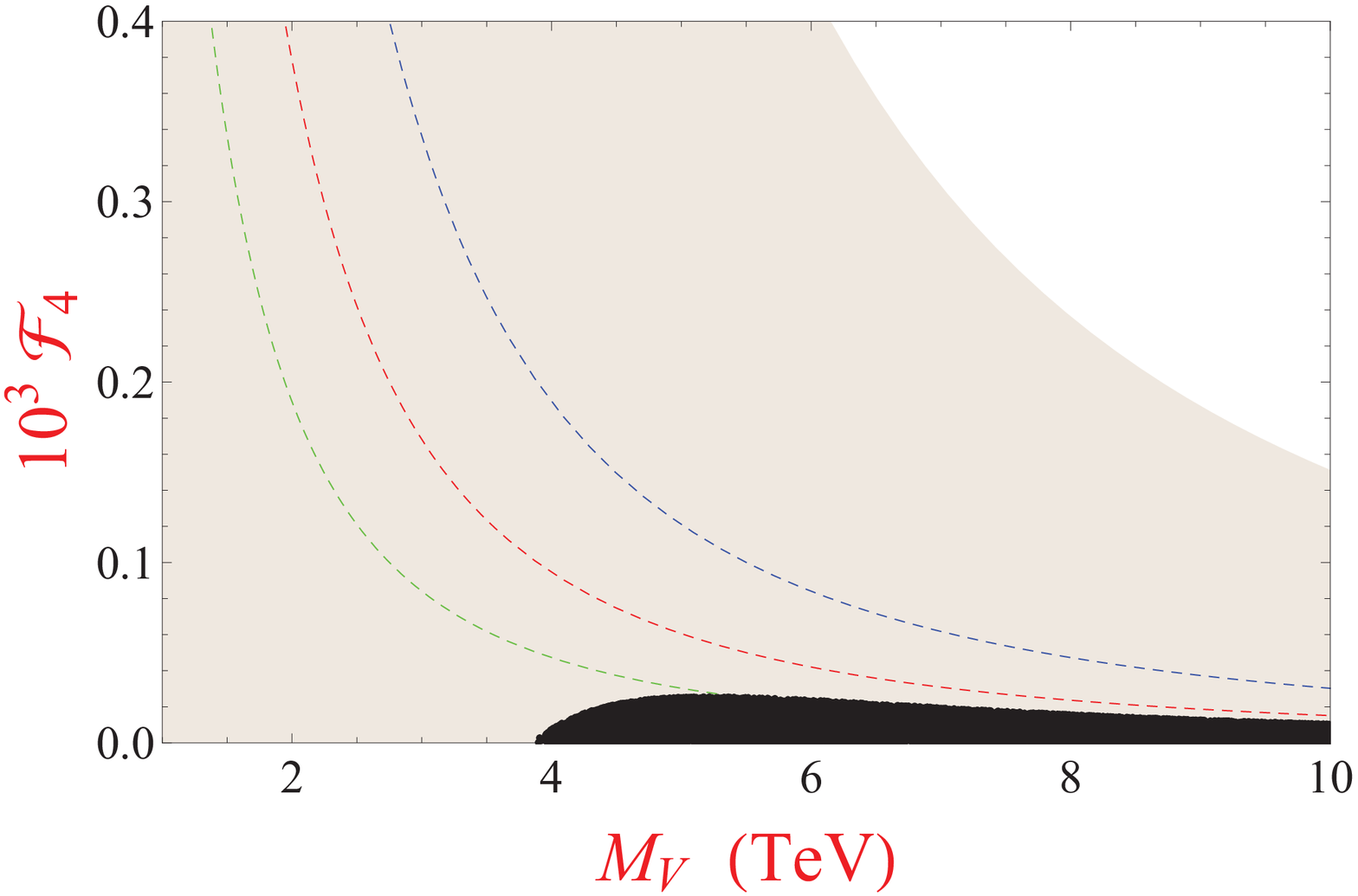}
\end{minipage}
\hskip .5cm
\begin{minipage}[c]{6cm}
\includegraphics[width=6cm]{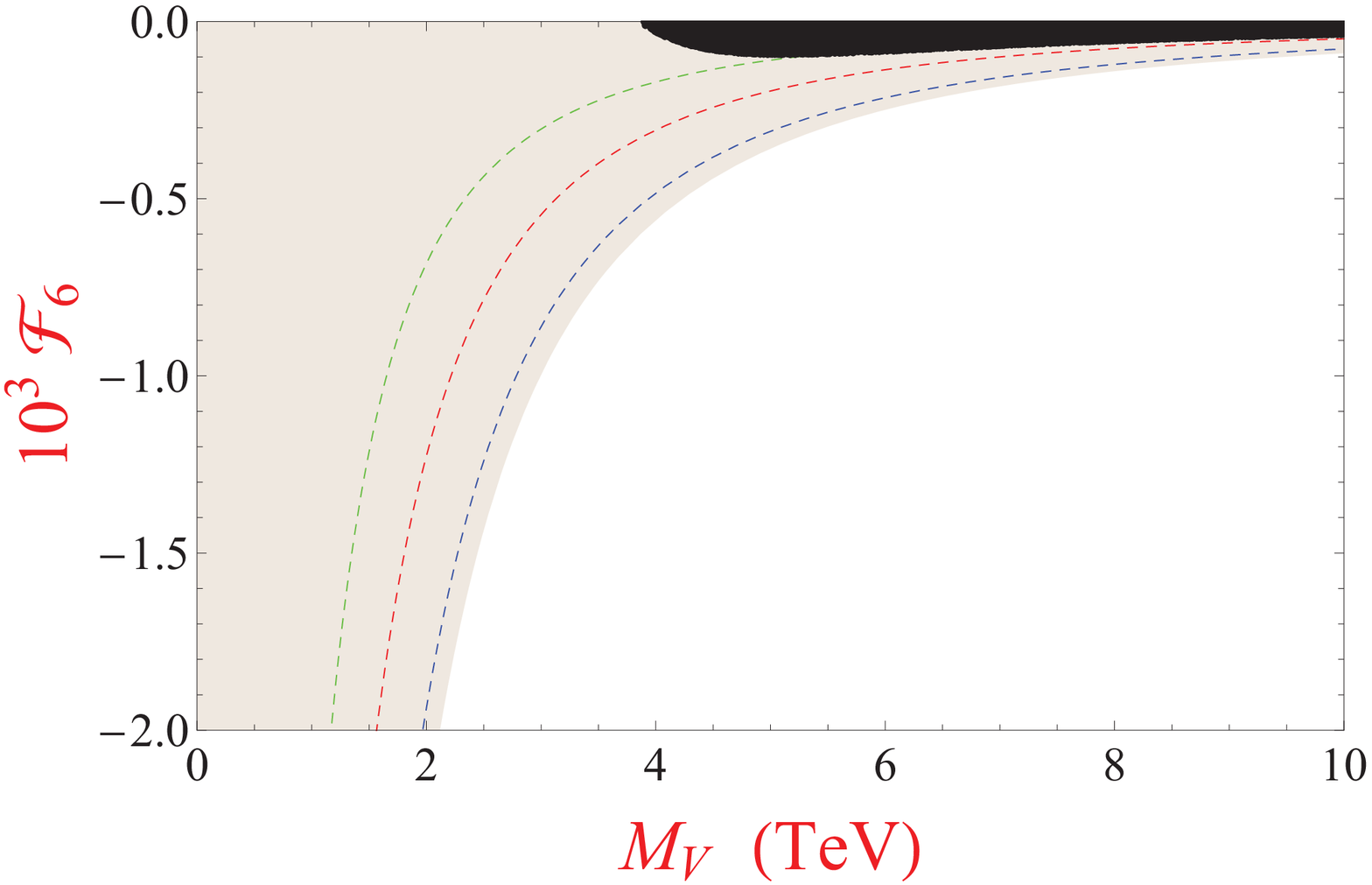}
\end{minipage}
\\[8pt]
\begin{minipage}[c]{6cm}
\includegraphics[width=6cm]{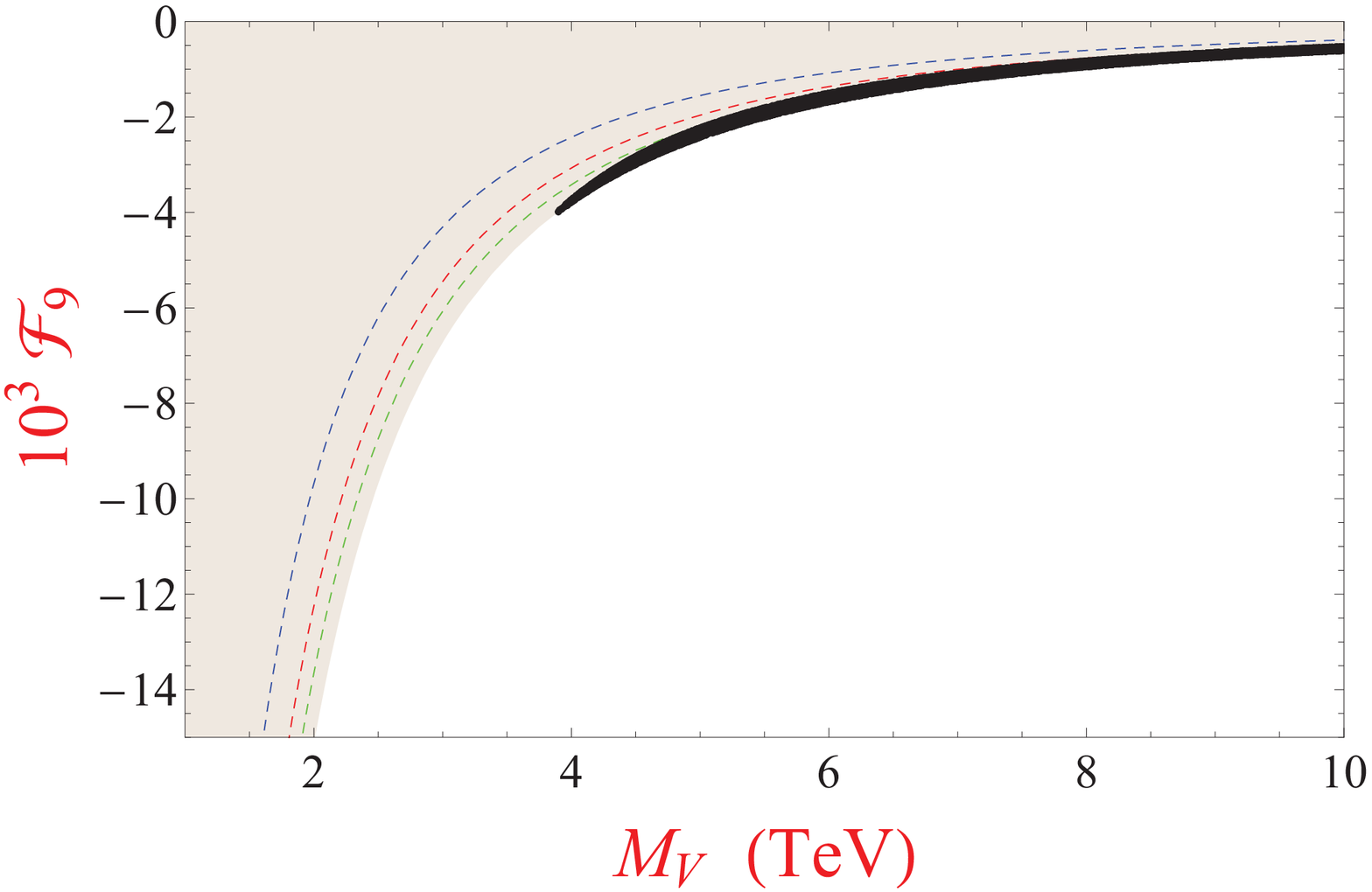}
\end{minipage}
\caption{Predicted values of the $\cO(p^4)$ LECs in asymptotically-free theories, as function of $M_V$ \cite{Pich:2015kwa}. The light-shaded regions cover all possible values with $M_A>M_V$, while the blue, red and green lines correspond to $M_V^2/M_A^2 = 0.8,\, 0.9$ and 0.95, respectively. $\cF_3$ does not depend on $M_A$. The $S$ and $T$ constraints restrict the allowed values to the dark areas.}
\label{fig:F1}
\vskip .6cm
\begin{minipage}[c]{6cm}
\includegraphics[width=6cm]{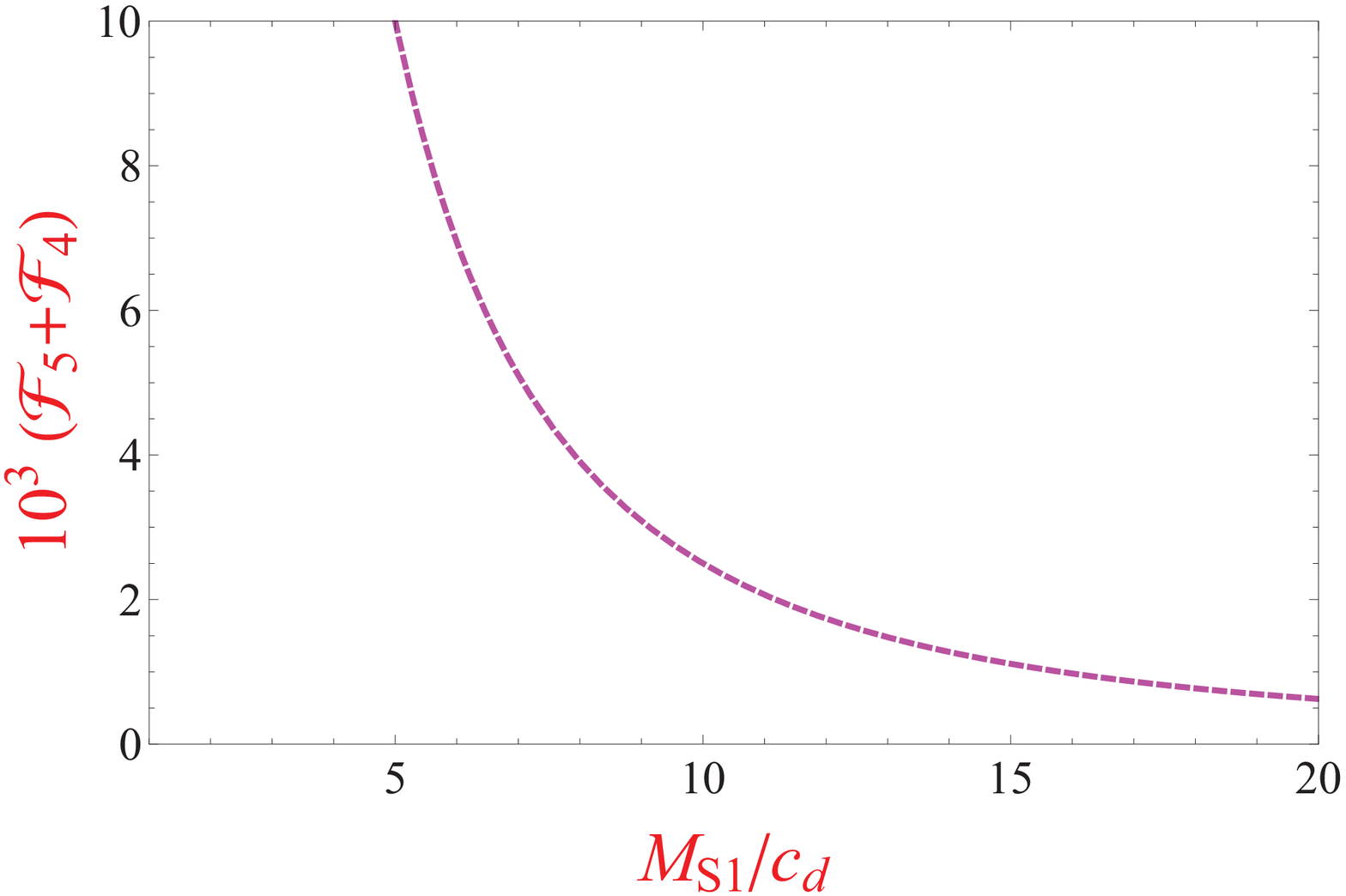}
\end{minipage}
\hskip .5cm
\begin{minipage}[c]{6cm}
\includegraphics[width=6cm]{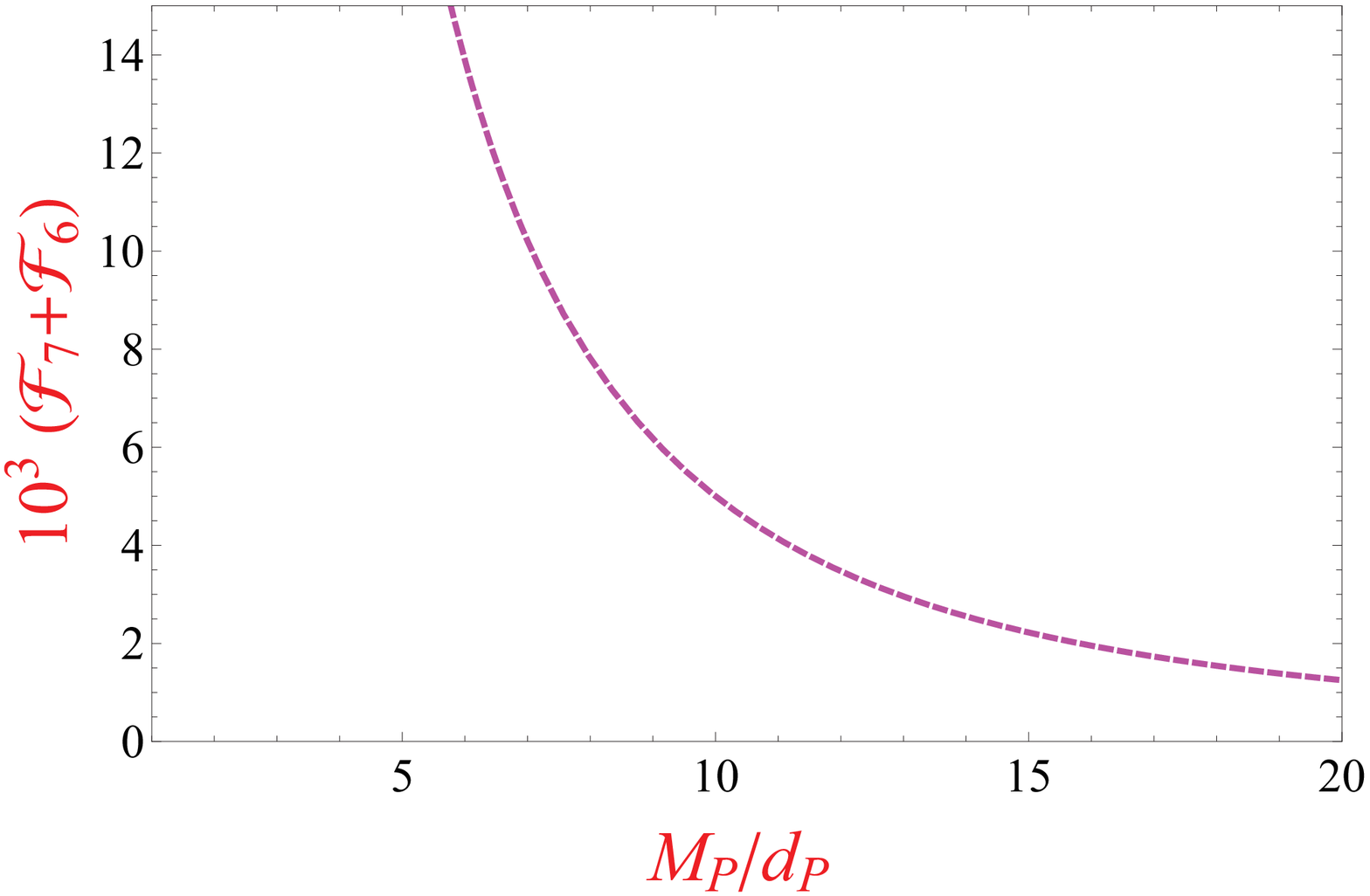}
\end{minipage}
\caption{Scalar and pseudoscalar contributions to $\cF_5$ and $\cF_7$, respectively \cite{Pich:2015kwa}.}
\label{fig:F2}
\end{center}
\end{figure*}

The numerical predictions for the different LECs $\cF_i\equiv \cF_{i,0}$, obtained with the short-distance constraints, are shown in Fig.~\ref{fig:F1} as functions of $M_V$. The light-shaded regions indicate all a priori possible values with $M_A > M_V$. The dashed blue, red and green lines correspond to $M_V^2/M_A^2 = 0.8,\, 0.9$ and 0.95, respectively. A single dashed purple curve is shown for $\cF_3$, which is independent of $M_A$.
The coupling $\cF_2$, which is not displayed in the figure, satisfies the inequality
\be 
\cF_2\, \le\, -\frac{v^2}{8 M_V^2}\, ,
\ee
but its absolute value cannot be bounded with the current information.

The scalar and pseudoscalar contributions can be isolated through the combinations $\cF_4+\cF_5$ and $\cF_6+\cF_7$ that depend only on the ratios $M_{S_1}/c_d$ and $M_P/d_P$, respectively.
The predicted values are shown in Fig.~\ref{fig:F2}.

\subsection{Gauge-boson self-energies}

\label{subsec:oblique}

The presence of massive resonance states coupled to the gauge bosons modifies the $Z$ and $W^\pm$ self-energies. The deviations with respect to the SM predictions are characterized through the so-called oblique parameters 
$S$ and $T$ and $U$~\cite{Kennedy:1990ib,Kennedy:1991sn,Peskin:1990zt,Peskin:1991sw} (or equivalently $\varepsilon_1$, $\varepsilon_2$ and $\varepsilon_3$ \cite{Altarelli:1990zd}), which are bounded by the electroweak precision data to the ranges~\cite{Haller:2018nnx}:
\bel{eq:oblique}
S = 0.04\pm 0.11\, ,
\qquad\qquad
T = 0.09\pm 0.14\, ,
\qquad\qquad
U = -0.02\pm 0.11\, .
\ee
$S$ measures the difference between the off-diagonal $W_3 B$ correlator and its SM value, while $T$ parametrizes the breaking of custodial symmetry, being related to the difference between the $W_3$ and $W^\pm$ self-energies. The parameter $U$ is not relevant for our discussion. These electroweak gauge self-ener\-gies are tightly constrained by the super-convergence properties of $\Pi_{LR}(t)$~\cite{Peskin:1990zt}. 

Al LO, the oblique parameter $T$ is identically zero, while $S$ receives tree-level contributions from vector and axial-vector exchanges \cite{Peskin:1991sw}:
\begin{equation}
S_{\mathrm{LO}}\, =\, -16\pi\,\cF_1\, =\,
4\pi \left( \frac{F_V^2}{M_V^2} - \frac{F_A^2}{M_A^2} \right)  
\; =\; \frac{4\pi v^2}{M_V^2}\,  \left( 1 + \frac{M_V^2}{M_A^2} \right)
\, .
\label{eq:S-LO}
\end{equation}
In the EWET, $S_{\mathrm{LO}}$ is governed by the LEC $\cF_1$. 
The prediction from $\cL_{\mathrm{EWR}\chi\mathrm{T}}^{\phantom{*}}$,
in terms of vector and axial-vector parameters, reproduces the value of $\cF_1$ in Table~\ref{tab:EW-LECs}. The expression on the right-hand side of \eqn{eq:S-LO} makes use of the two Weinberg sum rules \eqn{eq:WSRs}, which also imply $M_A>M_V$. Therefore, $S_{\rm   LO}$ is bounded 
to be in the range \cite{Pich:2012jv}
\begin{equation}
\frac{4\pi v^2}{M_V^2} \; < \; S_{\rm   LO} \; < \;   \frac{8 \pi v^2}{M_V^2} \, , \label{SLOtwoWSR}
\end{equation}
which puts a strong limit on the resonance mass scale: $M_V > 1.9$~TeV (90\% CL).

\begin{figure}[t]
\begin{center}
\includegraphics[width=.9\linewidth]{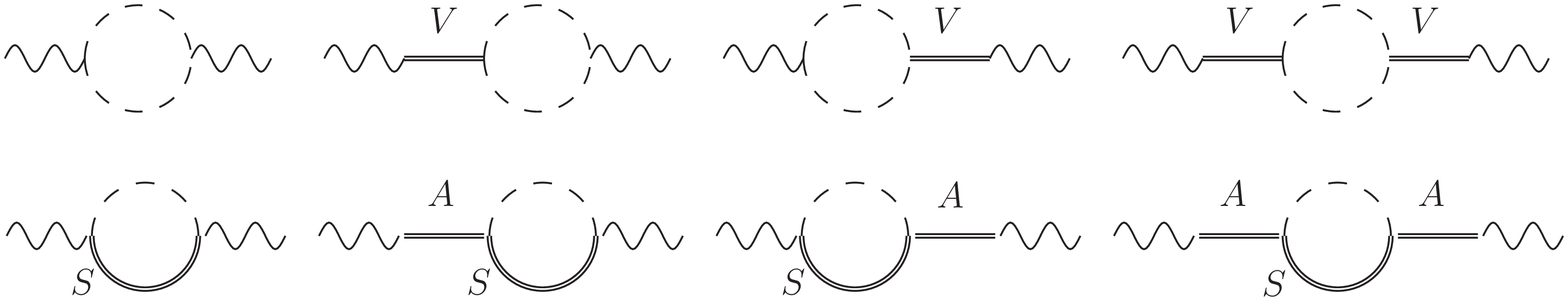}
\\[15pt]
\includegraphics[width=.7\linewidth]{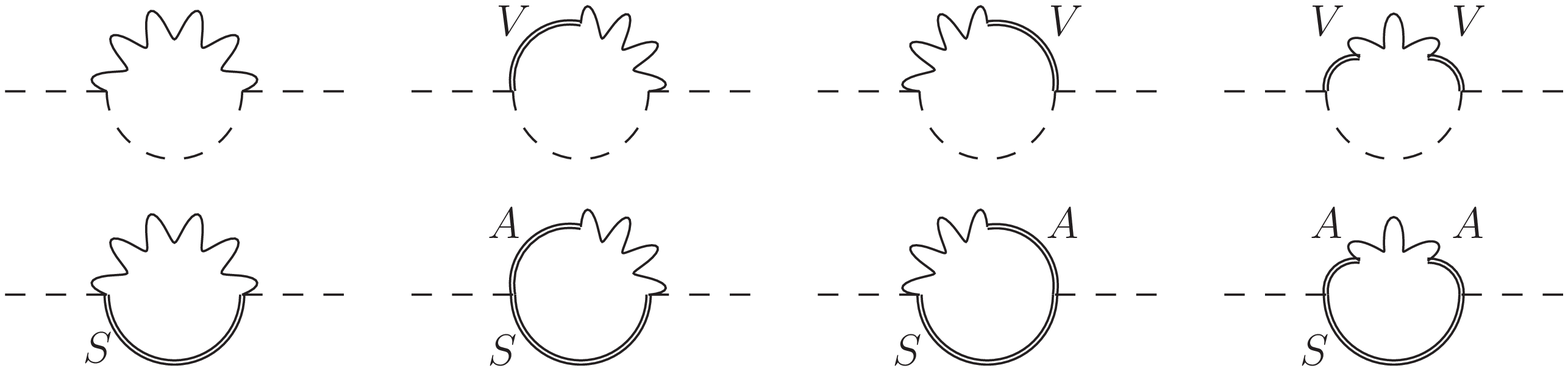}
\caption{NLO contributions to $S$ (first 2 lines) and $T$ (last 2 lines).
A dashed (double) line stands for a Nambu--Goldstone (resonance) boson and a wiggled line indicates a gauge boson.}
\label{fig:NLO_graphs}
\end{center}
\end{figure}

At NLO, $S$ and $T$ receive corrections from the one-loop diagrams displayed in Fig.~\ref{fig:NLO_graphs}. The calculation of $S$ can be simplified with the dispersive representation~\cite{Peskin:1991sw}
\begin{equation}
S\, =\, \frac{16 \pi}{g^2\tan\theta_W}\,
\int_0^\infty \, \frac{dt}{t} \; \left[ \rho_S(t) -  \rho_S(t)^{\rm SM} \right]\, ,
\end{equation}
where $\rho_S(t)$ is the spectral function of the $W_3 B$ correlator. Since $\rho_S(t)$ vanishes at short distances, the integral is convergent and, therefore, the dispersive relation does not require any subtractions. The relevant diagrams were computed in Refs.~\cite{Pich:2012dv,Pich:2013fea}, at LO in $g$ and $g'$. The dominant contributions originate from the lightest two-particle cuts, {\it i.e.}, two Nambu--Goldstone bosons or one Higgs plus one Nambu--Goldstone boson. The corrections generated by $\phi^a V$ and $\phi^a A$ intermediate states are suppressed by their higher mass thresholds.

Up to corrections of $\mathcal{O}(m_W^2/M_R^2)$, the parameter $T$ is related to the difference between the charged and neutral Nambu--Goldstone self-energies~\cite{Barbieri:1992dq}. Since the $SU(2)_L$ gauge coupling $g$ does not break custodial symmetry, the one-loop contributions to $T$ must involve the exchange of one $B$ boson. The dominant effects correspond again to the lowest two-particle cuts, {\it i.e.}, the $B$ boson plus one Nambu--Goldstone or one Higgs boson. Requiring the $W_3 B$ correlator to vanish at high energies implies also a good convergence of the Nambu--Goldstone self-energies, at least for the two-particle cuts considered. Therefore, their difference also obeys an unsubtracted dispersion relation, which enables to compute $T$ through the dispersive integral~\cite{Pich:2013fea}:
\be 
T \, =\, \frac{4 \pi}{g'^2 \cos^2\theta_W}\, \int_0^\infty \,\frac{dt}{t^2} \,
\left[  \rho_T(t)   -  \rho_T(t)^{\rm SM} \right] \, ,
\ee
with $\rho_T(t)$  the spectral function of the difference of the neutral and charged Nambu--Goldstone self-energies.

Neglecting terms of $\cO (m_h^2/M_{V,A}^2)$ and making use of the short-distance conditions derived previously, the one-loop calculation gives~\cite{Pich:2012dv,Pich:2013fea}
\begin{eqnarray}
S  &\, =\, &   4 \pi v^2 \left(\frac{1}{M_{V}^2}+\frac{1}{M_{A}^2}\right) + \frac{1}{12\pi} \left\{ \log\frac{M_V^2}{m_{h}^2}  -\frac{11}{6}
+\frac{M_V^2}{M_A^2}\log\frac{M_A^2}{M_V^2}
\quad\right.\nonumber  \\ &&\hskip 4.2cm\left.\mbox{}
- \frac{M_V^4}{M_A^4}\, \bigg(\log\frac{M_A^2}{m_h^2}-\frac{11}{6}\bigg) \right\} 
\label{eq.1+2WSR}
\end{eqnarray}
and~\cite{Pich:2013fea}
\begin{equation}
 T\; =\;  \frac{3}{16\pi \cos^2 \theta_W} \left\{ 1 + \log \frac{m_{h}^2}{M_V^2}
 - a^2 \left( 1 + \log \frac{m_{h}^2}{M_A^2} \right)  \right\}  \, .
\label{eq:T}
\end{equation}
%
\begin{figure}[t]
\begin{center}
\includegraphics[scale=0.6]{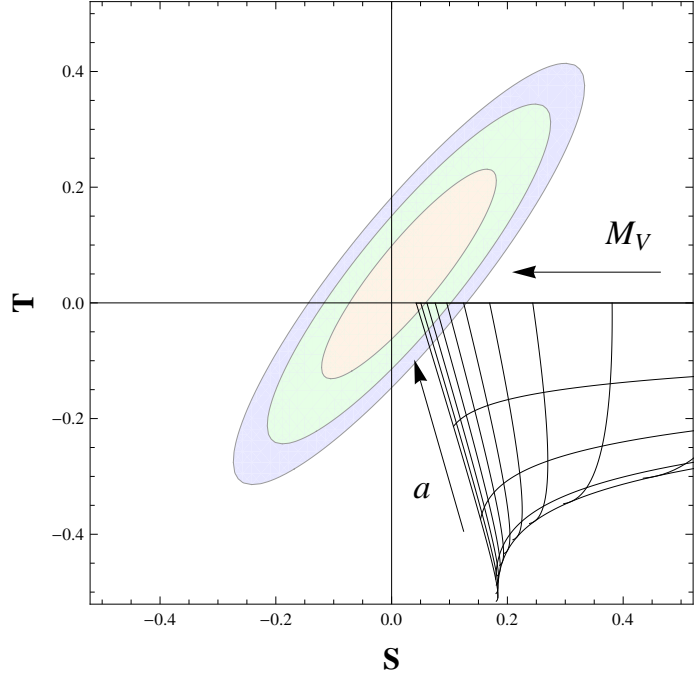}
\caption{NLO determinations of $S$ and $T$ in asymptotically-free theories~\cite{Pich:2012dv,Pich:2013fea}.
The approximately vertical lines correspond to constant values
of $M_V$, from $1.5$ to $6.0$~TeV at intervals of $0.5$~TeV.
The approximately horizontal curves have constant values
of $a = 0.00$, 0.25, 0.50, 0.75 and 1.00.
The arrows indicate the directions of growing  $M_V$ and $a$.
The ellipses give the experimentally allowed regions at 68\% (orange), 95\% (green) and 99\% (blue) CL.}
\label{fig:S-T_constraint}
\end{center}
\end{figure}
%
These NLO determinations are compared in Fig.~\ref{fig:S-T_constraint} with the experimental constraints. When $a= M_V^2/M_A^2 =1$, the parameter $T$ vanishes identically, as it should, while  the NLO prediction of $S$ reaches the LO upper bound in eqn~(\ref{SLOtwoWSR}). Smaller values of $a = M_V^2/M_A^2\to 0$ lead to increasingly negative values of $T$, outside its experimentally allowed range. Thus, the precision electroweak data
require that the Higgs-like scalar should have a $WW$ coupling very close to the SM one. At 95\% CL, one gets~\cite{Pich:2013fea}
%
\be 
0.94\, \le\, a\, \le\, 1\, ,
\ee
in nice agreement with the present LHC evidence \cite{Khachatryan:2016vau}, but much more restrictive. Moreover, the vector and axial-vector states should be very heavy (and quite degenerate)~\cite{Pich:2013fea}:
\be 
M_A\, \ge\, M_V\, >\, 4~\mathrm{TeV}\qquad (95\%\;\mathrm{C.L.})\, .
\ee

Combining these oblique constraints with the previous resonance-exchange predictions for the LECs of the EWET, one gets much stronger constraints on the $\cF_i$ couplings. At 95\% C.L., the allowed regions get reduced to the dark areas shown in Fig.~\ref{fig:F1}, which imply the limits~\cite{Pich:2013fea}:
\begin{displaymath}
-2\cdot 10^{-3}\, <\,\cF_1\, <\, 0 \, , 
\qquad\quad
0\, <\,\cF_4\, <\, 2.5\cdot 10^{-5} \, ,
\qquad\quad
-2\cdot 10^{-3}\, <\,\cF_3\, <\, 0 \, , 
\end{displaymath}
\bel{eq:Fi-limits}
-9\cdot 10^{-5}\, <\, \cF_6\, <\, 0\, , 
\qquad\qquad
-4\cdot 10^{-3}\, <\, \cF_9\, <\, 0\, .
\ee
These constraints are much more restrictive than the current direct bounds obtained from the Higgs signal strengths and anomalous gauge couplings~\cite{Buchalla:2015qju,deBlas:2018tjm,Falkowski:2015jaa,Falkowski:2016cxu}. The limits~\eqn{eq:Fi-limits} apply to any scenarios of new physics where the two Weinberg sum rules are satisfied, in particular in asymptotically-free theories. More generic limits that remain valid in models where only the first Weinberg sum rule is fulfilled have been derived in Ref.~\cite{Pich:2013fea}.

\section{Summary}
\label{sec:Summary}

EFT is a very adequate framework to describe the low-energy dynamics of Nambu--Goldstone degrees of freedom. They are massless fields, separated by an energy gap from the massive states of the theory, and their dynamics is highly constrained by the pattern of symmetry breaking. The effective Lagrangian can be organised as an expansion in powers of momenta over some characteristic scale associated with the spontaneous (or dynamical) symmetry breaking. The zero-order
term in this expansion vanishes identically because a constant shift of the Nambu--Goldstone coordinates amounts to a global rotation of the whole vacuum manifold that leaves the physics unchanged. Therefore, the Nambu--Goldstone bosons become free particles at zero momenta. Moreover, at the leading non-trivial order in the momentum expansion, the symmetry constraints are so restrictive that all the Nambu--Goldstone interactions are governed by a very low number of parameters, leading to a very predictive theoretical formalism.

In these lectures, we have concentrated the discussion on two particular applications of high relevance for fundamental physics: QCD and the electroweak theory. While the dynamical content of these two quantum field theories is very different, they share the same pattern of symmetry breaking (when only two light flavours are considered in QCD). Therefore, the low-energy dynamics of their corresponding Nambu--Goldstone modes is formally identical in the limit where explicit symmetry-breaking contributions are absent.

The low-energy EFT of QCD, $\chi$PT, has been thoroughly studied for many years and current phenomenological applications have reached a two-loop accuracy. The main limitation is the large number of LECs that appear at $\cO(p^6)$. These LECs encode the short-distance information from the underlying QCD theory and are, in principle, calculable functions of the strong coupling and the heavy quark masses. Owing to the non-perturbative nature of the strong interaction at low energies, the actual calculation is, however, a formidable task. Nevertheless a quite good dynamical understanding of these LECs has been already achieved through large-$N_C$ methods and numerical lattice simulations.

The EWET contains a much richer variety of ingredients, such as fermions and gauge symmetries, and a more involved set of explicit symmetry-breaking effects. Therefore, the effective Lagrangian has a more complex structure and a larger amount of LECs. The number of unknown couplings blows when the flavour quantum numbers are included, clearly indicating our current ignorance about the fundamental flavour dynamics. The electroweak effective Lagrangian parametrizes the low-energy effects of any short-distance dynamics compatible the SM symmetries and the assumed pattern of EWSB. The crucial difference with QCD is that the true underlying electroweak theory is unknown.

In the absence of direct discoveries of new particles, the only accessible
signals of the high-energy dynamics are hidden in the LECs of the EWET. High-precision measurements of scattering amplitudes among the known particles are sensitive to these LECs and could provide indirect evidence for new phenomena.
The pattern of LECs emerging from the experimental data could then give us precious hints about the nature of the unknown ultraviolet theory responsible for any observed ``anomalies''.

\section*{Acknowledgements}

I would like to thank the organisers for the charming atmosphere of this school and all the school participants for their many interesting questions and comments.
This work has been supported in part by the Spanish Government and ERDF funds from the EU Commission [Grants FPA2017-84445-P and FPA2014-53631-C2-1-P], by Generalitat Valenciana [Grant Prometeo/2017/053] and by the Spanish
Centro de Excelencia Severo Ochoa Programme [Grant SEV-2014-0398].

\begin{exercises}
\begin{enumerate}

\item Prove the Current-Algebra commutation relations in eqn~\eqn{eq:commutation_relations}.

\item
The quadratic mass term of the $\cO(p^2)$ $\chi$PT Lagrangian generates a small mixing between the $\phi_3$ and $\phi_8$ fields, proportional to the quark mass difference $\Delta m \equiv m_d-m_u$ (see footnote~\ref{foot:PiEtaMixing}). 

a) Diagonalize the neutral meson mass matrix and find out the correct $\pi^0$ and $\eta_{\raisebox{-1pt}{$\scriptstyle 8$}}$ mass eigenstates and their masses.

b) When isospin is conserved, Bose symmetry forbids the transition $\eta\to\pi^0\pi^+\pi^-$ (why?). Compute the decay amplitude to first-order in $\Delta m$.

\item 
a) Compute the axial current at $\cO(p^2)$ in $\chi$PT and check that $F_\pi = F$ at this order.

b) Expand the $\cO(p^2)$ axial current to $\cO(\Phi^3)$ and compute the 1-loop corrections to $F_\pi$. Remember to include the pion wave-function renormalization. 

c) Find the tree-level contribution of the $\cO(p^4)$ $\chi$PT Lagrangian to the axial current. Renormalize the UV loop divergences with the $\cO(p^4)$ LECs and obtain the $\cO(p^4)$ expression for $F_\pi$ in eqn~\eqn{eq:f_meson}.

\item
Compute the elastic scattering amplitude of two electroweak Nambu--Goldstone bosons at LO and obtain the $\cO(p^2)$ term in eqn~\eqn{eq:WWscattering-NLO}.

\item
Consider the two-point correlation function of a left-handed and a right-handed vector currents, either in R$\chi$T or the EWET with massive states (EWR$\chi$T), and compute the LO (tree-level) contributions. 

a) Obtain the result for $\Pi_{LR}(t)$ in eqn~\eqn{eq:WSR}. 

b) Demonstrate that in QCD $\Pi_{LR}(t)\sim 1/t^3$ at $t\to\infty$.

c) Derive the Weinberg sum rules in eqn~\eqn{eq:SD3}.

\item
Assume the existence of a hypothetical light Higgs which couples to quarks with the Yukawa interaction
\begin{center}
$\displaystyle
\cL_{h^0\bar q q}\; =\; -\frac{h^0}{v}\;\sum_q k_q\, m_q\, \bar q q\, .$
\end{center}

a) Determine at LO in $\chi$PT the effective Lagrangian describing the Higgs coupling to pseudoscalar mesons induced by  the light-quark Yukawas.

b) Determine the effective $h^0 G_a^{\mu\nu} G_{\mu\nu}^a$ coupling induced by heavy quark loops.

c) The $G_a^{\mu\nu} G_{\mu\nu}^a$ operator can be related to the trace of the energy-momentum tensor, in the 3-flavour QCD theory:
\begin{center}
$\displaystyle
\Theta^\mu_\mu \; =\; \frac{\beta_1 \alpha_s}{4\pi}\, G_a^{\mu\nu} G_{\mu\nu}^a
+ \bar q \cM q\, ,$
\end{center}
where $\beta_1 =-\frac{9}{2}$ \ is the first coefficient of the $\beta$ function.
Using this relation, determine the LO $\chi$PT Lagrangian incorporating the Higgs coupling to pseudoscalar mesons induced by the heavy-quark Yukawas.

d) Compute the decay amplitudes $h^0\to 2\pi$ and $\eta\to h^0\pi^0$. 

\end{enumerate}
\end{exercises}

\medskip
\bibliographystyle{OUPnum_notitle}
\bibliography{bibtexfile_pich}
%

\end{document}